\documentclass[11pt]{amsart}
\usepackage{setspace}
\usepackage{amsmath,amstext,amssymb,amsthm}
\usepackage{eurosym}
\usepackage{graphicx}
\usepackage{mathrsfs}
\usepackage{xcolor}
\usepackage{hyperref}
\usepackage{enumerate}   
\allowdisplaybreaks
\usepackage{geometry}
\usepackage{multirow}
\usepackage{multicol}
\usepackage{listings}
\usepackage[font=footnotesize]{caption}

\newtheorem{theorem}{Theorem}[section]

\newtheorem{proposition}[theorem]{Proposition}
\theoremstyle{definition}
\newtheorem{remark}[theorem]{Remark}

\newcommand{\griglia}{\mathbb{A}\times\mathbb{B}}
\newcommand{\im}{\mathrm{i}}

\title[Valuation of general GMWB annuities in a low interest rate environment]{Valuation of general GMWB annuities \\ in a low interest rate environment}

\author[C. Fontana]{Claudio Fontana}
\address{Department of Mathematics ``Tullio Levi - Civita'', University of Padova, Italy.}
\email{fontana@math.unipd.it}
\author[F. Rotondi]{Francesco Rotondi}
\address{Department of Mathematics ``Tullio Levi - Civita'', University of Padova, Italy.}
\email{francesco.rotondi@math.unipd.it}

\date{\today} 
\thanks{
The authors are thankful to an Editor and to two Reviewers for constructive feedback that helped to improve the paper and to Anna Rita Bacinello for useful discussions on the topic of the paper.
Financial support from the Europlace Institute of Finance (project ``Interest Rate Term Structures in a Low-Rate Environment'') and the University of Padova (research programmes BIRD190200/19 and STARS StG PRISMA - ``Probabilistic Methods for Information in Security Markets'') is gratefully acknowledged.}

\begin{document}

\begin{abstract}
Variable annuities with Guaranteed Minimum Withdrawal Benefits (GMWB) entitle the policy holder to periodic withdrawals together with a terminal payoff linked to the performance of an equity fund. In this paper, we consider the valuation of a general class of GMWB annuities, allowing for step-up, bonus and surrender features, taking also into account mortality risk and death benefits. When dynamic withdrawals are allowed, the valuation of GMWB annuities leads to a stochastic optimal control problem, which we address here by dynamic programming techniques. Adopting a Hull-White interest rate model, correlated with the equity fund, we propose an efficient tree-based algorithm. We perform a thorough analysis of the determinants of the market value of GMWB annuities and of the optimal withdrawal strategies. In particular, we study the impact of a low/negative interest rate environment. Our findings indicate that low/negative rates profoundly affect the optimal withdrawal behaviour and, in combination with step-up and bonus features, increase significantly the fair values of GMWB annuities, which can only be compensated by large management fees. 
\end{abstract}

\maketitle

\section{Introduction}
Variable annuities are widespread financial products sold by insurance companies to retail investors to complement their retirement plans. Along with the reimbursement of the nominal amount, a variable annuity provides returns linked to the performance of an underlying market index or fund. As the holder of a variable annuity is not directly exposed to the financial market while still benefiting from favorable movements of the underlying, these products are among the most popular types of annuities. According to the Secure Retirement Institute, variable annuities accounted for  45\% of the annuities sold in the US in 2020, corresponding to a total market value of \$98.8 billions.\footnote{See the 2020 U.S. Annuity Sales Report available at \url{https://www.limra.com/siteassets/newsroom/fact-tank/sales-data/2020/q4/final-4q-2020-annuity-sales-estimates-vfinal.pdf}.}

Variable annuities often carry additional guarantees, in the form of contractual riders that provide additional benefits to the holder (see \cite{BacinelloMillossovichMontealegre2016,ShevchenkoLuo2016} and \cite[Section 7.10]{OlivieriPitacco2015} for a complete overview of the possible features of variable annuities). 
In this paper, we focus on {\em Guaranteed Minimum Withdrawal Benefit} (GMWB) annuities. Typically, a GMWB annuity requires an initial premium paid as a lump sum by the policy holder (PH hereafter), which is then invested in an underlying equity index or fund. The annuity is characterized by two accounts: the investment account that tracks the movements of the underlying fund (deducting a management fee) and the benefit account, from which the PH is entitled to periodic withdrawals. Each withdrawal reduces the values of both accounts and is subject to a withdrawal penalty, unless it is lower than a fixed guaranteed amount. At inception, the values of the two accounts coincide with the initial premium paid by the PH, while at maturity (which usually ranges from 5 to 20 years) the PH receives the maximum of the residual values of the two accounts.

As far as the periodic withdrawals are concerned, one may consider  {\em static withdrawal strategies}, if the PH constantly withdraws the guaranteed amount, or {\em dynamic withdrawal strategies}, if the PH optimally determines her withdrawals. 
The case of static withdrawals has been studied in the seminal paper \cite{MilevskySalisbury2006}, also allowing for the possibility of surrender before maturity. 
In the case of dynamic withdrawals, the valuation of a GMWB annuity leads to a stochastic optimal control problem, as first considered in \cite{ChenForsyth2008} and \cite{DaiKwokZong2008} in the context of impulse stochastic control and singular stochastic control, respectively. These two pioneering works assume that the PH makes continuous withdrawals from the benefit account, in order to make easier the solution of the ensuing continuous-time stochastic optimal control problems. In this paper, in line with the more recent literature, we shall assume more realistically that the PH can withdraw from the benefit account only at predetermined dates. 
Due to the similarities with American options, four kinds of methods are commonly used for the valuation of GMWB annuities: PDE approaches (see, e.g., \cite{BauerKlingRuss2008,ChenVetzalForsyth2008,DonnellyJaimungalRubisov2014,GudkovIgnatievaZiveyi2019}), Monte Carlo methods inspired by the seminal work of \cite{LongstaffSchwartz2001} (see, e.g., \cite{BacinelloMillossovichOlivieriPitacco2011,HuangKong2016}), lattice methods (see, e.g., \cite{DaiYangLiu2015,Costabile2017,DongXuXwok2019}) and Fourier-based techniques (see, e.g., \cite{AlonsoGarciaWoodZiveyi2018,IgnatievaSongZiveyi2018}). Some of these approaches have been recently combined in \cite{GoudenegeMolentZanette2019}, where efficient hybrid numerical methods for the valuation of GMWB annuities are proposed.
We refer the reader to \cite{ShevchenkoLuo2016} for a comprehensive overview of numerical methods for the pricing of variable annuities with guarantees.

The contribution of this paper is threefold. First, we study a general type of GMWB annuity, including most of the features that are described in the literature. On top of standard features (such as a payoff in the case of death of the PH and the possibility of surrendering the contract), we consider additional features that have been jointly considered only in \cite{ChenForsyth2008} within a purely log-normal setting: a {\em step-up} feature and a {\em bonus} feature. The main consequence of these two features is that they might increase the value of the benefit account over the life of the annuity. Indeed, without step-up and bonus features, the benefit account is a non-increasing process, thus simplifying the valuation problem. On the contrary, due to the step-up feature, the benefit account is periodically matched to the investment account when the latter has a greater value, while the bonus feature rewards the PH is she decides not to withdraw from the benefit account. 
We show that the inclusion of step-up and bonus features modifies significantly the optimal withdrawal decisions, especially in a market environment characterized by low interest rates. 
To the best of our knowledge, this is the first work proposing a unified treatment of the different features that can affect the value of GMWB annuities, while allowing for dynamic withdrawals.\footnote{We mention that general frameworks for variable annuities with guarantees have been proposed in \cite{BauerKlingRuss2008} and \cite{BacinelloMillossovichOlivieriPitacco2011}. However, the first work considers a very specific type of step-up feature, which is more restrictive than ours, and presents valuation results only in a constant interest rate setting. On the other hand, \cite{BacinelloMillossovichOlivieriPitacco2011} do not allow for a bonus feature. Moreover, both in \cite{BauerKlingRuss2008} and in \cite{BacinelloMillossovichOlivieriPitacco2011} the case of fully dynamic withdrawals is only discussed as a theoretical possibility and is not investigated numerically.}

Second, in a stochastic interest rate setting and allowing for dynamic withdrawals, we propose a general valuation algorithm for GMWB annuities, based on a dynamic programming approach. Although our model involves two correlated risk factors, representing market and interest rate risks, the algorithm exploits a dimensionality reduction and is entirely based on the interest rate process. The algorithm is proved to be stable and reliable, achieves an accuracy comparable to competing methods proposed in the recent literature and can be easily applied to richer models.

Finally, we perform a thorough numerical analysis of the determinants of the market value of GMWB annuities and their optimal withdrawal strategies. In particular, we focus on the impact of a low/negative interest rate environment (as of 12/31/2021), which has characterized financial markets for several years in the recent past. We demonstrate that in this market scenario GMWB annuities can be sold at par only at the expense of very large management fees and withdrawal penalties, especially when additional features are included. Moreover, we show that low/negative interest rates affect significantly the optimal withdrawal decisions, in line with the results recently established in \cite{BattauzRotondi2021} for American equity options. The impact of low/negative interest rates is particularly striking when compared to the current market scenario (as of 12/30/2022), characterized by positive and increasing interest rates, as a consequence of the recent changes in the monetary policy. In this second interest rate scenario, the valuation and the optimal withdrawal strategies of GMWB annuities radically change and become in line with previous findings reported in the literature. Therefore, for the first time in the literature, our work sheds light on the consequences of low/negative rates on GMWB annuities.

We point out that more specific types of GMWB annuities have been already analyzed in a stochastic interest rate setting, see \cite{PengLeungKwok2012,DaiYangLiu2015,ShevchenkoLuo2017,IgnatievaSongZiveyi2018,GoudenegeMolentZanette2019,GoudenegeMolentZanette2021,DeAngelisDeMarchisMartireRusso2022}. 
In particular, from the modelling viewpoint, some of these works rely on a Hull-White model correlated with the underlying fund, as considered in the present paper.
However, apart from studying less general types of annuities and proposing different valuation algorithms, all these works (even the most recent contributions) do not analyse an interest rate setting that corresponds to a low/negative rate environment, considering instead interest rates floating around 2-5\%. As we document in the present paper, low/negative rates have a sizable impact on the valuation of GMWB annuities, especially when one takes into account the mid to long maturities of such contracts and possible additional features. From the viewpoint of insurance companies, this raises issues on the possibility of selling GMWB annuities with acceptable management fees, particularly when several features are included in the contract.

The paper is structured as follows. In Section \ref{sec:ValuationFramework}, we describe a general GMWB annuity, we present the stochastic model and frame the valuation problem into a dynamic programming one. Section \ref{sec:numericalAlgorithm} describes the numerical algorithm to solve the valuation problem. Section \ref{sec:NumericalAnalysis} contains all  numerical results, the analysis of the determinants of the market value of GMWB annuities, the sensitivity analysis and a description of optimal withdrawal strategies in two different interest rate scenarios calibrated to market data. Section \ref{sec:conclusion} concludes the paper, while the Appendix contains some details on the discretization of the interest rate process and a technical proof.

\section{General GMWB annuities and valuation framework}\label{sec:ValuationFramework}

We first describe the structure and the features of a general GMWB annuity (Section \ref{sec:theGMWB}). The market model is then presented in Section \ref{subsec:theMarketModel}, while in Section \ref{subsec:dynamicProgramming} the valuation of a general GMWB annuity is framed as a dynamic programming problem.

\subsection{The GMWB annuity}\label{sec:theGMWB}

We consider a general type of GMWB annuity, taking into account most of the features described in the survey \cite{ShevchenkoLuo2016}. The specification of a standard GMWB annuity goes back to the works of \cite{ChenForsyth2008,MilevskySalisbury2006} and has been more recently considered in \cite{DaiYangLiu2015,DongXuXwok2019,GoudenegeMolentZanette2019,GudkovIgnatievaZiveyi2019} among others.

When investing in a GMWB annuity at time $t=0$, the PH pays an initial premium $P$ to the insurance company as a lump sum. This amount is then invested in a fund, typically selected by the PH among a set of possible funds. Let $(S_t)_{t \geq 0}$ be the strictly positive price process of the chosen fund.
We let $\mathcal{T} := \{ 1, \ldots , N\}$ be a set of $N$ periodic anniversary dates, with $N=T$ corresponding to the maturity of the contract.\footnote{For simplicity of presentation and without loss of generality, we assume yearly anniversary dates $n=1,\ldots,N$.}
Similarly as in \cite{ShevchenkoLuo2016}, for each $n=1,\ldots,N$, we denote by $n^-$ (resp. $n^+$) the time instant just before (resp. after) a withdrawal at date $n$.

Every GMWB annuity is characterized by two accounts: the \emph{investment} (or \emph{primary}) account, whose value process is denoted by $(A_t)_{t \geq 0}$, and the \emph{benefit} (or \emph{secondary}) account, whose value process is denoted by $(B_t)_{t \geq 0}$.
The investment account evolves according to the return of the underlying fund, with the deduction of a proportional fee $\alpha$ (\emph{management fee})\footnote{Under some contract specifications, the management fee depends on the account value. For example, a fee can be deducted only if the value of the investment account is below a certain threshold (see, e.g., \cite{BacinelloZoccolan2019,BernardHardyMackay2014,Delong2014}). The possible extension of our valuation framework to state-dependent fees will be discussed below in Remark \ref{rem:algoWSU}.}:
\[ 
\mathrm{d} A_t = \frac{A_t}{S_t} \mathrm{d} S_t - \alpha A_t \mathrm{d}t,
\qquad A_0=P.
\]

The benefit account remains constant between any two consecutive anniversary dates (i.e., $B_{n^+} = B_{(n+1)^-}$, for all $n = 0, \ldots, N-1$) and at $t=0$ is set equal to the initial premium paid by the PH, so that $B_0 = P$. Withdraw/bonus/step-up events, which are assumed to take place at anniversary dates only, modify the value of the benefit account as described below.

At each anniversary date $n\in\mathcal{T}$ (maturity $T$ included), the PH can withdraw some funds from the benefit account (\emph{withdrawal event}). If the  withdrawal does not exceed a predetermined threshold $G$ (called \emph{guaranteed minimum benefit} and usually defined as $G=P/N$), no withdrawal penalty applies. On the contrary, if the PH withdraws more than $G$, a (time-dependent) penalty $\beta_n$ applies to the withdrawn amount in excess of $G$. Therefore, denoting by $w_n$ the withdrawal decision of the PH at $n$, the net cashflow $X_n(w_n)$ received at $n$ is given by
\[
X_n(w_n) := 
\begin{cases}
w_n, & \text{ if }w_n\leq G, \\
w_n -\beta_{n}(w_n - G), & \text{ if }w_n>G,
\end{cases}
\]
for $w_n \in [0,B_{n^-}]$, since the withdrawal cannot exceed the residual value of the benefit account. If the PH withdraws $w_n$, the values of  the investment and the benefit accounts decrease accordingly. For the benefit account, it holds that
\begin{equation}\label{eq:benefit_std}
B_{n^+} = B_{n^-} - w_n. 
\end{equation}
Since $w_n \in [0,B_{n^-}]$, the benefit account is always bounded from below by zero. On the contrary, after possible bad performances of the underlying fund, it may happen that $A_{n^-}<w_n \leq B_{n^-}$. In this case, a floor at zero is assumed for the investment account. Therefore, it holds that
\begin{equation}\label{eq:investment_std}
A_{n^+} = ( A_{n^-} - w_n )^+.
\end{equation}

In addition, we assume that at each anniversary date (before any other event takes place) the PH can fully surrender the contract (\emph{surrender event}) and receive the greater of the current values of the two accounts (with a penalty if the benefit account has a greater value). If the surrender decision takes place at date $n$, then the last cashflow received by the PH at $n$ is
\begin{equation}\label{eqn:fullSurrenderCashflow}
    X^S_n := \max \bigl\{ A_{n^-}; B_{n^-} -\beta_n ( B_{n^-} - G )^+ \bigr\}.
\end{equation}
We denote by $\tau^S$ the anniversary date at which the PH decides to surrender the contract.

If at an anniversary date $n\in\mathcal{T}$ the PH makes no withdrawals from the benefit account, a \emph{bonus} might be credited to it (\emph{bonus event}). In this case, $B_{n^+} = B_{n^-} +b$, where $b \geq 0$ is a deterministic quantity, usually specified as a percentage of the amount $P$.

\begin{remark}[Withdrawal strategies]
\label{rem:withdrawal}
In the following, we will consider three benchmark withdrawal strategies: a \emph{static withdrawal strategy} (S scheme), where the PH withdraws always the guaranteed amount $G$ at every anniversary date; a \emph{static withdrawal strategy with surrender} (S+S scheme), where at each anniversary date the possibility of surrendering the contract is added to the static withdrawal scheme; a \emph{dynamic withdrawal strategy} (D scheme), where at each anniversary date the PH optimally determines her withdrawal and the surrender decision by maximising expected future discounted payoffs. As pointed out in \cite{ShevchenkoLuo2016}, for a GMWB annuity the optimal dynamic withdrawal strategy does not necessarily coincide with the static one, nor it is of ``bang-bang'' type (as it is the case for GLWB annuities, see \cite{AzimzadehForsyth2015,BacinelloMaggistroZoccolan2022}). This fact will be confirmed by our analysis of optimal withdrawal strategies in Section \ref{subsec:optimalWithdrawalSection}.
\end{remark}

In this work, we also allow for the possibility of a \emph{step-up} or \emph{ratchet} feature (\emph{step-up event}): at each anniversary date, the benefit account value is matched to the investment account, if the latter has a higher value. Depending on the specification of the contract, the step-up event can take place either before or after the withdraw/bonus event (see \cite[Section 4.1]{ShevchenkoLuo2016}):
\begin{itemize}
    \item[(W-SU):] {\em withdraw/bonus first and then step-up}. In this case, it holds that
    $$ B_{n^+} = \max \left\{ \left( A_{n^-} - w_n \right)^+; B_{n^-} + b \mathbf{1}_{ \{ w_{n}=0\} } - w_n \right\};  $$
    \item[(SU-W):] {\em step-up first and then withdraw/bonus}. In this case, it holds that
\[
        B_{n^+} = \max \left\{ A_{n^-}; B_{n^-} \right\} + b \mathbf{1}_{ \{ w_{n}=0\} } - w_n.
\]
\end{itemize}

At maturity, after the last withdrawal $w_N$, the PH receives the maximum between the residual value of the benefit account (with penalty $\beta_N$) and the terminal value of the investment account, i.e., $\max\{ A_{N^+}, ( 1 -\beta_N )B_{N^+} \}$. According to the scheme (W-SU), assuming that no bonus is credited at maturity and considering the sum of the last withdrawal cashflow $X_N(w_N)$ and the payoff at maturity, the cashflow at $T=N$ can be represented as
\begin{align*}
&\max_{w_N \in [ 0, B_{N^-} ]} \Bigl[ w_N -\beta_N ( w_N - G )^+\Bigr. \\
&\qquad\qquad\qquad\Bigl.
+ \max \Bigl\{ ( A_{N^-} - w_N )^+; 
 ( 1 -\beta_N ) \max \left\{ (A_{N^-}-w_N)^+;B_{N^-} - w_N \right\} \Bigr\} \Bigr].
\end{align*}
According to the scheme (SU-W), we have instead that
\begin{align*}
&\max_{w_N \in [ 0, A_{N^-}\vee B_{N^-} ]} \Bigl[ w_N -\beta_N ( w_N - G )^+ \Bigr. \\
&\qquad\qquad\qquad\qquad\Bigl.
+ \max \Bigl\{ ( A_{N^-} - w_N )^+; ( 1 -\beta_N ) \bigl(\max \{ A_{N^-};B_{N^-} \} - w_N\bigr) \Bigr\} \Bigr].
\end{align*}
Despite their different structure, it can be easily checked that the last two expressions generate the same cashflow at date $N$, denoted by $X_N(w^*_N)$ with some abuse of notation:
\begin{equation}\label{eqn:lastCashflow}
   X_N(w^*_N) = \max \bigl\{ A_{N^-}; ( 1 - \beta_N )B_{N^-} + \beta_N \min \{ G, B_{N^-} \} \bigr\},
\end{equation}
where $w^*_N = \min\{ B_{N^-},G \}$, similarly to the case of a GMWB annuity with no step-up feature (compare with \cite[Section 2.5]{GoudenegeMolentZanette2019}).
Observe that for $n=N$ the surrender payoff \eqref{eqn:fullSurrenderCashflow} coincides with the terminal payoff \eqref{eqn:lastCashflow}. Hence, no surrender decision is assumed to take place at maturity.

\begin{remark}[Surrender and full withdrawal decisions]
\label{rem:surrender}
Under the scheme (SU-W), when the step-up event takes place before the withdraw/bonus event, a full withdrawal corresponds to surrendering the contract: indeed, if $w_n = \max \{ A_{n^-}, B_{n^-} \}$, then both $A_{n^+}$ and $B_{n^+}$ become equal to zero and no more cashflows can be generated by the contract. On the contrary, according to the scheme (W-SU), a full withdrawal does not imply surrender: if $A_{n^-}>B_{n^-}$, a full withdrawal would reduce the benefit account to zero but not the investment account. The subsequent step-up event, though, would make the benefit account strictly positive again.
\end{remark}

Finally, we take into account mortality risk. Let $\tau^D \in (0,+\infty)$ be the random residual lifetime of the PH at inception of the contract. If the PH deceases between anniversary dates $n-1$ and $n$ (i.e., if $\tau^D \in (n-1,n]$,  \emph{death event}), a death benefit $X^D_n$ is credited to the heirs at date $n$. We assume that no step-up/bonus events take place in the case of a death event. In line with the final payoff after the last withdrawal when the PH outlives the annuity's maturity, we set
\begin{equation}\label{eqn:deathEventCashFlow}
    X^D_n := \max \bigl\{ A_{n^-}; ( 1 - \beta_n )B_{n^-} \bigr\}.
\end{equation}

\subsection{The stochastic model}\label{subsec:theMarketModel}

We work on a filtered probability space $(\Omega, \mathcal{F}, \mathbb{F}=(\mathcal{F}_t)_{t\geq0}, \mathbb{Q})$, supporting all processes and random variables introduced in the following and where $\mathbb{Q}$ represents a pricing measure. 
In the following, we denote by $\mathbb{E}[\cdot]$ the expectation with respect to $\mathbb{Q}$, using for conditional expectations the notation $\mathbb{E}_t[\cdot]:=\mathbb{E}[\cdot|\mathcal{F}_t]$, for  $t\geq0$.
Let $(W^S_t)_{t\geq0}$ and $(W^r_t)_{t\geq0}$ be two Brownian motions with correlation $\rho\in[-1,1]$, representing respectively the sources of market risk and interest rate risk. 

We assume that the interest rate $(r_t)_{t\geq0}$ is stochastic and has mean-reverting dynamics of the Hull-White type. The market model is described by the following SDEs:
\begin{equation}    \label{eq:market_model}\begin{aligned}
\mathrm{d} S_t & = S_t(r_t-q)\,\mathrm{d}t + S_t\,\sigma_S \,\mathrm{d} W_t^S, \\
\mathrm{d} r_t & = \left( \theta(t) - a r_t \right) \mathrm{d} t + \sigma_r \,\mathrm{d}W_t^r,
\end{aligned}\end{equation}
with initial conditions $S_0>0$ and $r_0\in\mathbb{R}$ and where $q\in\mathbb{R}$ is the dividend yield of the underlying fund.
The function $\theta:\mathbb{R}_+\to\mathbb{R}$ corresponds to the time-varying long-run mean of $r$ and is chosen in such a way to match the yield curve observed at $t=0$. This represents an important feature for the calibration of the model. We denote by $p^M(0,T)$ the market price at $t=0$ of a zero coupon bond with maturity $T$. If $f^M(0,T) := - \partial_T \log  p^M(0,T)$ is the market instantaneous forward rate for maturity $T$, it holds that (see \cite[Section 3.3.1]{BrigoMercurio2006})
\begin{equation}\label{eqn:thetat}
    \theta(t) = \frac{\partial f^M(0,t)}{ \partial t} + af^M(0,t) + \frac{\sigma_r^2}{2 a} ( 1- e^{- 2a t} ).
\end{equation}

We denote by $(S^0_t)_{t \geq 0}$ the money market account, given by $S_t^0=\exp ( -\int_0^t r_s \mathrm{d}s )$, for all $t\geq0$, which serves as the num\'eraire associated to the pricing measure $\mathbb{Q}$. For all $T\geq t$, letting $p(t,T)$ be the price at time $t$ of a zero coupon bond with maturity $T$, it holds that
\[
p(t,T) = \mathbb{E}_t [S^0_t/S^0_T] = A(t,T) \exp \left(  -B(t,T)r_t \right),
\]
where $B(t,T)=(1-e^{-a(T-t)})/a$ and 
\[
A(t,T) =  \frac{p^M(0,T)}{p^M(0,t)} \exp \left(  B(t,T)f^M(0,t) - \frac{\sigma_r^2}{4a} \left( 1 - e^{-2a t} \right) B(t,T)^2 \right),
\]
see \cite[Section 3.3.2]{BrigoMercurio2006}.
By construction, it holds that $p(0,T)=p^M(0,T)$ for all $T\geq0$.

The market model \eqref{eq:market_model} is similar to the one adopted in \cite{DaiYangLiu2015,GoudenegeMolentZanette2019}. The adoption of a constant diffusive volatility for $(S_t)_{t\geq0}$ is motivated by the fact that our main objective consists in studying the impact of stochastic low/negative interest rates on general GMWB annuities. For the same reason, we have adopted a Hull-White model for the interest rate, rather than the Cox-Ingersoll-Ross (CIR) model. However, the dynamic programming approach described in Section \ref{subsec:dynamicProgramming} and the valuation algorithm introduced in Section \ref{sec:algo_structure} can be easily generalized to alternative and richer models (see Remark \ref{rem:extension_DP} and Section \ref{sec:extension_algo} below).

We assume that mortality risk is independent of market and interest risks, in line with most of the works on GMWB valuation in the presence of stochastic mortality and interest rates (see, e.g., \cite{Costabile2017,DaiYangLiu2015,GudkovIgnatievaZiveyi2019,ShevchenkoLuo2016,YangDai2013} among others). Under this assumption, the only quantities that are needed for the valuation of a GMWB annuity are the conditional survival probabilities $\pi_n := \mathbb{Q}_{n-1}(\tau^D>n|\tau^D>n-1)$, corresponding to the probability of survival until anniversary date $n$, conditionally on survival up to the previous anniversary date $n-1$ and on $\mathcal{F}_{n-1}$, for $n=1,\ldots,N-1$.
For simplicity, we assume that survival probabilities are retrieved from mortality tables using a cohort-based approach (see \cite{Pitacco2004}). We refer to Remark \ref{rem:extension_DP} and Section \ref{sec:extension_algo} for a discussion of the extension of our framework to stochastic mortality models.

\begin{remark}[On the independence on financial and mortality risks]
\label{rem:independence}
As pointed out in \cite{Dhaene_et_al2013}, the independence of financial and mortality risks under the risk-neutral measure $\mathbb{Q}$ does not automatically follow from their independence under the real-world measure $\mathbb{P}$. However, for the model considered in this section, if financial and mortality risks are assumed to be independent under $\mathbb{P}$, there always exists a risk-neutral measure $\mathbb{Q}$ that preserves independence. This follows from the fact that for the market model \eqref{eq:market_model} the minimal martingale measure\footnote{In incomplete markets, the minimal martingale measure represents the most natural candidate for a risk-neutral measure, see e.g. \cite{FollmerSchweizer2010}. In the market model \eqref{eq:market_model}, the minimal martingale measure also coincides with the Esscher martingale measure.} $\widehat{\mathbb{Q}}$ is characterized by a Girsanov kernel (corresponding to the market price of risk) that is independent of mortality risk, thereby preserving independence between financial and mortality risks under $\widehat{\mathbb{Q}}$.
\end{remark}

\subsection{Valuation by dynamic programming}\label{subsec:dynamicProgramming}

As mentioned in the introduction, we allow for dynamic withdrawal strategies. In this case, the valuation of a GMWB annuity leads to a  stochastic optimal control problem along the anniversary dates $\mathcal{T}$ (we refer to \cite{Bertsekas2005} for a detailed account of discrete-time  stochastic optimal control). We assume that the PH pursues an optimal withdrawal strategy under $\mathbb{Q}$. As pointed out in \cite{BacinelloMaggistroZoccolan2022}, this corresponds to the worst-case scenario from the point of view of the insurer who has to set up a hedging portfolio. 

The valuation of a GMWB amounts to solving the following stochastic control problem:
\begin{equation}\label{eqn:pricingProblemMaximization}
\begin{aligned}
V_0(A_0, B_0, r_0) &:= 
\sup_{ \tau^S, \{ w_n  \}_{n\in\mathcal{T}} } 
\mathbb{E} \Biggl[ \sum_{n=1}^{N \wedge \tau^S} \left( \mathbf{1}_{ \{ \tau^D \wedge \tau^S >n \} }\frac{X_n(w_n)}{S_n^0} \right.\Biggr.\\
&\qquad\qquad\qquad\qquad\qquad\Biggl.\left.
+ \mathbf{1}_{ \{ \tau^D >n, \tau^S = n \} } \frac{X^S_n}{S_n^0} + \mathbf{1}_{ \{ n-1<\tau^D\leq n \} }\frac{X^D_n}{S_n^0} \right) \Biggr],
\end{aligned}\end{equation}
where the supremum is taken over all stopping times $\tau^S$ with values in $\{1,\ldots,N-1\}$ and over all withdrawal strategies $\{w_n\}_{n\in\mathcal{T}}$ such that $w_n\in[0,B_{n^-}]$ for each $n=1,\ldots,N$, in the case (W-SU) (or $w_n\in[0,A_{n^-}\vee B_{n^-}]$ for each $n\in\mathcal{T}$, in the case (SU-W)).

For each $n=0,1,\ldots,N$, we denote by $V_n(A_{n^-}, B_{n^-}, r_n)$ the fair value of the GMWB annuity as a function of the current values of the two accounts and of the interest rate, before any event (withdrawal/surrender/bonus/step-up) takes place at date $n$ and conditionally on $\{\tau^D>n\}$. As will be clarified below, the fair value of the GMWB annuity does not depend on the current value of the underlying fund.

For the solution of problem \eqref{eqn:pricingProblemMaximization}, we adopt a backward induction approach. Starting at $N=T$, if the PH is still alive at the last anniversary date $N$ (i.e., conditionally on the event $\{\tau^D>N\}$), then by \eqref{eqn:lastCashflow} the value of the contract is given by
\begin{equation}\label{eqn:terminalValueT}
V_N(A_{N^-},B_{N^-},r_N) = \max \bigl\{ A_{N^-}; ( 1 - \beta_N )B_{N^-} + \beta_N \min \{ G, B_{N^-} \} \bigr\}.
\end{equation}

The recursive relationship (Bellman equation) between the value functions at two generic dates $n$ and $n-1$ depends on the specification of the contract. Indeed, as explained in Remark \ref{rem:surrender}, under the scheme (SU-W) the surrender decision (corresponding to the choice of the optimal stopping time $\tau^S$) is embedded in the withdrawal decision. On the contrary, under the scheme (W-SU), the surrender decision has to be considered separately from the withdrawal decision. 

Let us first consider the scheme (W-SU). In this case, exploiting the independence of $\tau^D$ with respect to the interest rate and market risk factors, it holds that
\begin{equation}\label{eqn:ValueFunctionSpecification1}\begin{aligned}
    &V_{n-1}(A_{(n-1)^-}, B_{(n-1)^-}, r_{n-1}) = \max \left\{ X^S_{n-1};  \sup_{w_{n-1} \in [0,B_{(n-1)^-}]} \bigg( X_{n-1}(w_{n-1}) \right. \\
    &\qquad \left. + \pi_n \,\mathbb{E}_{n-1} \left[ e^{-\int_{n-1}^{n} r_s \mathrm{d}s} \, V_n(A_{n^-}, B_{n^-}, r_n) \right] \right. + (1 - \pi_n) \,\mathbb{E}_{n-1} \left[ e^{-\int_{n-1}^{n} r_s \mathrm{d}s} X^D_n \right] \bigg) \Bigg\},
\end{aligned}\end{equation}
where, denoting by $\Delta t$ the time step between dates $n-1$ and $n$,
\begin{align}
    A_{n^-} & = ( A_{(n-1)^-} - w_{n-1} )^+ \exp \left( \int_{n-1}^{n} r_s \mathrm{d}s - \left(q+\alpha+\frac{\sigma_S^2}{2} \right) \Delta t + \sigma_S \left( W^S_n - W^S_{n-1} \right) \right), \label{eqn:recursiveExpressionA} \\
    B_{n^-} & = \max \left\{ B_{(n-1)^-} + b \mathbf{1}_{ \left\{ w_{n-1}=0 \right\} }  - w_{n-1} ; (A_{(n-1)^-} - w_{n-1})^+ \right\}.
\label{eqn:Bn-Specification1}
\end{align}
Equation \eqref{eqn:recursiveExpressionA} makes clear that in the valuation of the GMWB annuity, the current value of the underlying fund is not relevant: only its return over the time period $[n-1,n]$ is relevant.

Considering instead the scheme (SU-W), we have that
\begin{equation}\label{eqn:ValueFunctionSpecification2}\begin{aligned}
    & V_{n-1}(A_{(n-1)^-}, B_{(n-1)^-}, r_{n-1}) = \sup_{w_{n-1} \in [0,A_{(n-1)^-}\vee B_{(n-1)^-}]} \Bigg( X_{n-1}(w_{n-1})  \\
    & \qquad + \pi_n\, \mathbb{E}_{n-1} \left[ e^{-\int_{n-1}^{n} r_s \mathrm{d}s} \,V_n(A_{n^-}, B_{n^-}, r_n) \right] + (1 - \pi_n )\, \mathbb{E}_{n-1} \left[ e^{-\int_{n-1}^{n} r_s \mathrm{d}s} X^D_n \right] \Bigg), 
\end{aligned}\end{equation}
where $A_{n^-}$ evolves as in (\ref{eqn:recursiveExpressionA}) and
\begin{equation}\label{eqn:Bn-Specification2}
B_{n^-} = \max \{ A_{(n-1)^-}; B_{(n-1)^-} \} + b \mathbf{1}_{ \left\{ w_{n-1}=0 \right\} } - w_{n-1}. 
\end{equation}

By inspecting formulas \eqref{eqn:ValueFunctionSpecification1}--\eqref{eqn:ValueFunctionSpecification2}, we can see that the determination of the optimal withdrawal (and surrender) strategy and, therefore, the valuation of our general GMWB annuity requires the knowledge of the $\mathcal{F}_{n-1}$-conditional joint distribution of the triplet $(\int_{n-1}^nr_s\mathrm{d}s,r_n,W^S_n-W^S_{n-1})$, for each $n=1,\ldots,N$. In our setup, this distribution admits an explicit description, given in the following proposition (the proof is postponed to the Appendix).

\begin{proposition}\label{prop:ConditionalDistribution}
For every $n=1,\ldots,N$, it holds that
\begin{equation}	\label{eq:joint_distribution}
\left( \int_{n-1}^{n} r_s \mathrm{d} s , r_n ,W_n^S - W_{n-1}^S \right) \bigg|\mathcal{F}_{n-1} \sim \mathcal{N}(\mu_{n-1} , \Sigma),
\end{equation}
where
\[
    \mu_{n-1} := \left[ \begin{array}{c}
    \mu_{1,n-1} \\ \mu_{2,n-1} \\ 0
    \end{array} \right] 
    \quad\text{ and }\quad
     \Sigma := \left[ \begin{array}{ccc}
    \sigma_{11} & \sigma_{12} & \sigma_{13} \\
    \sigma_{12} & \sigma_{22} & \sigma_{23} \\
    \sigma_{13} & \sigma_{23} & \sigma_{33}
    \end{array} \right],
\]
with
\begin{align*}
\mu_{1,n-1} &:= \frac{1}{a} ( 1-e^{-a\Delta t} )\bigl( r_{n-1}-\alpha(n-1) \bigr) + \log \left( \frac{p^M(0,n-1)}{p^M(0,n)} \right) + \frac{1}{2}\bigl( V(n)-V(n-1) \bigr), \\
\mu_{2,n-1} &:= r_{n-1}e^{-a\Delta t} + \alpha(n)-\alpha(n-1)e^{-a \Delta t}, \\
\alpha(n) &:= f^M(0,n) + \frac{\sigma_r^2}{2a^2} \left( 1 - e^{-an} \right)^2, 
\qquad\quad
V(n) := \frac{\sigma_r^2}{a^2} \left( n + \frac{2}{a} e^{-an} - \frac{1}{2a} e^{-2an} - \frac{3}{2a} \right) \\
\sigma_{11} &:= V(\Delta t),
\qquad\qquad\qquad
\sigma_{22} := \frac{\sigma_r^2}{2a}\left( 1 - e^{-2a \Delta t} \right),
\qquad\qquad\qquad
\sigma_{33} := \Delta t \\
\sigma_{12} &:= \frac{\sigma_r^2}{2 a^2} \left( 1- e^{-a \Delta t} \right)^2,
\qquad
\sigma_{13} := \frac{\rho \sigma_r}{a} \left( \Delta t - \frac{1-e^{-a \Delta t}}{a} \right),
\qquad
\sigma_{23} := \frac{\rho \sigma_r}{a} \left( 1 - e^{-a \Delta t} \right).
\end{align*}
As a consequence, for every $n=1,\ldots,N$, it holds that
\begin{equation}\label{eq:cond_dsitribution}
W_n^S - W_{n-1}^S  \bigg|\,\mathcal{F}_{n-1}\vee\sigma\left( \int_{n-1}^{n} r_s \mathrm{d} s , r_n\right) \sim \mathcal{N}(\tilde{\mu}_{n-1} , \tilde{\sigma}^2),
\end{equation}
where
\[
\tilde{\mu}_{n-1} := \Sigma_{12} \,\Sigma_{22}^{-1} \left( \left[ \begin{array}{c} \int_{n-1}^{n} r_s \mathrm{d} s \\ r_{n}  \end{array} \right]- \left[ \begin{array}{c} \mu_{1,n-1} \\ \mu_{2,n-1} \end{array} \right] \right)
\quad\text{ and }\quad
 \tilde{\sigma}^2 := \Sigma_{33} - \Sigma_{12} \Sigma_{22}^{-1} \Sigma_{21},
\]
with
\[
    \Sigma_{22} := \left[ \begin{array}{cc}
    \sigma_{11} & \sigma_{21} \\
    \sigma_{21} & \sigma_{22}
    \end{array} \right], 
    \qquad
    \Sigma_{21} := \left[ \begin{array}{c}
    \sigma_{13} \\
    \sigma_{23}
    \end{array} \right], 
    \qquad
    \Sigma_{12} := \left[ \begin{array}{cc}
    \sigma_{31} & \sigma_{32}
    \end{array} \right], 
    \qquad
    \Sigma_{33} := \sigma_{33}.
\]
\end{proposition}

\begin{remark}	\label{rem:info}
Observe that, as a consequence of Proposition \ref{prop:ConditionalDistribution}, the only relevant information when conditioning on the $\sigma$-algebra $\mathcal{F}_{n-1}$ is represented by the current value $r_{n-1}$ of the interest rate. This fact will be exploited in the numerical algorithm developed in next section. 
\end{remark}

\begin{remark}  \label{rem:extension_DP}
While the result of Proposition \ref{prop:ConditionalDistribution} is specific to the market model \eqref{eq:market_model}, the dynamic programming equations \eqref{eqn:ValueFunctionSpecification1} and \eqref{eqn:ValueFunctionSpecification2} do not depend on the structure of the model. If the model is extended by introducing additional stochastic factors (such as stochastic volatility or stochastic mortality), then the value function will depend also on those factors, besides  the interest rate.
In a stochastic mortality model, the conditional probabilities $\pi_n$ can be explicitly computed if $\tau^D$ is modelled as a doubly stochastic random time with stochastic intensity, for instance driven by an affine process (see \cite{Biffis2005}).
We refer to Section \ref{sec:extension_algo} below for a more detailed discussion of possible extensions of the model.
\end{remark}

\section{The valuation algorithm}\label{sec:numericalAlgorithm}

In this section, we describe the numerical algorithm to solve the stochastic control problem \eqref{eqn:pricingProblemMaximization} by means of the dynamic programming approach described in Section \ref{subsec:dynamicProgramming}. For simplicity of presentation, we  consider the scheme (SU-W).\footnote{We point out in Remark \ref{rem:algoWSU} which features of the algorithm have to be modified under scheme (W-SU).}
The presence of two sources of risk might suggest that a full discretization of the bivariate process $(S,r)$ is needed, as considered in \cite{BattauzRotondi2021} and \cite{DeAngelisDeMarchisMartireRusso2022} for the pricing of contingent claims under interest rate risk. However, as pointed out in Section \ref{subsec:dynamicProgramming}, the value function associated to problem \eqref{eqn:pricingProblemMaximization} does not depend on the value of the underlying fund. Moreover, besides the current values of the two accounts, the only relevant information encoded in the filtration $\mathbb{F}$ is the current value of the interest rate (see Remark \ref{rem:info}). Therefore, similarly as in \cite{Chang2014}, we can reduce the dimension of the problem and work with a simple one-dimensional discretization for $r$, regardless of the presence of correlation between interest rate and market risk. 
In Section \ref{sec:algo_structure} we present the algorithm for the model adopted in Section \ref{subsec:theMarketModel}, while in Section \ref{sec:extension_algo} we describe its application to possible extensions of the model.

\subsection{Structure of the algorithm}
\label{sec:algo_structure}

Our valuation algorithm is structured as follows:
\begin{itemize}
\item[0.]
\begin{itemize}
\item[a)] Perform a one-dimensional binomial discretization of the interest rate process $r$ with $m$ uniform steps between each anniversary date (see Appendix \ref{sec:discretization} for details).
\item[b)] Construct a two-dimensional grid for the possible values of the  accounts $A$ and $B$:
\[
\mathbb{A}:=\{a_0,a_1,\ldots,a_{n_A}\}
\quad\text{ and }\quad
\mathbb{B}:=\{b_0,b_1,\ldots,b_{n_B}\},
\]
where $a_0=b_0=0$, $a_i-a_{i-1}=\Delta_A$ and $b_j-b_{j-1}=\Delta_B$, for all $i=1,\ldots,n_A$ and $j=1,\ldots,n_B$, where $n_A$ and $n_B$ represent the number of points in the grid, with $\Delta_A$ being a multiple of $\Delta_B$. The upper bounds of the grid are given by $\bar{A}:=a_{n_A}=n_A\Delta_A$ and $\bar{B}:=b_{n_B}=n_B\Delta_B$.
\end{itemize}
\item[1.] At the last anniversary date $N$, initialize the value function by specifying $V_N(a_i,b_j,r_N)$ as in \eqref{eqn:terminalValueT}, for every $(a_i,b_j)\in\griglia$. Note that, at maturity, the value function does not depend on the value $r_N$ of the interest rate.
\item[2.] Proceeding backwards, for each $n=N,N-1,\ldots,2$ and for every $(a_i,b_j)\in\griglia$:
\begin{itemize}
\item[a)] For each $w_{n-1}\in\{0,\Delta_B,\ldots,a_i\vee b_j\}$, compute 
\begin{equation}\label{eq:expectation_algo}\begin{aligned}
J_{n-1}(w_{n-1},a_i,b_j)&:= \pi_n\, \mathbb{E}\left[ e^{-\int_{n-1}^{n} r_s \mathrm{d}s} \,V_n(A_{n^-}, B_{n^-}, r_n) \Big | r_{n-1}\right]  \\
&\quad\; 
+ (1-\pi_n) \,\mathbb{E} \left[ e^{-\int_{n-1}^{n} r_s \mathrm{d}s} X^D_n \Big| r_{n-1}\right],
\end{aligned}\end{equation}
where $X^D_n$ is given by \eqref{eqn:deathEventCashFlow} and $A_{n-}$ and $B_{n-}$ are respectively given by equations \eqref{eqn:recursiveExpressionA} and \eqref{eqn:Bn-Specification2}, with $A_{(n-1)^-}$ and $B_{(n-1)^-}$ respectively replaced by $a_i$ and $b_j$.
\item[b)] Determine the optimal withdrawal $w^*_{n-1}$ by solving
\[
\max_{w_{n-1}\in\{0,\Delta_B,\ldots,a_i\vee b_j\}}\bigl\{X_{n-1}(w_{n-1})+J_{n-1}(w_{n-1},a_i,b_j)\bigr\}.
\]
\item[c)] Set $V_{n-1}(a_i,b_j,r_{n-1}):=X_{n-1}(w^*_{n-1})+J_{n-1}(w^*_{n-1},a_i,b_j)$.
\end{itemize}
\item[3.] The value of the GMWB annuity at $t=0$ is given by 
\[
V_0(A_0,B_0,r_0) = \pi_1 \,\mathbb{E}\left[ e^{-\int_0^1 r_s \mathrm{d}s} \,V_1(A_{1^-}, B_{1^-}, r_1)\right]   
+ (1 - \pi_1)\, \mathbb{E} \left[ e^{-\int_0^1 r_s \mathrm{d}s} X^D_1\right],
\]
with $A_0=B_0=P$.
\end{itemize}

A crucial step of the algorithm consists in the computation of the conditional expectations in \eqref{eq:expectation_algo}. This can be done efficiently by proceeding as follows. Consider a generic anniversary date $n-1$ and a value $r_{n-1}$ of the binomial tree discretization of the process $r$. 
Observe first that there are $2^m$ paths along the binomial tree that link $r_{n-1}$ to the $m+1$ compatible values of $r_n$. The risk-neutral probability of each path $k$ starting from $r_{n-1}$ at date $n-1$ can be explicitly computed and is denoted by $q_{k,n-1}$, for $k=1,\ldots,2^m$.
For each path $k$, we approximate the discount factor $\exp(-\int_{n-1}^{n} r_s \mathrm{d}s)$ by discretizing the integral as $\exp( -\frac{\Delta t}{m+1} \sum_{j=1}^{m+1} r^k_{n-1:j:n})$, where $(r^k_{n-1:j:n})_{j=1, \ldots, m+1}$ denotes the discretized values of $r$ along the $k$-th path, starting from the given value $r_{n-1}$ at date $n-1$. 
Applying iterated conditioning, we compute
\begin{align}
&\mathbb{E}\left[ e^{-\int_{n-1}^{n} r_s \mathrm{d}s} \,V_n(A_{n^-}, B_{n^-}, r_n) \Big | r_{n-1}\right] \nonumber\\
&\quad = \mathbb{E} \left[ e^{-\int_{n-1}^{n} r_s \mathrm{d}s} \, \mathbb{E} \left[  V_n(A_{n^-}, B_{n^-}, r_n) \bigg| \left( \int_{n-1}^n r_s \mathrm{d} s, r_n,r_{n-1} \right) \right] \Bigg| r_{n-1}\right]	\nonumber\\
&\quad\approx \sum_{k=1}^{2^m} q_{k,n-1} \exp \biggl( -\frac{\Delta t}{m+1} \sum_{j=1}^{m+1} r^k_{n-1:j:n} \biggr) \mathbb{E} \Biggl[  V_n(A_{n^-}, B_{n^-}, r^k_n) \Bigg| \biggl( \frac{\Delta t}{m+1} \sum_{j=1}^{m+1} r^k_{n-1:j:n} , r^k_n,r_{n-1} \biggr) \Biggr], 
\label{eq:approx1}
\end{align}
where $A_{n^-}$ is determined from $A_{(n-1)^-}$ via the recursive relation \eqref{eqn:recursiveExpressionA}, replacing the term $\int_{n-1}^{n} r_s \mathrm{d}s$ by $\frac{\Delta t}{m+1} \sum_{j=1}^{m+1} r^k_{n-1:j:n}$.
Proceeding backwards, suppose that $V_n(a_h,b_l,r^k_n)$ has been already determined, for all $(a_h,b_l)\in\griglia$ and for all possible $m+1$ values $r^k_n$ that can be reached from $r_{n-1}$. 
Then, for each $(a_i,b_j)\in\griglia$ (representing the possible values of the pair $(A_{(n-1)^-},B_{(n-1)^-})$) and for each $w_{n-1}\in\{0,\Delta_B,\ldots,a_i-\Delta_B\}$, the conditional expectation appearing on the last line of \eqref{eq:approx1} can be computed as follows, relying on Proposition \ref{prop:ConditionalDistribution}:
\begin{align}
&	\mathbb{E} \Biggl[  V_n(A_{n^-}, B_{n^-}, r^k_n) \Bigg| \biggl( \frac{\Delta t}{m+1} \sum_{j=1}^{m+1} r^k_{n-1:j:n} , r^k_n,r_{n-1} \biggr) \Biggr]	\nonumber\\
&\quad\approx \sum_{h=1}^{n_A}V_n(a_h,B_{n-},r^k_n) \, \mathbb{Q} \Biggl(a_h-\frac{\Delta_A}{2} \leq A_{n^-} < a_h + \frac{\Delta_A}{2} \Bigg| \biggl( \frac{\Delta t}{m+1} \sum_{j=1}^{m+1} r^k_{n-1:j:n} , r^k_n,r_{n-1} \biggr)  \Biggr)	\nonumber\\
&\quad\approx \sum_{h=1}^{n_A}V_n(a_h,B_{n-},r^k_n) \,\Bigl(\Phi\bigl(a^+_{h,k,n}(a_i,w_{n-1})\bigr)-\Phi\bigl(a^-_{h,k,n}(a_i,w_{n-1})\bigr)\Bigr),
\label{eq:approx2}
\end{align}
with $A_{n-}$ and $B_{n-}$ respectively given by equations \eqref{eqn:recursiveExpressionA} and \eqref{eqn:Bn-Specification2}, with $A_{(n-1)^-}$ and $B_{(n-1)^-}$ respectively replaced by $a_i$ and $b_j$, and where $\Phi$ denotes the distribution function of a $\mathcal{N}(0,1)$ random variable and the quantities $a^{\pm}_h(a_i,w_{n-1})$ are defined as follows:
\[
a^{\pm}_{h,k,n}(a_i,w_{n-1}) := \frac{1}{\tilde{\sigma}\,\sigma_S} \Biggl( \log \biggl(\frac{a_h \pm \frac{\Delta_A}{2}}{a_i-w_{n-1}}\biggr) - \frac{\Delta t}{m+1} \sum_{j=1}^{m+1} r^k_{n-1:j:n} + \left( q + \alpha + \frac{\sigma_S^2}{2} \right)\Delta t \Biggr) - \frac{\tilde{\mu}_{n-1}}{\tilde{\sigma}},
\]
with $\tilde{\mu}_{n-1}$ and $\tilde{\sigma}$ as in Proposition \ref{prop:ConditionalDistribution}.
For $w_{n-1}\in\{a_i,\ldots,a_i\vee b_j\}$, we simply set $A_{n^-}=0$.
The computation of the conditional expectation appearing in the second line of \eqref{eq:expectation_algo} is performed analogously, relying on equation \eqref{eqn:deathEventCashFlow} which defines the death benefit.

\begin{remark}
\label{rem:algoWSU}
(1) Under the alternative scheme (W-SU), the algorithm has to be modified only in step 2. Indeed, in step 2.a the withdrawal $w_{n-1}$ must belong to $\{0,\Delta_B,\ldots,b_j\}$ and $B_{n-}$ is determined by equation \eqref{eqn:Bn-Specification1}, with $A_{(n-1)^-}$ and $B_{(n-1)^-}$ respectively replaced by $a_i$ and $b_j$. In step 2.b, the optimal withdrawal $w^*_{n-1}$ is chosen in the set $\{0,\Delta_B,\ldots,b_j\}$. Finally, step 2.c has to be modified taking into account the possibility of surrender, by setting
\[
V_{n-1}(a_i,b_j,r_{n-1}):= \max\bigl\{X^S_{n-1};X_{n-1}(w^*_{n-1})+J_{n-1}(w^*_{n-1},a_i,b_j)\bigr\},
\]
where $X^S_{n-1}$ is given by \eqref{eqn:fullSurrenderCashflow}, with $A_{(n-1)^-}$ and $B_{(n-1)^-}$ respectively replaced by $a_i$ and $b_j$.

(2) In the absence of step-up/bonus features, step 2 of the algorithm is modified by restricting $w_{n-1}\in\{0,\Delta_B,\ldots,b_j\}$ and using equations \eqref{eq:benefit_std}-\eqref{eq:investment_std} for the evolution of the two accounts.

(3) Step 2.b of the algorithm can be easily modified in order to account for sub-optimal withdrawal behaviors. This can be done by introducing suitable  constraints in the maximization problem or, as considered in  \cite{ChenVetzalForsyth2008}, by assuming that the amount withdrawn is equal to the static withdrawal $G$ unless the optimal withdrawal $w^*_{n-1}$ yields a sufficiently higher value. The adoption of a sub-optimal withdrawal policy reduces the fair value of a GMWB annuity.

(4) Under some contract specifications, the management fee $\alpha$ depends on the current value of the investment account. This possibility can be easily implemented in our algorithm, letting $\alpha$ depend on $A_{(n-1)^-}$ in equation \eqref{eqn:recursiveExpressionA}, which is used in step 2.a of the algorithm.
\end{remark}

\begin{figure}[ht]
    \centering
    \includegraphics[width = \textwidth]{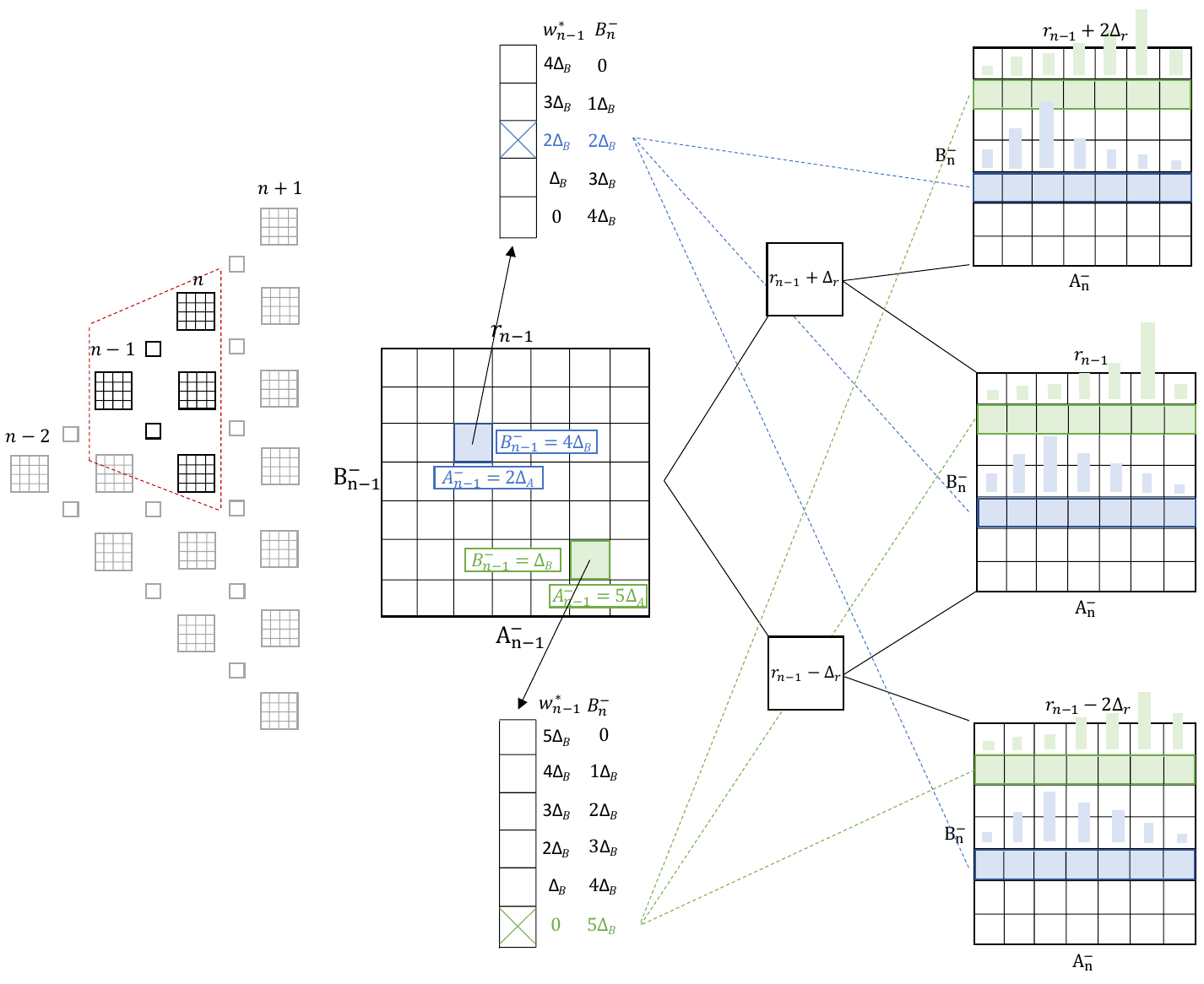}
    \caption{Graphical illustration of the backward induction in the valuation algorithm.}
    \label{fig:algorithmScheme}
\end{figure}

Figure \ref{fig:algorithmScheme} provides an illustration of step 2 of the algorithm, displaying a portion of the binomial discretization of $r$ from date $n-2$ to $n+1$. Nodes at anniversary dates are equipped with the  grid $\griglia$, represented in the figure by matrices. The $m-1$ intermediate nodes between consecutive anniversary dates contain only the discretized value of $r$. Let us focus on the backward induction from $n$ to $n-1$. Proceeding backwards,  $V_n(A_{n^-},B_{n^-},r_n)$ is already known, for all nodes of $r_n$ (in Figure \ref{fig:algorithmScheme}, this corresponds to knowing all matrices associated to date $n$). We need to fill the matrix associated to a certain value $r_{n-1}$ at date $n-1$. Consider for instance the blue cell (corresponding to a specific pair $(a_i,b_j)\in\griglia$, in this case $(2 \Delta_A, 4 \Delta_B)$). When considering the step-up feature, it is practical to set $\bar{A} = \bar{B}$ and $n_A = n_B$, so that $\Delta_A = \Delta_B$.  In the blue cell, the step-up feature does not take place, since $4 \Delta_B \geq 2 \Delta_A$. The PH can make five different withdrawals $w_{n-1}$: $0$, $\Delta_B$, $2 \Delta_B$, $3 \Delta_B$ or $4 \Delta_B$.\footnote{For illustration, in the case of the light green cell in Figure \ref{fig:algorithmScheme}, the step-up feature increases $B_{(n-1)^-} = \Delta_B$ to $A_{(n-1)^-} = 5\Delta_A$, generating six possible withdrawals $w_{n-1}\in\{0,\Delta_B,2 \Delta_B,3 \Delta_B,4 \Delta_B,5 \Delta_B\}$.} For each possible withdrawal, we compute the quantity in \eqref{eq:expectation_algo}. For illustration, $w_{n-1} = 2 \Delta_B$ is considered in the figure. In this case, we have that $B_{n^-} = B_{(n-1)^-} - w_{n-1} = 2 \Delta_B$. We then consider the matrices associated to the values of $r_n$ at date $n$. These $m+1$ matrices (3, in the case of Figure \ref{fig:algorithmScheme}) are linked to the matrix at date $n-1$ under consideration through $2^m$ paths (4, in the case of Figure \ref{fig:algorithmScheme}). Using \eqref{eq:approx1}, we compute the discounted expectation of $V_{n}(A_{n^-},B_{n^-},r_{n})$ as the weighted average of $2^m$ terms. While the row to consider is determined by $B_{n^-}=2\Delta$, we average out the columns (corresponding to the possible values of $A_{n^-}$), using conditional probabilities computed by relying on Proposition $\ref{prop:ConditionalDistribution}$. In Figure \ref{fig:algorithmScheme}, the conditional probabilities of the discretized values of $A_{n^-}$ are represented as light blue histograms within each matrix. An analogous computation is performed for the conditional expectation on the second line of \eqref{eq:expectation_algo}, thus completing step 2.a of the algorithm. This procedure is repeated for all possible withdrawals $w_{n-1}\in\{0,\Delta_B,2 \Delta_B,3 \Delta_B,4 \Delta_B\}$, in order to determine the optimal withdrawal $w^*_{n-1}$, as for step 2.b. Finally, the value  $V_{n-1}(a_i,b_j,r_{n-1})$ associated to the blue cell is determined as in step 2.c.
After having filled the matrices associated to all values $r_{n-1}$ at date $n-1$, one proceeds backwards at $n-2$.

\subsection{Extensions of the model}
\label{sec:extension_algo}

In this subsection, we discuss the applicability of the valuation algorithm introduced in Section \ref{sec:algo_structure} to possible extensions of the model. As can be deduced from the above discussion, the algorithm relies on the following three properties:
\begin{enumerate}
\item[(i)] the interest rate process admits a binomial discretization;
\item[(ii)] for each $n=1,\ldots,N$, the distribution of the fund return $S_n/S_{n-1}$ conditionally on $\mathcal{F}_{n-1}$ only depends on $r_{n-1}$;
\item[(iii)] for each $n=1,\ldots,N$, the distribution of the fund return $S_n/S_{n-1}$ conditionally on $(\int_{n-1}^nr_s\mathrm{d} s,r_n,r_{n-1})$, or conditionally on the interest rate path $(r_t;t\in[n-1,n])$, can be efficiently computed.
\end{enumerate}
As shown in Section \ref{sec:algo_structure}, these three properties are clearly satisfied by the model of Section \ref{subsec:theMarketModel}. More generally, property (i) holds for a wide class of Markovian diffusion models for the interest rate. Property (ii) is always satisfied if the interest rate is a Markov process and there is no serial dependence in the fund returns (under the risk-neutral probability). Finally, property (iii) depends on the specific model under consideration.

\subsubsection*{Jump-diffusion dynamics.}
The generalization of the model \eqref{eq:market_model} to jump-diffusion dynamics with L\'evy-type jumps in the returns does not impact on properties (i)-(ii). Concerning property (iii), we recall that a pure jump L\'evy process is necessarily independent from any Brownian motion defined on the same filtered probability space. Therefore, the conditional expectation in \eqref{eq:approx1} can be computed by first integrating with respect to the jump density\footnote{The jump density is explicitly known for several L\'evy processes, including compound Poisson processes with a known density for the jump sizes. For other L\'evy processes, the density can be retrieved by Fourier inversion from the characteristic function, which is known in closed form for all L\'evy processes (see, e.g., \cite{ContTankov}).} and then making use of Proposition \ref{prop:ConditionalDistribution}, thus retaining analytical tractability.

\subsubsection*{Stochastic volatility.}
As pointed out in Remark \ref{rem:extension_DP}, the introduction of stochastic volatility in the fund returns introduces an additional stochastic factor (here denoted by $v$), which has to be treated together with the interest rate. In this case, property (i) corresponds to a binomial discretization of the two-dimensional process $(r,v)$, while property (ii) requires that the distribution of $S_{n}/S_{n-1}$ conditionally on $\mathcal{F}_{n-1}$ only depends on $(r_{n-1},v_{n-1})$. In this form, property (i) holds for a sufficiently large class of Markovian diffusion models for $(r,v)$, while property (ii) is almost always satisfied. Property (iii) would require that the distribution of $S_{n}/S_{n-1}$ conditionally on the path $((r_t,v_t);t\in[n-1,n])$ can be efficiently computed. To this effect, the ``decoupling method'' of \cite{KirkbyNguyenCui2017} and \cite[Lemma 1]{Kirkby2023} can be helpful. Finally, the algorithm requires the transition probabilities of $(r,v)$, which can be determined for some diffusion processes.

\subsubsection*{Stochastic mortality.}
As in the case of stochastic volatility, the extension of the model to stochastic mortality introduces an additional stochastic factor. However, unlike in the case of stochastic volatility, stochastic mortality does not enter directly into the dynamics of the fund returns, thereby simplifying the analysis. The analysis becomes further simplified by the assumption of independence between financial and mortality risks (see Remark \ref{rem:independence}), with no major change to the algorithm described in Section \ref{sec:algo_structure}.

\begin{remark}[On the discretization of $r$]
We point out that, at the expense of a greater computational effort, one may avoid the binomial discretization of the process $r$ in step 1 of the algorithm. Indeed, if the possible values of $r$ are quantized into a grid $\mathbb{I}$ and, at each date $n$, the value function is approximated by interpolating its values at the points of the grid $\griglia\times\mathbb{I}$, then the conditional expectation in \eqref{eq:expectation_algo} can be computed by three-dimensional numerical integration over the conditional density of $(\int_{n-1}^nr_s\mathrm{d}s,r_n,W^S_n-W^S_{n-1})$, which is given in Proposition \ref{prop:ConditionalDistribution}.
We have chosen to work with a binomial discretization in order to make our algorithm potentially applicable to other Markovian dynamics for the interest rate process, for which an explicit description of the joint conditional density might not be possible. We illustrate this point below in the case of the CIR interest rate model.
As shown in the next section, our algorithm yields accurate results even with a very parsimonious binomial discretization of the process $r$.
\end{remark}

\begin{remark}[The CIR interest rate model]    
\label{rem:CIR}
The market model \eqref{eq:market_model} can be modified by considering a CIR process for the interest rate (see, e.g.,  \cite[Section 3.2.3]{BrigoMercurio2006}):
\[
\mathrm{d}r_t = (\theta-ar_t)\mathrm{d}t + \sigma_r\sqrt{r_t}\,\mathrm{d}W_t^r,
\qquad r_0>0.
\]
The process $r$ remains strictly positive if the Feller condition $\theta\geq\sigma^2_r/2$ holds. For $\rho\neq0$ an explicit description of the joint distribution \eqref{eq:cond_dsitribution} is not available. Nevertheless, our algorithm remains applicable. Indeed, by relying on \cite[Lemma 1]{Kirkby2023}, the fund return $S_n/S_{n-1}$ can be decomposed into two terms: a first term that is measurable with respect to $(r_t;t\in[n-1,n])$ and a Gaussian term that is independent of $(r_t;t\in[n-1,n])$. By conditioning with respect to $(r_t;t\in[n-1,n])$, one can then compute the conditional expectation \eqref{eq:approx1} by following the same procedure of Section \ref{sec:algo_structure}, leading to an explicit formula of the form \eqref{eq:approx2}. The probabilities $q_{k,n-1}$ can be computed in closed form as shown in \cite[Section 1.4]{nelsonRamaswamy90}. We remark that this technique is applicable to any diffusion interest rate model with a strictly positive volatility function.
\end{remark}

\section{Numerical Results}\label{sec:NumericalAnalysis}

In this section, we present and discuss our numerical results. First, in order to assess the reliability of the valuation algorithm, we compare it to some previous results available in the literature (Section \ref{subsec:preliminaryCheck}). In Section \ref{subsec:pricingAndFairFees}, we calibrate our model to two different market scenarios and determine fair fees, highlighting the impact of low/negative interest rates. We proceed by studying the determinants of the value of GMWB annuities (Section \ref{subsec:determinantsOfThePrice}). Finally, in Section \ref{subsec:optimalWithdrawalSection}, we describe the optimal withdrawal strategies in the two interest rate scenarios considered.

The results reported in this section have been obtained using Matlab R2021a, on a quad-core Intel i7 CPU at 1.80 GHz with 16 GB of physical memory. The computational time of the algorithm described in Section \ref{sec:numericalAlgorithm} depends on the withdrawal scheme, on the maturity of the contract, on the presence of the step-up feature and on the choice of the discretization parameters $m$, $n_A$, $n_B$. We report here the computational times for the benchmark choice of these parameters, for maturity $T=10$. If the annuity involves no step-up, we set $n_A = 30$, $n_B = 10$, $m=2$ and the algorithm takes 2.5 minutes for the S withdrawal scheme (6 minutes for the D scheme). If the step-up feature is included, we set $n_A = 30$, $n_B = 30$, $m=2$ and the algorithm takes 18 minutes for the S withdrawal scheme (85 minutes for the D scheme).

\subsection{Algorithm validation}\label{subsec:preliminaryCheck}

To validate our valuation algorithm, we first compute the fair value of a GMWB annuity without step-up and bonus features considering the S and D withdrawal schemes (see Remark \ref{rem:withdrawal}). In the absence of step-up and bonus features, GMWB annuities have been recently analyzed in \cite{GoudenegeMolentZanette2019}, which is adopted here as validation benchmark for our algorithm under the S and D schemes. Afterwards, we compare the values under the S and S+S schemes obtained through our algorithm to the ones obtained through the Least Square Method (LSM, henceforth) of \cite{LongstaffSchwartz2001}.

\begin{remark}[On the LSM for variable annuities pricing] The LSM is a widespread method for solving optimal stopping problems and, hence, it is also applicable to the pricing of several types of variable annuities. The LSM turns out to be especially useful and efficient when dealing with high-dimensional models, provided that the PH has a limited set of possible actions. The latter property holds whenever the optimal withdrawal strategy is of bang-bang type, namely when the set of withdrawal strategies can be reduced to zero withdrawal, a withdrawal equal to $G$, or a full surrender of the contract (see Remark \ref{rem:withdrawal}). This is typically the case of GLWB annuities, for which the LSM has been successfully applied in \cite{HuangKwok16} and \cite{WeiZhu22} (see \cite{BacinelloMaggistroZoccolan2022} for a proof of the bang-bang property in GLWB annuities).
When, on the contrary, the model is analytically tractable but the optimal withdrawal strategy cannot be reduced to a limited set of actions (as is the case of GMWB annuities, as pointed out by \cite{LuoShevchenko2015} and as we document in Section \ref{subsec:optimalWithdrawalSection}), a lattice-based method as considered in the present work offers greater transparency and allows for an efficient sensitivity analysis.
\end{remark}

Starting from the S scheme, we choose the same parameters reported in \cite[Table 3]{GoudenegeMolentZanette2019} and aim at replicating the numerical results reported in their Table 4.\footnote{In our notation, the parameters listed in \cite[Table 3]{GoudenegeMolentZanette2019} are given by: $\sigma = 20\%$, $r_0 = 5\%$, $a = 1$, $\sigma_r = 20\%$, $\rho = -50\%$.} 
Since the fair fees vary slightly according to the four different valuation methods proposed by \cite{GoudenegeMolentZanette2019}, we consider the average ($\bar{\alpha}$) of the four fair fees, their minimum ($\alpha_{\min}$) and their maximum ($\alpha_{\max}$). We test our algorithm by checking whether, for these values of fees, the GMWB annuity is priced at par. As an additional check, we compute 95\% confidence intervals for the fair values of the GMWB annuity with static withdrawals and fee $\bar{\alpha}$ by means of Monte Carlo simulations. Table \ref{tab:test1_static} reports the numerical results of this first test, showing that our algorithm always delivers a fair value close to $P=100$, with an accuracy comparable to the different methods used in \cite{GoudenegeMolentZanette2019}.

We then test our algorithm under the D scheme, replicating the numerical results shown in \cite[Table 11]{GoudenegeMolentZanette2019}, considering fees determined analogously to above. In this case, due to the optimization involved, no standard Monte Carlo benchmark is available. Table \ref{tab:test1_dynamic} displays the results of this test, confirming the reliability of our algorithm.

\begin{table}[ht]
    \centering
    \begin{tabular}{c|ccc|cc|cc}
         $T$ & $\bar{\alpha}$ & fair price & MC fair price & $\alpha_{\min}$ & fair price & $\alpha_{\max}$ & fair price \\ \hline
         5 & 1.915\% & 100.22 & 99.68 $\pm$ 0.51 & 1.908\% & 100.49 & 1.918\% & 100.11 \\
         10 & 0.795\% & 100.34 & 100.54 $\pm$ 0.79 & 0.793\% & 100.40 & 0.798\% & 100.25 \\
         20 & 0.248\% & 100.35 & 99.82 $\pm$ 0.87 & 0.247\% & 100.38 & 0.263\% & 99.85 \\
    \end{tabular}
    \caption{Numerical results for the first validation test: static withdrawals. Simulation parameters for $T=5,10$: $m=2$, $n_A = 30$, $n_B = 10$; for $T=20$: $m=2$, $n_A = 30$, $n_B = 20$. MC simulation based on 10000 paths.}
    \label{tab:test1_static}
\end{table}

\begin{table}[ht]
    \centering
    \begin{tabular}{c|cc|cc|cc}
         $T$ & $\bar{\alpha}$ & fair price & $\alpha_{\min}$ & fair price & $\alpha_{\max}$ & fair price \\ \hline
         5 & 2.783\% & 99.72 & 2.262\% & 100.46 & 2.826\% & 99.65 \\
         10 & 1.582\% & 99.75 & 1.284\% & 100.35 & 1.635\% & 99.52 \\
         20 & 0.829\% & 99.49 & 0.655\% & 100.36 & 0.907\% & 99.38 \\
    \end{tabular}
    \caption{Numerical results for the first validation test: dynamic withdrawals. Simulation parameters for $T=5,10$: $m=2$, $n_A = 30$, $n_B = 10$; for $T=20$: $m=2$, $n_A = 30$, $n_B = 20$.}
    \label{tab:test1_dynamic}
\end{table}

We now benchmark our algorithm to the LSM method under the S+S withdrawal scheme. Along with the processes $r$ and $S$, we also need to simulate the evolution of the accounts $A$ and $B$ and take into account  the intermediate cashflows equal to $G$ at each anniversary date. For this test, we use the parameters reported in Table \ref{tab:parameters2021} below, corresponding to the 2021 market scenario and we set $\alpha = 13.51\%$ and $\beta_n = 15\%$, for all $n=1,\ldots,N$.\footnote{As explained in Section \ref{subsec:pricingAndFairFees}, in the 2021 scenario this specification of $\alpha$ and $\beta$ ensures that a GMWB annuity without the step-up feature is sold at par (i.e., it has an initial market value of $P=100$).} The results of this test are reported in Table \ref{tab:testLSM}. We can observe that our algorithm produces fair values that are not statistically different from those obtained through the LSM-based algorithm.

\begin{table}[ht]
    \centering
    \begin{tabular}{cc|c|c|}
         $T$ & method & without step-up & with step-up \\ \hline
         \multirow{3}{*}{5} & S+S & 96.02 & 98.14 \\
         & \multirow{2}{*}{LSM} & 95.97 & 98.24 \\
         & & $^{\pm 0.07}$ & $^{\pm 0.09}$ \\ \hline
         \multirow{3}{*}{10} & S+S & 99.32 & 101.95 \\
         & \multirow{2}{*}{LSM} & 99.26 & 101.82 \\
         & & $^{\pm 0.10}$ & $^{\pm 0.15}$ \\ \hline
         \multirow{3}{*}{20}& S+S & 101.10 & 102.84 \\
         & \multirow{2}{*}{LSM} & 101.22 & 102.98 \\
         & & $^{\pm 0.14}$ & $^{\pm 0.21}$ \\ \hline
    \end{tabular}
    \caption{Numerical results for the second validation test: static withdrawals and surrender option. S+S values are computed using our algorithm; LSM values are computed using the Least Squares Method of \cite{LongstaffSchwartz2001}. Simulation parameters: $T=5,10$: $m=2$, $n_A = 30$, $n_B = 10$; for $T=20$: $m=2$, $n_A = 30$, $n_B = 20$.}
    \label{tab:testLSM}
\end{table}

Finally, we test the numerical stability of our algorithm, assessing the impact of the discretization parameters ($m$, $n_A$, $n_B$), introducing the step-up feature and allowing for dynamic withdrawal strategies. We test several combinations of parameters $m$, $n_A$, $n_B$, choosing them in such a way that the guaranteed minimum benefit $G$ is a multiple of the step of the grid.
Table \ref{tab:stabilityTest2021} reports the results. All fair values differ only if not rounded to the second decimal figure. In particular, we observe that increasing $m$, the number of nodes between two consecutive anniversary dates along the binomial tree for $r$, yields negligible improvements in the accuracy. This is due to the fact that the process $r$ has a limited variability and it can be represented efficiently by a binomial tree with a small number of nodes.
We can conclude that the algorithm is numerically stable and the resulting fair values do not depend significantly on the discretization parameters.

\begin{remark}[On the construction of the grid $\mathbb{A}\times\mathbb{B}$] 
As Table \ref{tab:stabilityTest2021} shows, choosing a finer grid $\mathbb{A}\times\mathbb{B}$ (i.e., choosing higher values for $n_A$ and $n_B$) does not improve significantly the accuracy of the algorithm. We acknowledge here that choosing a uniform grid does not always represent the most efficient choice, nor is necessary for our algorithm to work. More specifically, if there is no step-up nor bonus feature only $\mathbb{B}$ has to be uniform, with $n_B$ chosen in such a way that $G$ is a multiple of $\Delta_B$, while a non-uniform grid $\mathbb{A}$ (like the one proposed in \cite{GoudenegeMolentZanette2019}) can be adopted for the investment account. However, if step-up or bonus features are included, then $G$ must be a multiple of $\Delta_A$ as well. For this reason, we shall work with a uniform grid $\mathbb{A}\times\mathbb{B}$.
\end{remark}

\begin{table}[]
    \centering
    \begin{tabular}{cc|c|c|c|}
         $\bar{A} = \bar{B}$ & $n_A = n_B$ & $m=2$ & $m=3$ & $m=4$ \\ \hline
         \multirow{2}{*}{200} & 21 & 102.6023 & 102.5968 & 102.5996 \\
         & 41 & 102.6026 & 102.6016 & 102.6014 \\ \hline
         \multirow{2}{*}{300} & 31 & 102.5976 & 102.6031 & 102.6032  \\
         & 61 & 102.6012  & 102.6002  & 102.6028  \\ \hline
         400 & 41 & 102.5977 & 102.5998 & 102.5998  \\ \hline
    \end{tabular}
    \caption{Fair value of a GMWB with step-up with dynamic withdrawals for different values of $m$, $\bar{A} = \bar{B}$ and $n_A = n_B$. $T=10$, $\alpha = 13.51\%$, $\beta_n = \beta = 15\%$, $n = 1,\ldots,N$, and remaining parameters as in Table \ref{tab:parameters2021}.}
    \label{tab:stabilityTest2021}
\end{table}

\subsection{Pricing of GMWB annuities and interest rate scenarios}\label{subsec:pricingAndFairFees}

In this section, we present our numerical results on the valuation of GMWB annuities in two different interest rate scenarios. 

\subsubsection{Parameter specification}	\label{sec:parameters}

\begin{figure}[]
    \centering
    \includegraphics[width=0.5\textwidth]{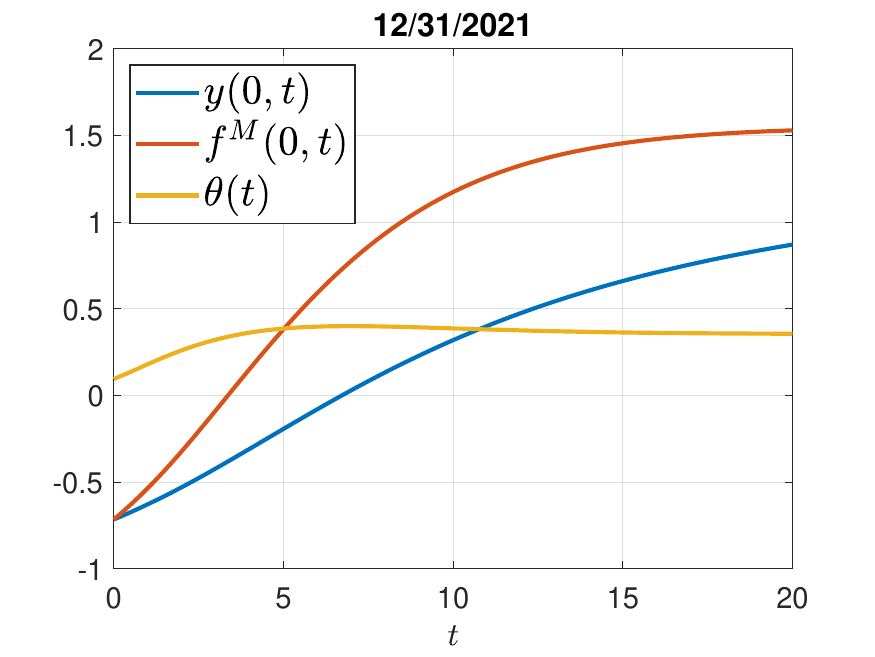}
    \caption{Yield curve, forward rate curve and long-run mean in percentage points as of 12/31/2021.}
    \label{fig:ThetaAndFwd_2021}
\end{figure}

The parameters characterizing the market model described in Section \ref{subsec:theMarketModel} are reported in Table \ref{tab:parameters2021}. These parameters have been chosen to reflect the current European market, considering the EuroStoxx50 as the underlying equity fund.
We set $\sigma_S = 20\%$\footnote{Value taken from the ECB annual report on the index volatility (available at \url{https://sdw.ecb.europa.eu}).} and $q = 2\%$\footnote{This figure corresponds to the weighted average of the annualized dividend yields of the index's constituents, which can be found at \url{http://www.dividendsranking.com/Index/EURO-STOXX-50.php}.}.
Since the cashflows of the GMWB annuity depend solely on the return of the index (and not on its level), we can normalize its initial value to $S_0=100$ with no loss of generality. 
The yield curve is modelled according to the ECB's methodology, adopting Svensson parametrization. In the notation of Section \ref{subsec:theMarketModel}, the market instantaneous forward rate at $t=0$ is given by
\[ 
f^M(0,T) = \beta_0 + \beta_1 e^{-\frac{T}{\tau_1}} + T\frac{\beta_2}{\tau_1}e^{-\frac{T}{\tau_1}} + T\frac{\beta_3}{\tau_2}e^{-\frac{T}{\tau_2}}, 
\]
for all $T\geq0$, using for the parameters $\beta_i$, $i=0,\ldots,3$, $\tau_j$, $j=1,2$, the official estimates of the ECB.\footnote{See \url{https://www.ecb.europa.eu/stats/financial_markets_and_interest_rates/euro_area_yield_curves/html/index.en.html}.}
The corresponding yield curve $y(0,T)$, defined by $y(0,T) := \frac{1}{T} \int_0^T f^M(0,t) \mathrm{d}t$, is then explicitly given by
\[
y(0,T) = \beta_0 + \left( \beta_1 + \beta_2 \right) \tau_1 \left( \frac{1 - e^{-\frac{T}{\tau_1}}}{T} \right) - \beta_2 e^{-\frac{T}{\tau_1}} + \beta_3 \left( \tau_2 \frac{1 - e^{-\frac{T}{\tau_2}}}{T} - e^{-\frac{T}{\tau_2}} \right).
\]
For the Hull-White time-independent parameters $a$ and $\sigma_r$, we take the values calibrated in the benchmark analysis of \cite{RussoTorri2019}, with $a = 10\%$ and $\sigma_r = 2\%$. We also set $\rho=0.5$.\footnote{As will be documented below, the impact of the parameters $a$, $\sigma_r$, $\rho$ is of second order with respect to the other model parameters and fair values are not significantly impacted by reasonable variations of these parameters.}
Figure \ref{fig:ThetaAndFwd_2021} shows the yield curve, the instantaneous forward rate curve and the resulting long-run mean of $r$ ($\theta(t)$, determined by \eqref{eqn:thetat}). As we can see, at 12/31/2021 the interest rate is markedly negative and is expected to stay negative for a significant time.

\begin{table}[ht]
    \centering
    \begin{tabular}{c|c|c}
        underlying & short-term & instantaneous \\
        fund, $S$ & rate, $r$ & forward, $f^M(0,t)$ \\ \hline
         $S_0=100$ & $r_0 = -0.67 \%$ & $\beta_0 = 0.3202$ \\
         $\sigma_S = 20\%$ & $a = 10\%$ & $\beta_1 = -1.0501$ \\
         $ q = 2\%$ & $\sigma_r = 2\%$ & $\beta_2=13.2616$ \\
         & & $\beta_3=-14.7208$ \\
         $ \rho = 50\%$ & & $\tau_1 = 1.8168$ \\
         & & $\tau_2 = 1.8656$
    \end{tabular}
    \caption{Parameters of the financial market model as of 12/31/2021.}
    \label{tab:parameters2021}
\end{table}

In line with the literature, we assume that the PH is a 65 years old man at inception of the contract and we account for mortality risk adopting the survival probabilities published by the United States Social Security Administration.\footnote{The corresponding mortality table is available at \url{https://www.ssa.gov/oact/STATS/table4c6.html}}

As far as the GMWB annuity is concerned, we consider a maturity $T=10$ and we set $P=100$. This implies that the guaranteed minimum amount that can be withdrawn at each anniversary date without paying any withdrawal penalty is equal to $G=10$. 

\subsubsection{Fair fees and pricing of GMWB annuities}

Table \ref{tab:optimalFees2021} reports the fair prices of the GMWB annuity for different values of the management fee and of the withdrawal penalty. We assume a constant withdrawal penalty, setting all $\beta_n$'s equal to a constant $\beta$. This simplifying assumption enables us to analyze in a clear way the determinants of the fair value of a GMWB annuity\footnote{{In some GMWB contracts, withdrawal penalties are specified as decreasing over time. As described in the previous sections, our valuation framework is applicable to generic time-dependent penalties.}.
Moreover, the analysis of Section \ref{subsec:optimalWithdrawalSection} will show that, in a low/negative interest rate environment, withdrawals exceeding the guaranteed amount occur more rarely than in the case of positive rates. For this reason, the following results are not significantly affected by the assumption of a constant withdrawal penalty $\beta$. This is also in line with the findings of \cite[Section 4.5]{ChenVetzalForsyth2008}.

The left panel of Table \ref{tab:optimalFees2021} considers a GMWB annuity without step-up feature, which is instead included in the right panel. For each combination of $\alpha$ and $\beta$ we consider the three main withdrawal schemes: static withdrawals without surrender option (scheme S), static withdrawals with surrender option (scheme S+S), dynamic withdrawals (scheme D). Obviously, adding the option of a full surrender and then of dynamic withdrawals increases the value of the annuity. The difference across the three withdrawal schemes is decreasing in $\beta$, because large withdrawal penalties reduce significantly the benefits of a full surrender or of the dynamic withdrawal strategy}. Coherently, we also observe that the fair values in the static case are much less sensitive to $\beta$ than in the other two cases. Indeed, under static withdrawals, $\beta$ affects only the terminal cashflow of the annuity. The step-up feature always increases the value of the GMWB annuity. The increment in the price due to the step-up feature is decreasing in $\alpha$, since the step-up feature is less likely to bring a real benefit if the management fee is high and the expected net growth of the investment account is therefore low.

We now determine fair management fees. By definition, the fair management fee is the value $\alpha$ such that the fair value of the GMWB annuity coincides with its nominal amount ($P=100$), for a suitable penalty $\beta$. In the market setting described in Section \ref{sec:parameters}, it turns out to be very difficult to match the fair value of a GMWB annuity with $P$, especially when step-up and bonus features are included. As explained in more detail below, this is specifically due to the negative interest rate scenario. Therefore, we choose to determine fair fees in such a way that the fair value of a standard version of the GMWB annuity (i.e., without step-up and bonus features, but allowing for dynamic withdrawals) is equal to $P=100$, whereas the value of additional features is assumed to be paid at inception as an extra premium on top of $P$. 

Looking at the left panel of Table \ref{tab:optimalFees2021}, we see that   $P=100$ is almost matched in the dynamic withdrawal case when $\alpha = 15\%$ and $\beta = 15\%$. Fixing $\beta=15\%$, the fair fee results equal to $\alpha=13.51\%$.\footnote{A penalty of $\beta=15\%$ is greater than what usually considered in the literature. As shown below, smaller values of $\beta$ lead to acceptable fair management fees only in a positive interest rate scenario. In the 31/12/2021 scenario, the fair fee $\alpha$ associated to $\beta=10\%$ would be larger than $50\%$, which is clearly unfeasible in practice.} We take this combination of $\alpha$ and $\beta$ as our benchmark configuration. With these values of $\alpha$ and $\beta$ the value in the static withdrawal case turns out to be 99.13, which is very close to $P$. This is due to the large value of $\beta$, which penalizes excess withdrawals and, therefore, reduces significantly the difference between static and dynamic withdrawal strategies. This effect is clearly visible from the analysis of optimal withdrawal strategies reported in Section \ref{subsec:optimalWithdrawalSection}. In particular, Figure \ref{fig:optimalWithdrawals_2021} shows that there are very large regions where the optimal withdrawal strategy coincides with $G=10$, while large optimal withdrawals that would differentiate the dynamic scheme from the static one are located only on very extreme, and therefore unlikely, regions of the plane.

Including the step-up feature, the fair value in the static (resp. dynamic) case increases to 101.36 (resp. 102.60). Equivalently, taking the standard GMWB annuity (without step-up and bonus features) as benchmark, the extra premium to be paid at inception for the step-up feature would amount to 2.60. Keeping $\beta=15\%$, including the step-up feature in the nominal amount would yield a fair fee $\alpha=27.58\%$, which represents a unrealistically large value in practice.

 \begin{table}[]
    \centering
        \resizebox{\textwidth}{!}{\begin{tabular}{c|c|c|c|c|c|c|c|}
        \multicolumn{3}{c|}{2021: Without} & \multicolumn{5}{c|}{$\alpha$} \\ \cline{4-8}
        \multicolumn{3}{c|}{step-up} & 0\% & 5\% & 10\% & 15\% & 20\% \\ \hline
        \multirow{15}{*}{$\beta$} & \multirow{3}{*}{0\%} & S & 110.53 & 103.53 & 101.10 & 100.43 & 100.27 \\ 
        & & S+S & 111.81 & 106.07 & 104.89 & 104.77 & 104.75 \\ 
        & & D & 119.48 & 111.46 & 109.03 & 108.36 & 108.18 \\ \cline{2-8}
        & \multirow{3}{*}{5\%} & S & 110.21 & 103.13 & 100.64 & 99.94 & 99.76 \\ 
        & & S+S & 110.61 & 104.07 & 102.22 & 101.92 & 101.87 \\ 
        & & D & 115.63 & 107.35 & 104.85 & 104.14 & 103.95 \\ \cline{2-8}
        & \multirow{3}{*}{10\%} & S & 109.88 & 102.74 & 100.19 & 99.45 & 99.26 \\ 
        & & S+S & 110.01 & 103.03 & 100.79 & 100.28 & 100.17 \\ 
        & & D & 112.88 & 104.65 & 102.22 & 101.48 & 101.27 \\ \cline{2-8}
        & \multirow{3}{*}{15\%} & S & 109.59 & 102.37 & 99.77 & 98.99 & 98.77 \\ 
        & & S+S & 109.62 & 102.41 & 99.88 & 99.18 & 99.01 \\ 
        & & D & 111.13 & 103.18 & 100.64 & 99.86 & 99.62 \\ \cline{2-8}
        & \multirow{3}{*}{20\%} & S & 109.32 & 102.02 & 99.36 & 98.53 & 98.29 \\ 
        & & S+S & 109.34 & 102.02 & 99.37 & 98.56 & 98.33 \\ 
        & & D & 110.00 & 102.32 & 99.71 & 98.89 & 98.63 \\ \hline
        \end{tabular}

        \begin{tabular}{c|c|c|c|c|c|c|c|}
        \multicolumn{3}{c|}{2021: With} & \multicolumn{5}{c|}{$\alpha$} \\ \cline{4-8}
        \multicolumn{3}{c|}{step-up} & 0\% & 5\% & 10\% & 15\% & 20\% \\ \hline
        \multirow{15}{*}{$\beta$} & \multirow{3}{*}{0\%} & S & 120.82 & 110.09 & 105.1 & 102.67 & 101.35 \\ 
        & & S+S & 123.08 & 115.07 & 110.03 & 107.56 & 106.24 \\ 
        & & D & 131.22 & 118.83 & 113.40 & 110.87 & 109.56 \\ \cline{2-8}
        & \multirow{3}{*}{5\%} & S & 119.70 & 109.35 & 104.36 & 102.01 & 100.82 \\ 
        & & S+S & 122.12 & 112.60 & 107.40 & 104.80 & 103.43 \\ 
        & & D & 126.35 & 114.14 & 108.97 & 106.55 & 105.28 \\ \cline{2-8}
        & \multirow{3}{*}{10\%} & S & 118.48 & 108.47 & 103.68 & 101.42 & 100.24 \\ 
        & & S+S & 121.34 & 110.87 & 105.64 & 103.03 & 101.67 \\ 
        & & D & 122.40 & 110.89 & 106.06 & 103.78 & 102.57 \\ \cline{2-8}
        & \multirow{3}{*}{15\%} & S & 117.37 & 107.69 & 103.01 & 100.84 & 99.68 \\ 
        & & S+S & 118.97 & 108.57 & 104.06 & 101.77 & 100.42 \\ 
        & & D & 119.37 & 108.80 & 104.31 & 102.07 & 100.87 \\ \cline{2-8}
        & \multirow{3}{*}{20\%} & S & 116.68 & 106.90 & 102.47 & 100.32 & 99.15 \\ 
        & & S+S & 116.83 & 107.25 & 102.93 & 100.90 & 99.60 \\ 
        & & D & 117.10 & 107.49 & 103.16 & 100.98 & 99.81 \\ \hline
        \end{tabular}}
        \newline
        \vspace*{0.5 cm}
        \newline
        \small
        \begin{tabular}{c|c|c|}
            \multicolumn{2}{c|}{2021: Without step-up} & $\alpha = 13.51\%$ \\ \hline
             \multirow{3}{*}{$\beta = 15\%$} & S & 99.13  \\
             & S+S & 99.32 \\
             & D & 100.00 \\ \hline
        \end{tabular}
        \quad
        \begin{tabular}{c|c|c|}
            \multicolumn{2}{c|}{2021: With step-up} & $\alpha = 13.51\%$ \\ \hline
             \multirow{3}{*}{$\beta = 15\%$} & S & 101.36  \\
             & S+S & 101.95 \\
             & D & 102.60 \\ \hline
        \end{tabular}
    \caption{Fair value of a GMWB annuity at $t=0$ for different values of  $\alpha$ and $\beta$. $T=10$ and market model parameters as in Table \ref{tab:parameters2021}, as of 12/31/2021.  Parameters: $m=2$, $n_A = 30$, $n_B = 10$ for the left panel, $n_B = 30$ for the right panel.}
    \label{tab:optimalFees2021}
\end{table}

\subsubsection{Impact of negative interest rates}

\begin{figure}
    \centering
    \includegraphics[width=0.48\textwidth]{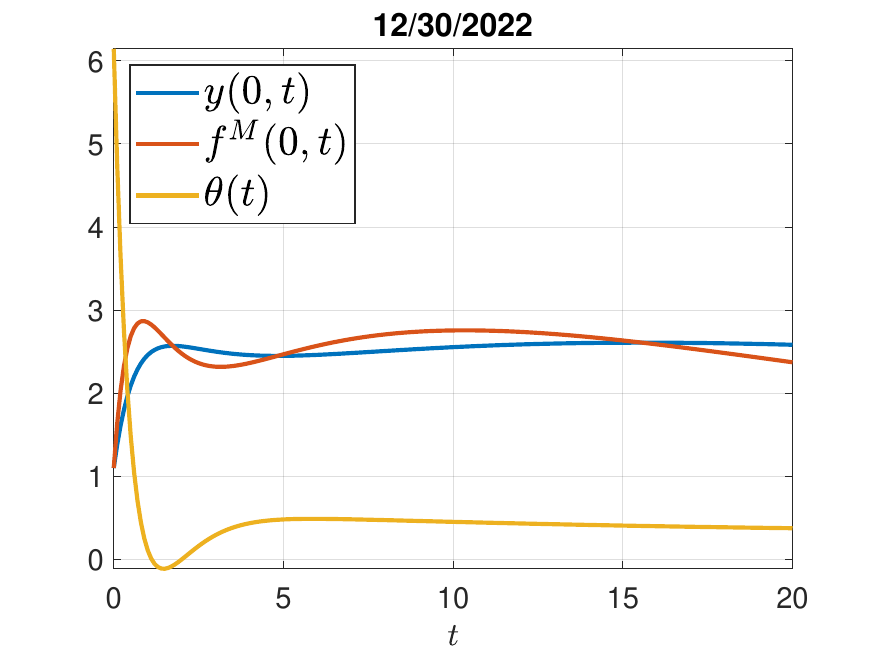}
    \caption{Yield curve, forward rate curve and long-run mean in percentage points as of 12/30/2022.}\label{fig:thetaAndFwd_2010}
\end{figure}

To fully analyze the consequences of the negative interest rate environment on the pricing of GMWB annuities, we consider an alternative interest rate scenario, calibrated to the current market scenario, as of end of 2022\footnote{The parameters of the forward rate curve as of 12/30/2022 are $\beta_0 = 1.2109$, $\beta_1 = -0.1090$, $\beta_2 = 4.3116$, $\beta_3 = 3.9468$, $\tau_1 = 10.0911$, $\tau_2 = 0.7052$. For sake of comparability, we leave unchanged the other parameters, $\sigma_r$, $a$, $\sigma_S$, $\rho$ and $q$. The analysis of an additional interest rate scenario, as of end of 2015, is reported in the Supplementary Material.}.
As a consequence of the recent changes in the monetary policy, the 2021 and 2022 interest rate scenarios exhibit striking differences on the level and on the outlook of the interest rate. Indeed, the current level of the interest rate is positive ($r_0 = 2.19\%$) and, as shown in Figure \ref{fig:thetaAndFwd_2010}, is expected to remain positive in the long term.

The level and the shape of the term structure of interest rates are relevant determinants of the value of GMWB annuities. Table \ref{tab:optimalFees2010} reports the fair values and fees for the 2022 scenario. We immediately observe that, even with a smaller penalty $\beta=10\%$, fair fees are significantly smaller than in the 2021 scenario ($\alpha=0.6270\%$ in the 2022 scenario).
The finding that fair fees are decreasing in the level (and in the outlook) of the interest rate is consistent with the results of \cite{ChenVetzalForsyth2008} and \cite{ShevchenkoLuo2016}. Moreover, the magnitude of fair fees in the current 2022 scenario is in line with the results of \cite{GoudenegeMolentZanette2019}. Indeed, as mentioned in the introduction, most of the previous related works assume positive interest rates, typically ranging from 2\% to 5\%. In such interest rate environment, considering a withdrawal penalty of about $10\%$, the fair fees are typically expressed in basis points.

\begin{table}[]
    \centering
        \resizebox{\textwidth}{!}{\begin{tabular}{c|c|c|c|c|c|c|c|}
        \multicolumn{3}{c|}{2022: Without} & \multicolumn{5}{c|}{$\alpha$} \\ \cline{4-8}
        \multicolumn{3}{c|}{step-up} & 0\% & 1.25\% & 2.5\% & 3.75\% & 5\% \\ \hline
        \multirow{15}{*}{$\beta$} & \multirow{3}{*}{0\%} & S & 99.65 & 96.14 & 93.17 & 90.68 & 88.64 \\ 
        & & S+S & 102.55 & 100.12 & 98.51 & 97.55 & 97.01 \\ 
        & & D & 107.05 & 104.24 & 102.25 & 100.83 & 99.87 \\ \cline{2-8}
        & \multirow{3}{*}{2.5\%} & S & 99.57 & 96.05 & 93.06 & 90.57 & 88.52 \\ 
        & & S+S & 101.60 & 98.93 & 97.05 & 95.79 & 95.11 \\ 
        & & D & 105.29 & 102.35 & 100.35 & 98.89 & 97.85 \\ \cline{2-8}
        & \multirow{3}{*}{5\%} & S & 99.49 & 95.96 & 92.96 & 90.46 & 88.40 \\ 
        & & S+S & 100.81 & 97.94 & 95.70 & 94.25 & 93.31 \\ 
        & & D & 103.78 & 100.72 & 98.61 & 97.05 & 95.91 \\ \cline{2-8}
        & \multirow{3}{*}{7.5\%} & S & 99.41 & 95.87 & 92.86 & 90.35 & 88.28 \\ 
        & & S+S & 100.28 & 97.08 & 94.69 & 92.84 & 91.69 \\ 
        & & D & 102.66 & 99.46 & 97.11 & 95.43 & 94.22 \\ \cline{2-8}
        & \multirow{3}{*}{10\%} & S & 99.32 & 95.78 & 92.76 & 90.24 & 88.16 \\ 
        & & S+S & 99.83 & 96.53 & 93.81 & 91.78 & 90.27 \\ 
        & & D & 101.76 & 98.47 & 95.90 & 94.01 & 92.67 \\ \hline
        \end{tabular}

        \begin{tabular}{c|c|c|c|c|c|c|c|}
        \multicolumn{3}{c|}{2022: With} & \multicolumn{5}{c|}{$\alpha$} \\ \cline{4-8}
        \multicolumn{3}{c|}{step-up} & 0\% & 1.25\% & 2.5\% & 3.75\% & 5\% \\ \hline
        \multirow{15}{*}{$\beta$} & \multirow{3}{*}{0\%} & S & 109.86 & 106.11 & 102.77 & 99.84 & 97.27 \\ 
        & & S+S & 116.22 & 113.12 & 110.46 & 108.26 & 106.53 \\ 
        & & D & 119.81 & 115.52 & 112.12 & 109.28 & 107.08 \\ \cline{2-8}
        & \multirow{3}{*}{2.5\%} & S & 109.27 & 105.55 & 102.25 & 99.34 & 96.80 \\ 
        & & S+S & 114.5 & 111.35 & 108.65 & 106.36 & 104.53 \\ 
        & & D & 117.6 & 113.27 & 109.87 & 107.06 & 104.83 \\ \cline{2-8}
        & \multirow{3}{*}{5\%} & S & 108.68 & 104.99 & 101.72 & 98.85 & 96.34 \\ 
        & & S+S & 112.87 & 109.65 & 106.90 & 104.54 & 102.58 \\ 
        & & D & 115.52 & 111.16 & 107.77 & 105.01 & 102.85 \\ \cline{2-8}
        & \multirow{3}{*}{7.5\%} & S & 108.10 & 104.44 & 101.20 & 98.36 & 95.88 \\ 
        & & S+S & 111.41 & 108.07 & 105.26 & 102.88 & 100.83 \\ 
        & & D & 113.66 & 109.36 & 106.00 & 103.39 & 101.24 \\ \cline{2-8}
        & \multirow{3}{*}{10\%} & S & 107.54 & 103.91 & 100.7 & 97.88 & 95.43 \\ 
        & & S+S & 110.11 & 106.71 & 103.78 & 101.34 & 99.27 \\ 
        & & D & 112.05 & 107.88 & 104.59 & 101.99 & 99.84 \\ \hline

        \end{tabular}}
        \newline
        \vspace*{0.5 cm}
        \newline
        \small
        \begin{tabular}{c|c|c|}
            \multicolumn{2}{c|}{2022: Without step-up} & $\alpha = 0.6270\%$ \\ \hline
             \multirow{3}{*}{$\beta = 10\%$} & S & 97.48  \\
             & S+S & 98.09 \\
             & D & 100.00 \\ \hline
        \end{tabular}
        \quad
        \begin{tabular}{c|c|c|}
            \multicolumn{2}{c|}{2022: With step-up} & $\alpha = 0.6270\%$ \\ \hline
             \multirow{3}{*}{$\beta = 10\%$} & S & 105.66  \\
             & S+S & 108.35 \\
             & D & 109.84 \\ \hline
        \end{tabular}
    \caption{Fair value of a GMWB annuity at $t=0$ for different values of  $\alpha$ and $\beta$. $T=10$ and market model parameters as of 12/31/2022. Parameters: $m=2$, $n_A = 30$, $n_B = 10$ for the panels with no step up feature, $n_B = 30$ for the panels with the step up feature.}
    \label{tab:optimalFees2010}
\end{table}

An important conclusion that we can draw from our analysis is that, in a low interest rate environment, selling a GMWB at par comes necessarily with large withdrawal penalties and management fees. Indeed, the role of the interest rate is twofold: on the one hand, it determines the discount factor; on the other hand, it represents the expected rate of return of the underlying index under the pricing measure. Negative interest rates increase the present value of future cashflows, while the  protective features of the GMWB annuity mitigate the adverse effect on the expected rate of return of the index, thereby yielding overall higher fair values.

Table \ref{tab:CBsAndPriceDecomposition} shows a decomposition of the fair value of a GMWB annuity with step-up in the two interest rate scenarios. The percentages are expressed with respect to the fair value of a GMWB annuity with step-up in the dynamic withdrawal case and the column ``CB'' refers to the present value of ten annual constant cashflows equal to $G$. The column ``time value'' reports the ratio between CB and the total value of the GMWB annuity with step-up, while the columns ``GMWB'' and ``step-up'' report the proportion of  value due to the GMWB and step-up features, respectively.
We can observe that the value decomposition changes significantly across the two interest rate scenarios. In particular, the 2021 scenario is characterized by very low interest rates and, therefore, the time value alone accounts for more than 97\% of the total value of the annuity. Additional features such as GMWB and step-up can only be included at the expense of large penalties and management fees, thereby reducing their contribution to the total value of the annuity. On the contrary, in the current 2022 scenario, both the GMWB and the step-up feature represent significant portions of the total value of the GMWB annuity with step-up.

\begin{table}
\small
    \centering
    \begin{tabular}{c|c|c|c|c|}
         scenario & CB & time value & GMWB value & step-up value \\ \hline
         2021 & 99.91 & 97.38\% & 0.09\% & 2.53\% \\
         2022 & 86.10 & 78.44\% & 12.60\% & 8.96\% \\ \hline
    \end{tabular}
    \caption{Decomposition of the fair value of the GMWB annuity with the step-up feature under the dynamic withdrawal scheme. CB is the present value of ten annual constant cashflows equal to $G$. Parameters as in Tables \ref{tab:optimalFees2021} and \ref{tab:optimalFees2010}.}
    \label{tab:CBsAndPriceDecomposition}
\end{table}

\subsubsection{Value of the surrender feature}

The different interest rate environment impacts also the value of the surrender option when comparing the value of a GMWB annuity under the S and S+S schemes.
Indeed, considering Tables \ref{tab:optimalFees2021} and \ref{tab:optimalFees2010} and restricting to the cases $\alpha \in \{ 0\%, 5\% \}$ and $\beta \in \{ 0\%, 5\%, 10\% \}$ for which the fair value of the annuity is reported in both tables, we can verify that the percentage increase in the value of the annuity when moving from the S to the S+S scheme is much larger in the current 2022 scenario rather than in the 2021 one. For example, for $\alpha = \beta = 5\%$, in the 2021 scenario with no step-up (resp. with step-up) the annuity is worth 103.13 (resp. 109.35) under S and 104.07 (resp. 112.60) under S+S scheme, so that the surrender feature yields a 0.91\% (resp. 2.97\%) value increase. The same percentage in the 2022 scenario reads instead 5.56\% (resp. 6.47\%). This is due to the fact that in a low/negative interest rate environment there is little incentive to surrender the contract before maturity.

\subsection{Determinants of the GMWB annuity's price}\label{subsec:determinantsOfThePrice}

In this section, we analyze the impact of mortality risk, we perform a sensitivity analysis with respect to the model parameters and we determine the additional value due to the bonus feature.

\subsubsection{Mortality risk}

Several previous works (from the seminal paper \cite{ChenForsyth2008} to the more recent work of \cite{GoudenegeMolentZanette2019}, for instance) do not consider mortality risk claiming that it is a diversifiable risk that should not be priced. To check whether mortality risk has a sizable impact, we compute the fair values for the benchmark cases considered in Tables \ref{tab:optimalFees2021} and \ref{tab:optimalFees2010} removing mortality risk. The results are shown in Table \ref{tab:optimalFeesNOMortalityRisk}.
We can observe that removing mortality risk yields a small increase of the fair values in the 2021 scenario, while it yields a small decrease in the current 2022 scenario. This is possible since the cashflow \eqref{eqn:deathEventCashFlow} of the death benefit does not  coincide with the one associated to a voluntary surrender. Therefore, there is no strict dominance of the case without mortality risk over the one which accounts for it, nor the other way round. The main factor determining the results of Table \ref{tab:optimalFeesNOMortalityRisk} is again the term structure of interest rates. Indeed, in the 2021 scenario interest rates are negative and this implies that, on average, the present value of future cashflows is higher than their nominal amounts: if a sudden death event takes place, the PH cannot benefit anymore from this situation and this reduces the value of the annuity. In the current scenario, instead, interest rates are positive and the present value of future cashflows is strictly smaller than their nominal value. In this case, a sudden death event might interrupt the discounting effect, thereby yielding a higher fair value.

\begin{table}[]
\small \begin{center}
    \begin{tabular}{c|c|c|}
            \multicolumn{2}{c|}{2021: Without step-up,} & \multirow{2}{*}{$\alpha = 13.51\%$} \\
            \multicolumn{2}{c|}{no mortality risk} & \\ \hline
             \multirow{3}{*}{$\beta = 15\%$} & S & 100.58  \\
             & S+S & 101.27 \\
             & D & 101.62 \\ \hline
        \end{tabular}
        \quad
        \begin{tabular}{c|c|c|}
            \multicolumn{2}{c|}{2021: With step-up,} & \multirow{2}{*}{$\alpha = 13.51\%$} \\
            \multicolumn{2}{c|}{no mortality risk} & \\ \hline
             \multirow{3}{*}{$\beta = 15\%$} & S & 103.55  \\
             & S+S & 104.10 \\
             & D & 104.69 \\ \hline
        \end{tabular}
    \end{center}
    \vspace*{0.5 cm}
    \begin{center}
    \begin{tabular}{c|c|c|}
        \multicolumn{2}{c|}{2022: Without step-up,} & \multirow{2}{*}{$\alpha = 0.6270\%$} \\
        \multicolumn{2}{c|}{no mortality risk} & \\ \hline
        \multirow{3}{*}{$\beta = 10\%$} & S & 96.75  \\
        & S+S & 97.39 \\
        & D & 99.60 \\ \hline
        \end{tabular}
        \quad
    \begin{tabular}{c|c|c|}
        \multicolumn{2}{c|}{2022: With step-up,} & \multirow{2}{*}{$\alpha = 0.6270\%$} \\
        \multicolumn{2}{c|}{no mortality risk} & \\ \hline
        \multirow{3}{*}{$\beta = 10\%$} & S & 105.55  \\
        & S+S & 108.12 \\
        & D & 109.67 \\ \hline
    \end{tabular}
    \end{center}
    \caption{Fair values of a GMWB annuity at $t=0$, with no mortality risk. Parameters as in Tables \ref{tab:optimalFees2021} and \ref{tab:optimalFees2010}.}
    \label{tab:optimalFeesNOMortalityRisk}
\end{table}

\subsubsection{Impact of the parameters of the market model}

We now analyse the sensitivity of the fair value of a GMWB annuity with respect to the parameters of the model described in Section \ref{subsec:theMarketModel}. Table \ref{tab:sensitivity2021} reports the sensitivity of the fair value of the GMWB annuity with respect to changes in $\sigma_S$, $\sigma_r$ and $\rho$ in the 2021 scenario, while the results for the current 2022 scenario are reported in Table \ref{tab:sensitivity2010}.
In Table \ref{tab:sensitivity2021}, we also report the results of an additional numerical test, consisting in the computation by Monte Carlo simulations of the 95\% confidence interval of the fair value of a  GMWB annuity (with and without step-up feature, with static withdrawals). This test aims at assessing the reliability of our algorithm across different configurations of the model parameters. Out of the 100 prices computed by relying on our algorithm, only 7 do not belong to the Monte Carlo confidence interval. The largest absolute difference  between the value obtained through our algorithm and the Monte Carlo estimate is equal to 0.26, while the average of all differences is 0.00, implying that our valuation algorithm has no systematic bias.

\begin{table}[]
    \centering
        \resizebox{\textwidth}{!}{\begin{tabular}{c|c|c|c|c|c|c|c|}
        \multicolumn{3}{c|}{2021: Without} & \multicolumn{5}{c|}{$\sigma_S$} \\ \cline{4-8}
        \multicolumn{3}{c|}{step-up} & 10\% & 15\% & 20\% & 25\% & 30\% \\ \hline
        \multirow{20}{*}{$\sigma_r$} & \multirow{4}{*}{1\%} & \multirow{2}{*}{S (MC)} & 98.60 & 98.60 & 98.91 & 99.35 & 99.67 \\ 
        & & & $^{\pm0.039}$ & $^{\pm0.039}$ & $^{\pm0.041}$ & $^{\pm0.048}$ & $^{\pm0.068}$ \\
        & & S & 98.57 & 98.67 & 98.89 & 99.26 & 99.70 \\ 
        & & D & 98.62 & 98.74 & 98.99 & 99.41 & 99.94 \\ \cline{2-8}
        & \multirow{4}{*}{1.5\%} & \multirow{2}{*}{S (MC)} & 98.67 & 98.78 & 99.03 & 99.46 & 99.81 \\ 
        & & & $^{\pm0.056}$ & $^{\pm0.056}$ & $^{\pm0.057}$ & $^{\pm0.062}$ & $^{\pm0.074}$ \\
        & & S & 98.63 & 98.75 & 99.00 & 99.39 & 99.84 \\ 
        & & D & 98.95 & 99.08 & 99.35 & 99.79 & 100.32 \\ \cline{2-8}
        & \multirow{4}{*}{2\%} & \multirow{2}{*}{S (MC)} & 98.72 & 98.90 & 99.11 & 99.48 & 100.11 \\ 
        & & & $^{\pm0.073}$ & $^{\pm0.073}$ & $^{\pm0.074}$ & $^{\pm0.077}$ & $^{\pm0.086}$ \\
        & & S & 98.71 & 98.85 & 99.13 & 99.54 & 100.00 \\ 
        & & D & 99.57 & 99.71 & 100.00 & 100.45 & 101.00 \\ \cline{2-8}
        & \multirow{4}{*}{2.5\%} & \multirow{2}{*}{S (MC)} & 98.74 & 98.90 & 99.24 & 99.63 & 100.20 \\ 
        & & & $^{\pm0.091}$ & $^{\pm0.09}$ & $^{\pm0.09}$ & $^{\pm0.093}$ & $^{\pm0.103}$ \\
        & & S & 98.82 & 98.98 & 99.29 & 99.72 & 100.18 \\ 
        & & D & 100.55 & 100.69 & 100.99 & 101.44 & 101.99 \\ \cline{2-8}
        & \multirow{4}{*}{3\%} & \multirow{2}{*}{S (MC)} & 99.00 & 99.12 & 99.46 & 99.90 & 100.28 \\ 
        & & & $^{\pm0.109}$ & $^{\pm0.108}$ & $^{\pm0.108}$ & $^{\pm0.11}$ & $^{\pm0.118}$ \\
        & & S & 98.96 & 99.14 & 99.48 & 99.92 & 100.38 \\ 
        & & D & 101.79 & 101.94 & 102.23 & 102.67 & 103.21 \\ \hline
        \end{tabular}
        \begin{tabular}{c|c|c|c|c|c|c|c|}
        \multicolumn{3}{c|}{2021: With} & \multicolumn{5}{c|}{$\sigma_S$} \\ \cline{4-8}
        \multicolumn{3}{c|}{step-up} & 10\% & 15\% & 20\% & 25\% & 30\% \\ \hline
        \multirow{20}{*}{$\sigma_r$} & \multirow{4}{*}{1\%} & \multirow{2}{*}{S (MC)} & 98.69 & 99.57 & 101.33 & 103.53 & 106.31 \\ 
        & & & $^{\pm0.04}$ & $^{\pm0.043}$ & $^{\pm0.056}$ & $^{\pm0.079}$ & $^{\pm0.115}$ \\
        & & S & 98.70 & 99.62 & 101.27 & 103.54 & 106.34 \\ 
        & & D & 98.79 & 99.77 & 101.49 & 103.78 & 106.42 \\ \cline{2-8}
        & \multirow{4}{*}{1.5\%} & \multirow{2}{*}{S (MC)} & 98.74 & 99.59 & 101.31 & 103.54 & 106.32 \\ 
        & & & $^{\pm0.056}$ & $^{\pm0.058}$ & $^{\pm0.065}$ & $^{\pm0.086}$ & $^{\pm0.118}$ \\
        & & S & 98.68 & 99.65 & 101.29 & 103.54 & 106.30 \\ 
        & & D & 99.13 & 100.15 & 101.90 & 104.21 & 106.89 \\ \cline{2-8}
        & \multirow{4}{*}{2\%} & \multirow{2}{*}{S (MC)} & 98.74 & 99.68 & 101.38 & 103.62 & 106.29 \\ 
        & & & $^{\pm0.073}$ & $^{\pm0.073}$ & $^{\pm0.079}$ & $^{\pm0.094}$ & $^{\pm0.122}$ \\
        & & S & 98.70 & 99.70 & 101.36 & 103.61 & 106.24 \\ 
        & & D & 99.77 & 100.82 & 102.60 & 104.94 & 107.62 \\ \cline{2-8}
        & \multirow{4}{*}{2.5\%} & \multirow{2}{*}{S (MC)} & 98.80 & 99.72 & 101.37 & 103.63 & 106.28 \\ 
        & & & $^{\pm0.091}$ & $^{\pm0.091}$ & $^{\pm0.094}$ & $^{\pm0.107}$ & $^{\pm0.129}$ \\
        & & S & 98.71 & 99.67 & 101.46 & 103.54 & 106.27 \\ 
        & & D & 100.75 & 101.82 & 103.63 & 105.96 & 108.64 \\ \cline{2-8}
        & \multirow{4}{*}{3\%} & \multirow{2}{*}{S (MC)} & 98.81 & 99.75 & 101.38 & 103.59 & 106.26 \\ 
        & & & $^{\pm0.109}$ & $^{\pm0.108}$ & $^{\pm0.11}$ & $^{\pm0.121}$ & $^{\pm0.143}$ \\
        & & S & 98.87 & 99.80 & 101.36 & 103.53 & 106.27 \\ 
        & & D & 102.01 & 103.08 & 104.89 & 107.22 & 109.88 \\ \hline
        \end{tabular}
        }
        \newline
        \vspace*{0.5 cm}
        \newline
        \resizebox{\textwidth}{!}{\begin{tabular}{c|c|c|c|c|c|c|c|}
        \multicolumn{3}{c|}{2021: Without} & \multicolumn{5}{c|}{$\sigma_r$} \\ \cline{4-8}
        \multicolumn{3}{c|}{step-up} & 1\% & 1.5\% & 2\% & 2.5\% & 3\% \\ \hline
        \multirow{20}{*}{$\rho$} & \multirow{4}{*}{-90\%} & \multirow{2}{*}{S (MC)} & 98.82 & 98.83 & 98.87 & 98.86 & 98.99 \\ 
        & & & $^{\pm0.135}$ & $^{\pm0.118}$ & $^{\pm0.109}$ & $^{\pm0.102}$ & $^{\pm0.093}$ \\
        & & S & 98.78 & 98.81 & 98.86 & 98.94 & 99.05 \\ 
        & & D & 98.83 & 99.12 & 99.71 & 100.68 & 101.90 \\ \cline{2-8}
        & \multirow{4}{*}{-50\%} & \multirow{2}{*}{S (MC)} & 98.87 & 98.86 & 98.90 & 98.93 & 99.03 \\ 
        & & & $^{\pm0.134}$ & $^{\pm0.12}$ & $^{\pm0.111}$ & $^{\pm0.103}$ & $^{\pm0.095}$ \\
        & & S & 98.80 & 98.84 & 98.90 & 98.99 & 99.10 \\ 
        & & D & 98.87 & 99.17 & 99.76 & 100.72 & 101.95 \\ \cline{2-8}
        & \multirow{4}{*}{0} & \multirow{2}{*}{S (MC)} & 98.88 & 98.98 & 99.02 & 99.02 & 99.13 \\ 
        & & & $^{\pm0.134}$ & $^{\pm0.121}$ & $^{\pm0.109}$ & $^{\pm0.1}$ & $^{\pm0.098}$ \\
        & & S & 98.84 & 98.91 & 99.00 & 99.12 & 99.26 \\ 
        & & D & 98.93 & 99.26 & 99.88 & 100.85 & 102.09 \\ \cline{2-8}
        & \multirow{4}{*}{50\%} & \multirow{2}{*}{S (MC)} & 98.91 & 99.08 & 99.08 & 99.23 & 99.41 \\ 
        & & & $^{\pm0.13}$ & $^{\pm0.115}$ & $^{\pm0.106}$ & $^{\pm0.098}$ & $^{\pm0.09}$ \\
        & & S & 98.89 & 99.00 & 99.13 & 99.29 & 99.48 \\ 
        & & D & 98.99 & 99.35 & 100.00 & 100.99 & 102.23 \\ \cline{2-8}
        & \multirow{4}{*}{90\%} & \multirow{2}{*}{S (MC)} & 98.85 & 99.16 & 99.23 & 99.50 & 99.81 \\ 
        & & & $^{\pm0.125}$ & $^{\pm0.115}$ & $^{\pm0.104}$ & $^{\pm0.094}$ & $^{\pm0.088}$ \\
        & & S & 98.93 & 99.08 & 99.25 & 99.45 & 99.68 \\ 
        & & D & 99.03 & 99.42 & 100.09 & 101.08 & 102.31 \\ \hline
        \end{tabular}
        \begin{tabular}{c|c|c|c|c|c|c|c|}
        \multicolumn{3}{c|}{2021: With} & \multicolumn{5}{c|}{$\sigma_r$} \\ \cline{4-8}
        \multicolumn{3}{c|}{step-up} & 1\% & 1.5\% & 2\% & 2.5\% & 3\% \\ \hline
        \multirow{20}{*}{$\rho$} & \multirow{4}{*}{-90\%} & \multirow{2}{*}{S (MC)} & 101.42 & 101.16 & 101.23 & 101.10 & 100.93 \\ 
        & & & $^{\pm0.264}$ & $^{\pm0.234}$ & $^{\pm0.209}$ & $^{\pm0.195}$ & $^{\pm0.183}$ \\
        & & S & 101.23 & 101.22 & 101.14 & 101.20 & 101.06 \\ 
        & & D & 101.87 & 102.17 & 102.83 & 103.79 & 104.99 \\ \cline{2-8}
        & \multirow{4}{*}{-50\%} & \multirow{2}{*}{S (MC)} & 101.36 & 101.17 & 101.29 & 101.30 & 101.18 \\ 
        & & & $^{\pm0.23}$ & $^{\pm0.209}$ & $^{\pm0.19}$ & $^{\pm0.176}$ & $^{\pm0.17}$ \\
        & & S & 101.28 & 101.22 & 101.22 & 101.23 & 101.21 \\ 
        & & D & 101.57 & 101.90 & 102.58 & 103.57 & 104.81 \\ \cline{2-8}
        & \multirow{4}{*}{0} & \multirow{2}{*}{S (MC)} & 101.22 & 101.27 & 101.20 & 101.28 & 101.38 \\ 
        & & & $^{\pm0.184}$ & $^{\pm0.171}$ & $^{\pm0.162}$ & $^{\pm0.155}$ & $^{\pm0.147}$ \\
        & & S & 101.23 & 101.26 & 101.30 & 101.27 & 101.30 \\ 
        & & D & 101.49 & 101.85 & 102.54 & 103.55 & 104.80 \\ \cline{2-8}
        & \multirow{4}{*}{50\%} & \multirow{2}{*}{S (MC)} & 101.29 & 101.12 & 101.40 & 101.41 & 101.44 \\ 
        & & & $^{\pm0.151}$ & $^{\pm0.141}$ & $^{\pm0.133}$ & $^{\pm0.132}$ & $^{\pm0.129}$ \\
        & & S & 101.30 & 101.29 & 101.38 & 101.28 & 101.43 \\ 
        & & D & 101.49 & 101.90 & 102.60 & 103.62 & 104.89 \\ \cline{2-8}
        & \multirow{4}{*}{90\%} & \multirow{2}{*}{S (MC)} & 101.29 & 101.37 & 101.60 & 101.53 & 101.63 \\ 
        & & & $^{\pm0.128}$ & $^{\pm0.118}$ & $^{\pm0.115}$ & $^{\pm0.111}$ & $^{\pm0.112}$ \\
        & & S & 101.28 & 101.33 & 101.39 & 101.44 & 101.59 \\ 
        & & D & 101.57 & 102.09 & 102.83 & 103.90 & 105.21 \\ \hline
        \end{tabular}
        }
    \caption{Fair value of a GMWB annuity at $t=0$ for different values of $\sigma_S$ and $\sigma_r$ (resp. $\sigma_r$ and $\rho$) in the top (resp. bottom) panel. Rows labelled by S (resp. D) refer to the case of static (resp. dynamic) withdrawals. When static withdrawals are considered, rows labelled by (MC) display the Monte Carlo estimate with 100000 paths and the 95\% confidence interval. $T=10$, $\alpha = 13.51\%$, $\beta = 15\%$ and remaining parameters as in Table \ref{tab:parameters2021}, as of 12/30/2021. Parameters: $m=2$, $n_A = 30$, $n_B = 10$ for the left panel with no step-up feature, $n_B = 30$ for the right panel  with the step-up feature.}
    \label{tab:sensitivity2021}
\end{table}

\begin{table}[]
    \centering
        \resizebox{\textwidth}{!}{\begin{tabular}{c|c|c|c|c|c|c|c|}
        \multicolumn{3}{c|}{2022: Without} & \multicolumn{5}{c|}{$\sigma_S$} \\ \cline{4-8}
        \multicolumn{3}{c|}{step-up} & 10\% & 15\% & 20\% & 25\% & 30\% \\ \hline
        \multirow{10}{*}{$\sigma_r$} & \multirow{2}{*}{1\%} & S & 92.42 & 94.63 & 96.25 & 96.98 & 96.93 \\ 
        & & D & 93.31 & 95.5 & 97.51 & 99.07 & 100.00 \\ \cline{2-8}
        & \multirow{2}{*}{1.5\%} & S & 93.13 & 95.72 & 97.43 & 98.03 & 97.72 \\ 
        & & D & 94.13 & 96.81 & 99.26 & 101.10 & 102.10 \\ \cline{2-8}
        & \multirow{2}{*}{2\%} & S & 93.36 & 95.95 & 97.48 & 97.82 & 97.25 \\ 
        & & D & 94.51 & 97.38 & 100.00 & 101.85 & 102.82 \\ \cline{2-8}
        & \multirow{2}{*}{2.5\%} & S & 93.35 & 95.73 & 96.96 & 97.01 & 96.20 \\ 
        & & D & 94.70 & 97.64 & 100.25 & 102.01 & 102.90 \\ \cline{2-8}
        & \multirow{2}{*}{3\%} & S & 93.15 & 95.20 & 96.07 & 95.84 & 94.82 \\ 
        & & D & 94.80 & 97.71 & 100.22 & 101.86 & 102.64 \\ \hline
        \end{tabular}
        \begin{tabular}{c|c|c|c|c|c|c|c|}
        \multicolumn{3}{c|}{2022: With} & \multicolumn{5}{c|}{$\sigma_S$} \\ \cline{4-8}
        \multicolumn{3}{c|}{step-up} & 10\% & 15\% & 20\% & 25\% & 30\% \\ \hline
        \multirow{10}{*}{$\sigma_r$} & \multirow{2}{*}{1\%} & S & 94.91 & 98.96 & 102.66 & 105.58 & 107.65 \\ 
        & & D & 95.87 & 100.18 & 104.95 & 109.44 & 113.14 \\ \cline{2-8}
        & \multirow{2}{*}{1.5\%} & S & 96.07 & 100.87 & 105.13 & 108.42 & 110.73 \\ 
        & & D & 97.24 & 102.58 & 108.29 & 113.66 & 118.18 \\ \cline{2-8}
        & \multirow{2}{*}{2\%} & S & 96.50 & 101.42 & 105.66 & 108.88 & 111.12 \\ 
        & & D & 97.88 & 103.69 & 109.84 & 115.59 & 120.50 \\ \cline{2-8}
        & \multirow{2}{*}{2.5\%} & S & 96.58 & 101.31 & 105.28 & 108.26 & 110.30 \\ 
        & & D & 98.24 & 104.28 & 110.61 & 116.56 & 121.65 \\ \cline{2-8}
        & \multirow{2}{*}{3\%} & S & 96.42 & 100.78 & 104.38 & 107.04 & 108.85 \\ 
        & & D & 98.47 & 104.59 & 111.00 & 117.02 & 122.19 \\ \hline
        \end{tabular}}
        \newline
        \vspace*{1 cm}
        \newline
        \resizebox{\textwidth}{!}{\begin{tabular}{c|c|c|c|c|c|c|c|}
        \multicolumn{3}{c|}{2022: Without} & \multicolumn{5}{c|}{$\sigma_r$} \\ \cline{4-8}
        \multicolumn{3}{c|}{step-up} & 1\% & 1.5\% & 2\% & 2.5\% & 3\% \\ \hline
        \multirow{10}{*}{$\rho$} & \multirow{2}{*}{-90\%} & S & 96.53 & 90.55 & 87.09 & 84.48 & 82.29 \\ 
        & & D & 97.93 & 92.99 & 90.79 & 89.46 & 88.54 \\ \cline{2-8}
        & \multirow{2}{*}{-50\%} & S & 97.39 & 93.58 & 91.09 & 89.14 & 87.50 \\ 
        & & D & 98.94 & 95.80 & 94.14 & 93.01 & 92.14 \\ \cline{2-8}
        & \multirow{2}{*}{0} & S & 97.29 & 96.00 & 94.77 & 93.56 & 92.37 \\ 
        & & D & 98.86 & 98.12 & 97.50 & 96.95 & 96.44 \\ \cline{2-8}
        & \multirow{2}{*}{50\%} & S & 96.25 & 97.43 & 97.48 & 96.96 & 96.07 \\ 
        & & D & 97.51 & 99.26 & 100.00 & 100.25 & 100.22 \\ \cline{2-8}
        & \multirow{2}{*}{90\%} & S & 94.80 & 98.04 & 99.22 & 99.24 & 98.57 \\ 
        & & D & 95.60 & 99.59 & 101.69 & 102.75 & 103.22 \\ \hline
        \end{tabular}
        \begin{tabular}{c|c|c|c|c|c|c|c|}
        \multicolumn{3}{c|}{2022: With} & \multicolumn{5}{c|}{$\sigma_r$} \\ \cline{4-8}
        \multicolumn{3}{c|}{step-up} & 1\% & 1.5\% & 2\% & 2.5\% & 3\% \\ \hline
        \multirow{10}{*}{$\rho$} & \multirow{2}{*}{-90\%} & S & 104.49 & 95.76 & 91.42 & 88.44 & 86.09 \\ 
        & & D & 108.13 & 99.98 & 96.49 & 94.44 & 93.02 \\ \cline{2-8}
        & \multirow{2}{*}{-50\%} & S & 105.22 & 99.81 & 96.63 & 94.29 & 92.40 \\ 
        & & D & 108.45 & 103.58 & 101.12 & 99.56 & 98.40 \\ \cline{2-8}
        & \multirow{2}{*}{0} & S & 104.66 & 103.14 & 101.68 & 100.26 & 98.86 \\ 
        & & D & 107.52 & 106.69 & 106.00 & 105.39 & 104.84 \\ \cline{2-8}
        & \multirow{2}{*}{50\%} & S & 102.66 & 105.13 & 105.66 & 105.28 & 104.38 \\ 
        & & D & 104.95 & 108.29 & 109.84 & 110.61 & 111.00 \\ \cline{2-8}
        & \multirow{2}{*}{90\%} & S & 99.79 & 105.84 & 108.49 & 109.21 & 108.83 \\ 
        & & D & 101.59 & 108.68 & 112.72 & 115.11 & 116.60 \\ \hline
        \end{tabular}}
    \caption{Fair value of a GMWB at $t=0$ for different values of $\sigma_S$ and $\sigma_r$ (resp. $\sigma_r$ and $\rho$) in the top (resp. bottom) panels. Rows labelled by S (resp. D) refer to the case of static (resp. dynamic) withdrawals. $T=10$, $\alpha = 0.6720\%$, $\beta = 10\%$ and remaining parameters as of 12/30/2022. Parameters: $m=2$, $n_A = 30$, $n_B = 10$ for the left panel with no step up feature, $n_B = 30$ for the right panel with the step up feature.}\label{tab:sensitivity2010}
\end{table}

The sensitivity analysis reveals the following roles of the model parameters:
\begin{itemize}
    \item[$\sigma_S$:] Fair values are always increasing in the volatility $\sigma_S$ of the underlying fund, across all interest rate scenarios and withdrawal policies, with and without the step-up feature. This finding is in line with the literature and reflects the optionality of the GMWB annuity. The sensitivity is small when there is no step-up feature, because in this case the return of the underlying fund determines only the terminal cashflow. Including the step-up feature, the sensitivity is instead quite large, since the return of the underlying fund might lead to an increase of the benefit account at each anniversary date.
    \item[$\sigma_r$:] Fair values are almost always increasing in the volatility $\sigma_r$ of the interest rate, across all scenarios and withdrawal policies, with and without the step-up feature. This finding is coherent with, e.g., \cite[Section 5.2]{KangZiveyi2018} and can be explained as follows. A larger $\sigma_r$ makes extreme realizations of $r$ more likely. On the one hand, high values of $r$ are associated to higher values of the expected return of $S$, with a positive effect on the GMWB annuity's cashflows, but with a negative effect due to the stronger discounting. On the other hand, low values of $r$ decrease the expected return of $S$, with a negative effect on the cashflows, but with a positive effect due to a milder discounting. There is therefore a trade-off in the role of $\sigma_r$, taking also into account that the optionality of the GMWB annuity and its guarantees shield only the negative impact of $r$ on the cashflows, whereas the impact on the discounting is unaffected.  We can also observe that fair values are more sensitive to changes in $\sigma_r$ in the dynamic withdrawal case.
    \item[$\rho$:] The impact of the correlation $\rho$ between market and interest rate risks is almost always positive\footnote{We acknowledge that small instabilities, due to numerical imprecision, arise when the fair price varies very little with the correlation, like in the 2021 scenario with static withdrawals when the step-up feature is included.}, with a more pronounced effect for larger values of $\sigma_r$, coherently with the findings reported in \cite[Section 4.1.2]{GudkovIgnatievaZiveyi2019}. However, the sensitivity is rather small and of second order if compared to the volatility parameters, in line with \cite[Section 5.2]{KangZiveyi2018}. 
    \item[$a$:] The impact of the speed of mean reversion parameter $a$ is of the same order of magnitude of $\rho$. Table \ref{tab:sensitivity_a_2021_2010} reports the fair values for different values of $a$. The variations in the fair values are quite limited and the monotonicity  varies across different withdrawal schemes, interest rate scenarios and depending also on the presence of the step-up feature.
\end{itemize}

\begin{table}[ht]
\small
    \centering
    \resizebox{\textwidth}{!}{\begin{tabular}{c|c|c|c|c|c|}
         2021: Without & \multicolumn{5}{c|}{$a$} \\ \cline{2-6}
         step-up & 5\% & 10\% & 15\% & 20\% & 25\% \\ \hline
         S & 99.16 & 99.13 & 99.10 & 99.08 & 99.06 \\
         D & 100.68 & 100.00 & 99.61 & 99.39 & 99.25 \\ \hline
    \end{tabular}
    \begin{tabular}{c|c|c|c|c|c|}
         2021: With & \multicolumn{5}{c|}{$a$} \\ \cline{2-6}
         step-up & 5\% & 10\% & 15\% & 20\% & 25\% \\ \hline
         S & 101.25 & 101.36 & 101.37 & 101.46 & 101.47 \\
         D & 103.25 & 102.60 & 102.22 & 102.01 & 101.88 \\ \hline
    \end{tabular}}
    \newline
    \vspace*{0.5 cm}
    \newline
    \resizebox{\textwidth}{!}{\begin{tabular}{c|c|c|c|c|c|}
         2022: Without & \multicolumn{5}{c|}{$a$} \\ \cline{2-6}
         step-up & 5\% & 10\% & 15\% & 20\% & 25\% \\ \hline
         S & 95.8 & 96.75 & 97.77 & 99.05 & 100.82 \\ 
        D & 99.06 & 99.6 & 100.3 & 101.36 & 102.99 \\ \hline
    \end{tabular}
    \begin{tabular}{c|c|c|c|c|c|}
         2022: With & \multicolumn{5}{c|}{$a$} \\ \cline{2-6}
         step-up & 5\% & 10\% & 15\% & 20\% & 25\% \\ \hline
         S & 104.17 & 105.66 & 107.14 & 108.8 & 110.84 \\ 
        D & 108.95 & 109.84 & 110.93 & 112.39 & 114.46 \\ \hline
    \end{tabular}}    
    \caption{Fair value of a GMWB at $t=0$ for different values of $a$. Parameters as in Tables \ref{tab:optimalFees2021} and \ref{tab:optimalFees2010}.}\label{tab:sensitivity_a_2021_2010}
    \label{tab:my_label}
\end{table}

\subsubsection{Valuation of the bonus feature}

We now analyze a GMWB annuity with step-up and bonus features. As already mentioned, the step-up feature alone delivers a significant benefit, with the consequent difficulty of selling at par a GMWB with step-up, unless significant penalties and fees are applied. We consider a relatively small bonus equal to $b = 2.5$ (i.e., $b = 2.5\%$ of $P$).
Table \ref{tab:pricesWithBonus} reports the fair values of the benchmark GMWB annuity in the two interest rate scenarios. Clearly, when the bonus feature is included, the fair values increase significantly. This is due to the fact that, when the bonus feature is present, the PH is prone  to substitute moderate withdrawals with zero withdrawals in order to gain the bonus. This possibility is even more valuable in the 2021 scenario, when interest rate are negative. The impact of the bonus feature on the optimal withdrawal strategies is illustrated in Figure \ref{fig:optimalWithdrawals_bonus} in the next section.

\begin{table}
\small
    \centering
    \begin{tabular}{c|c|c|c|c|c|}
         scenario & $\alpha$ & $\beta$ & no step-up & step-up only & step-up + bonus  \\ \hline
         2021 & 13.51\% & 15\% & 100.00 & 102.60 & 112.55 \\
         2022 & 0.6720\% & 10\% & 100.00 & 107.54 & 114.98 \\ \hline
    \end{tabular}
    \caption{Fair value of a GMWB annuity with $b=2.5$, dynamic withdrawals. Parameters: $m=2$, $n_A = n_B = 80$.}
    \label{tab:pricesWithBonus}
\end{table}

\subsection{Optimal withdrawal strategies}\label{subsec:optimalWithdrawalSection}

\begin{figure}
    \centering
    \begin{tabular}{p{0.47\textwidth}|p{0.47\textwidth}}
    \begin{center} Without step-up \end{center} & \begin{center} With step-up \end{center} \\
    \hspace{20pt} Panel a) & \hspace{20pt} Panel d) \\
    \includegraphics[width=0.48\textwidth]{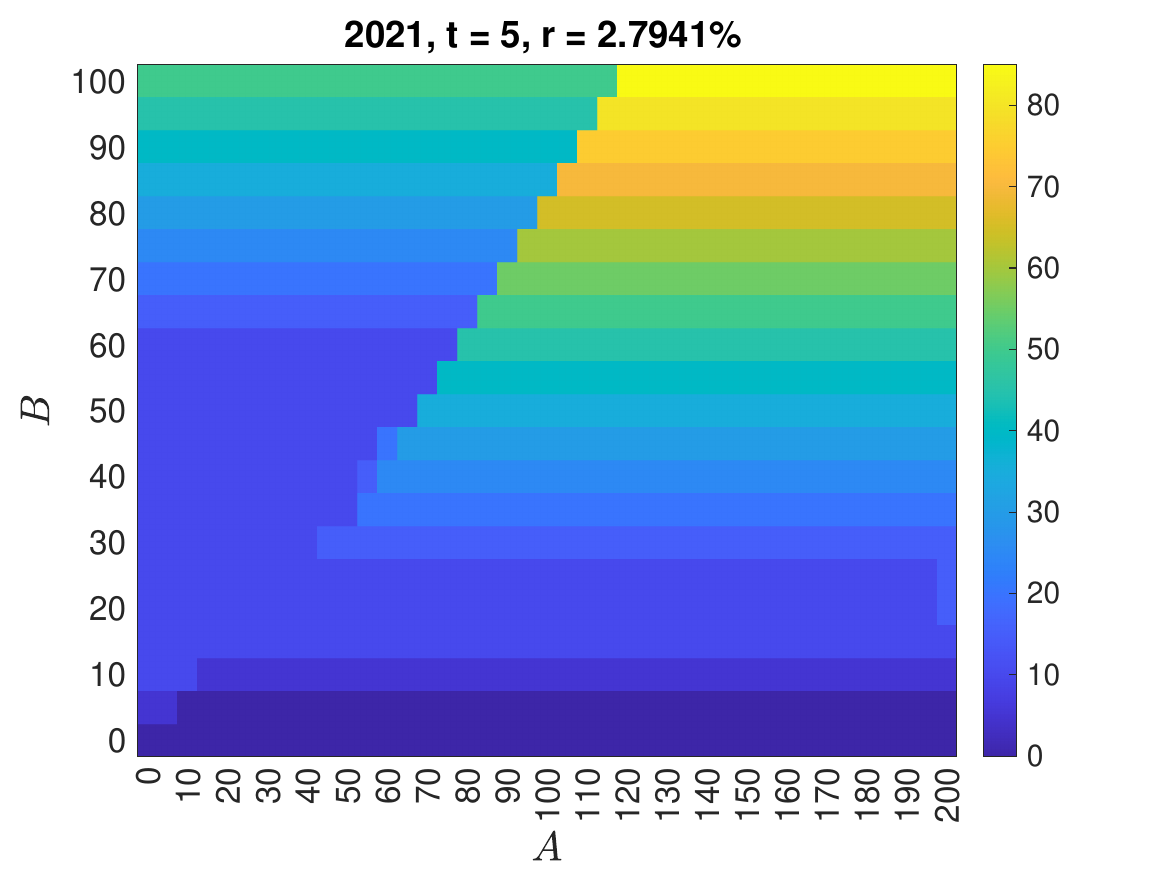} &
    \includegraphics[width=0.48\textwidth]{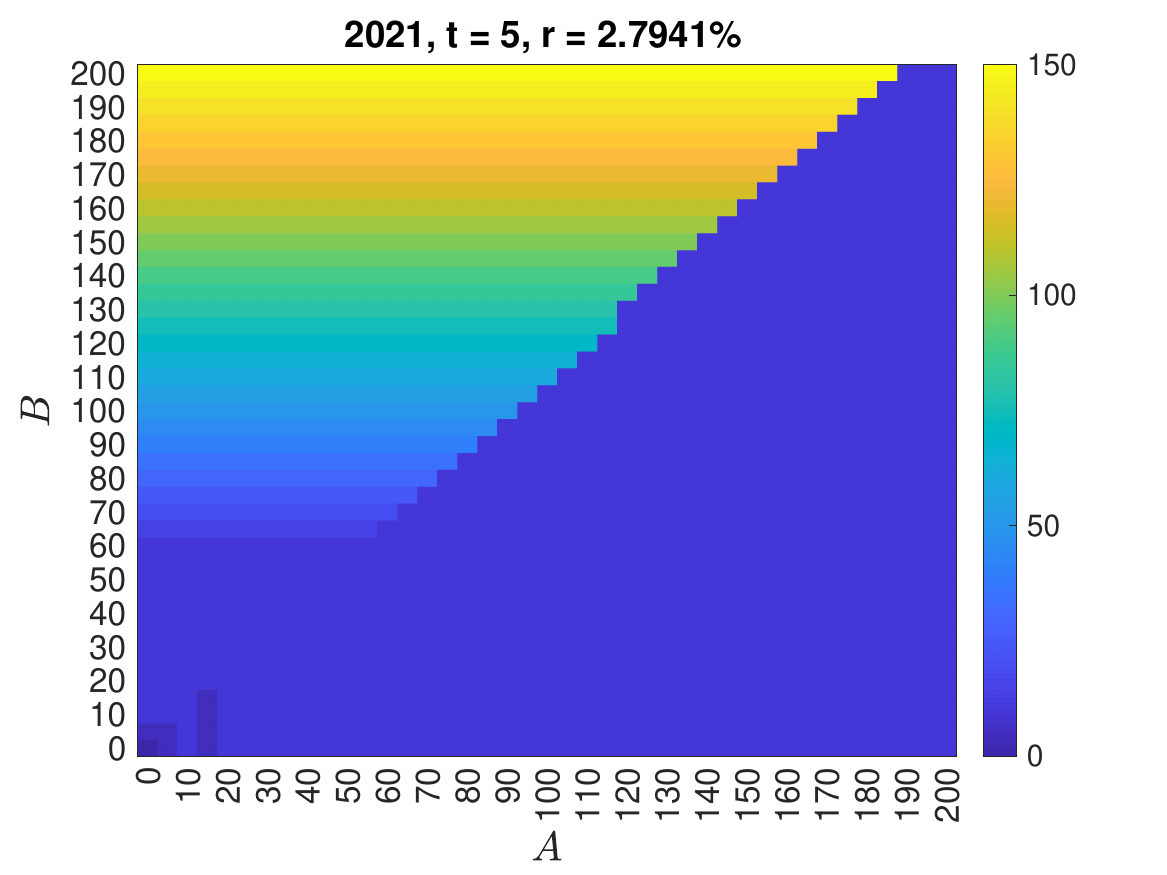} \\
    \hspace{20pt} Panel b) & \hspace{20pt} Panel e) \\
    \includegraphics[width=0.48\textwidth]{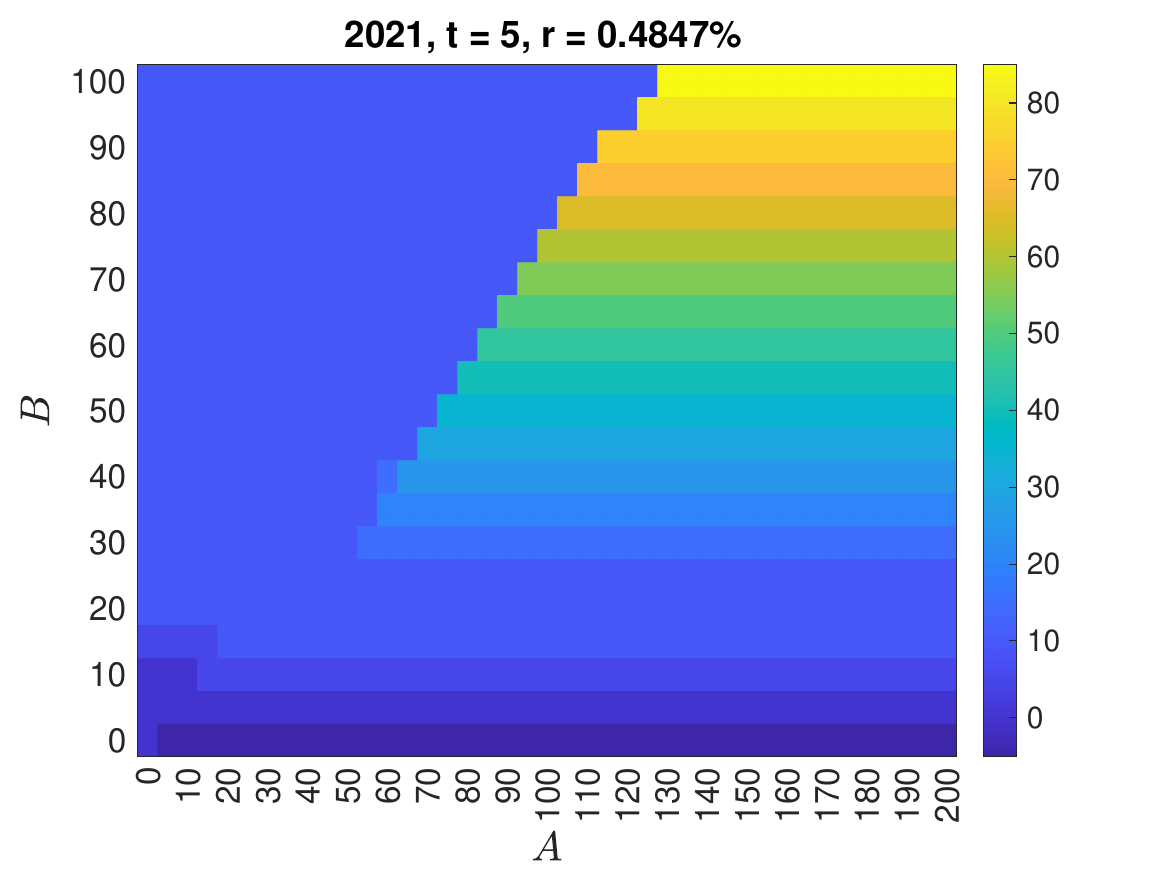} & 
    \includegraphics[width=0.48\textwidth]{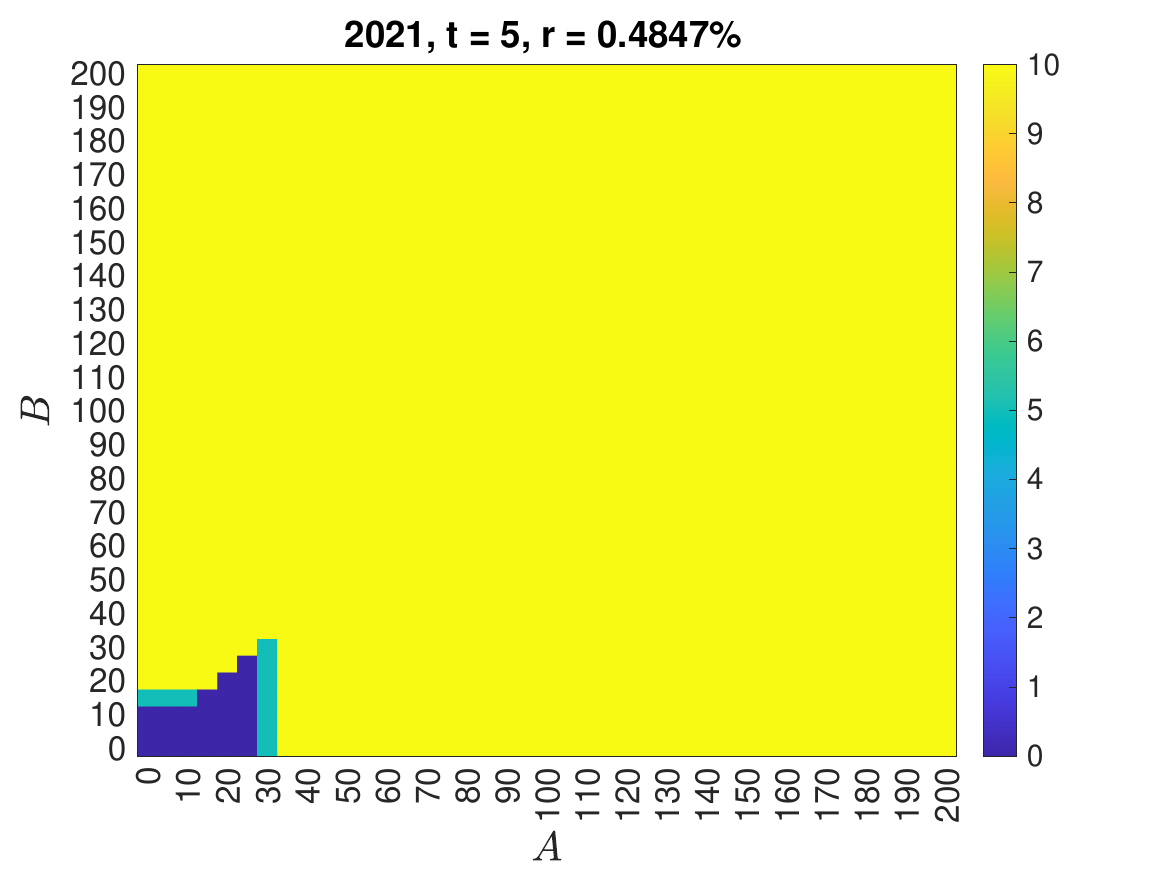} \\
    \hspace{20pt} Panel c) & \hspace{20pt} Panel f) \\
    \includegraphics[width=0.48\textwidth]{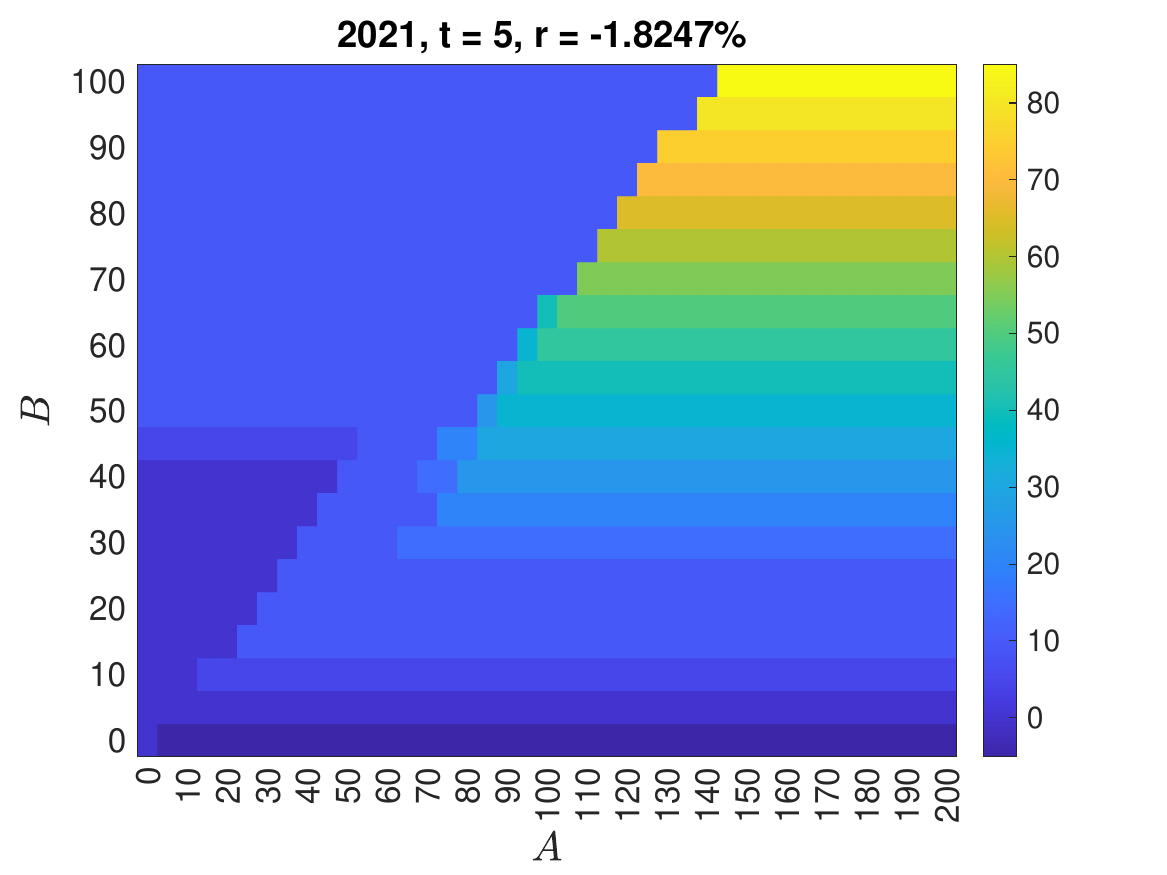} &
    \includegraphics[width=0.48\textwidth]{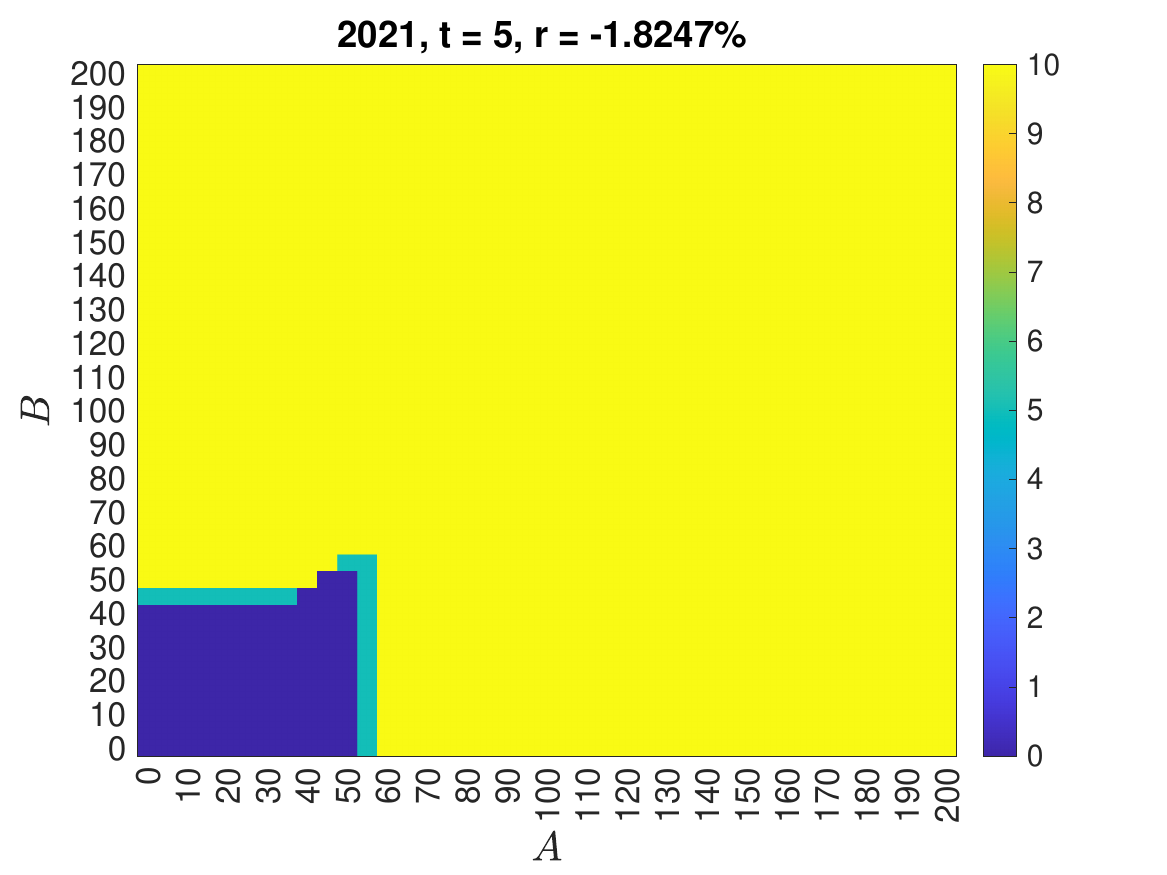}
    \end{tabular}
    \caption{Optimal withdrawal policies at $t=5$ for different level of the interest rate. $T=10$ and model parameters as of 12/31/2021. Parameters: $m=3$, $n_A = 40$, $n_B = 20$ for the panels with no step-up feature, $n_B = 40$ for the panels with the step-up feature.}
    \label{fig:optimalWithdrawals_2021}
\end{figure}

\begin{figure}
    \centering
    \begin{tabular}{p{0.47\textwidth}|p{0.47\textwidth}}
    \begin{center} Without step-up \end{center} & \begin{center} With step-up \end{center} \\
    \hspace{20pt} Panel g) & \hspace{20pt} Panel j) \\
    \includegraphics[width=0.48\textwidth]{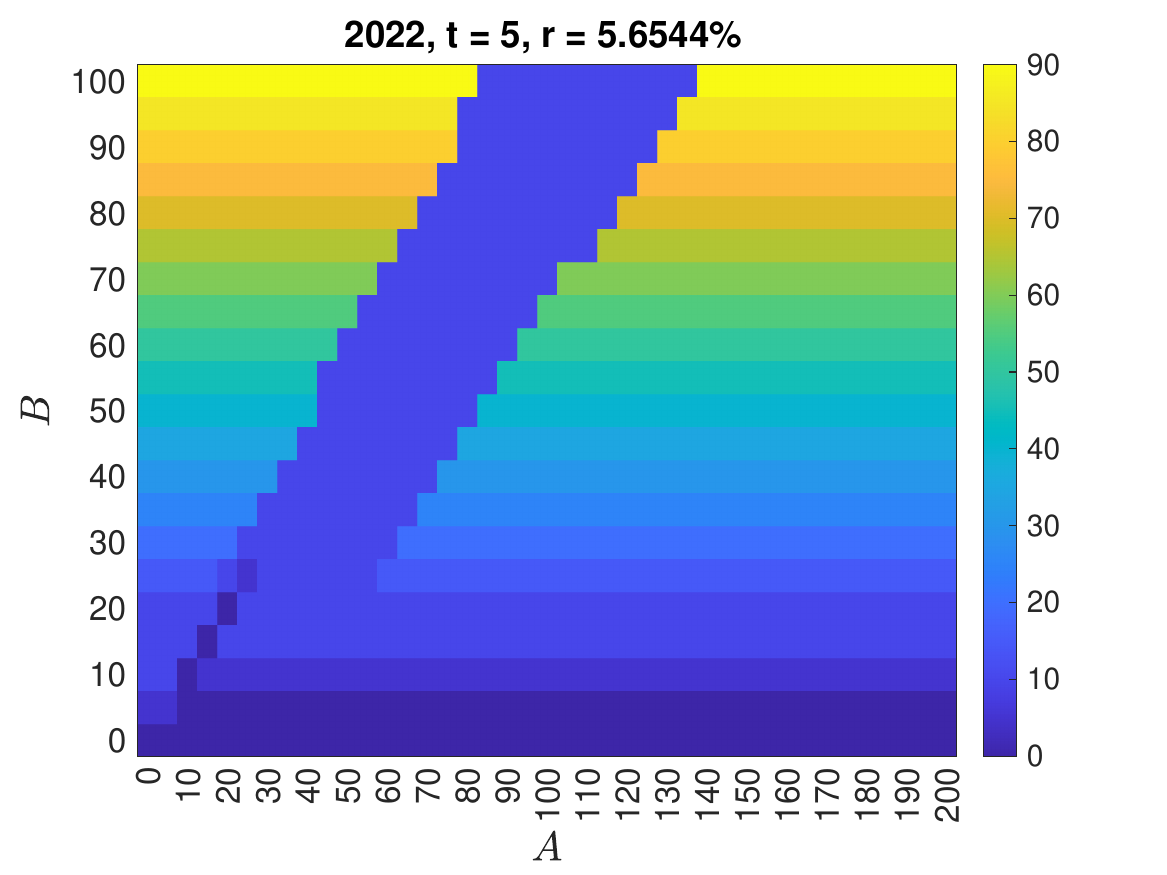} &
    \includegraphics[width=0.48\textwidth]{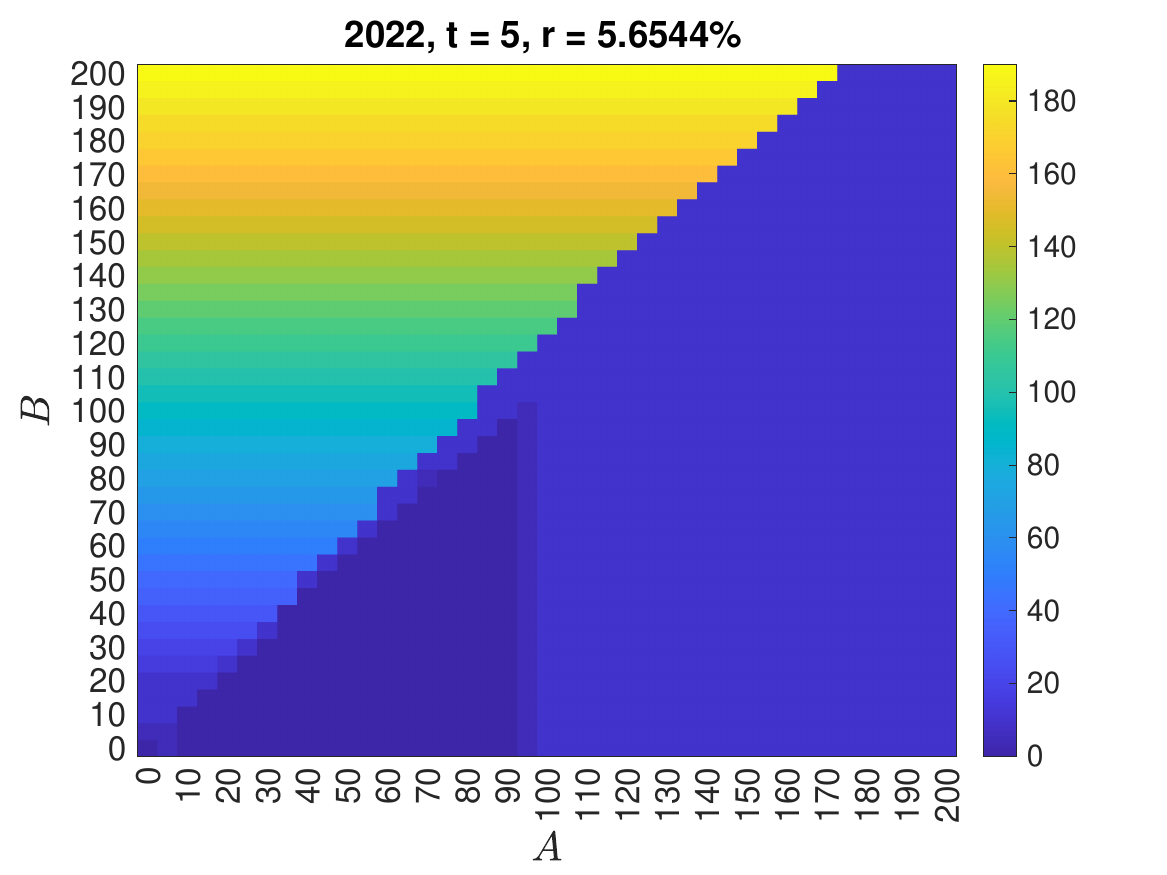} \\
    \hspace{20pt} Panel h) & \hspace{20pt} Panel k) \\
    \includegraphics[width=0.48\textwidth]{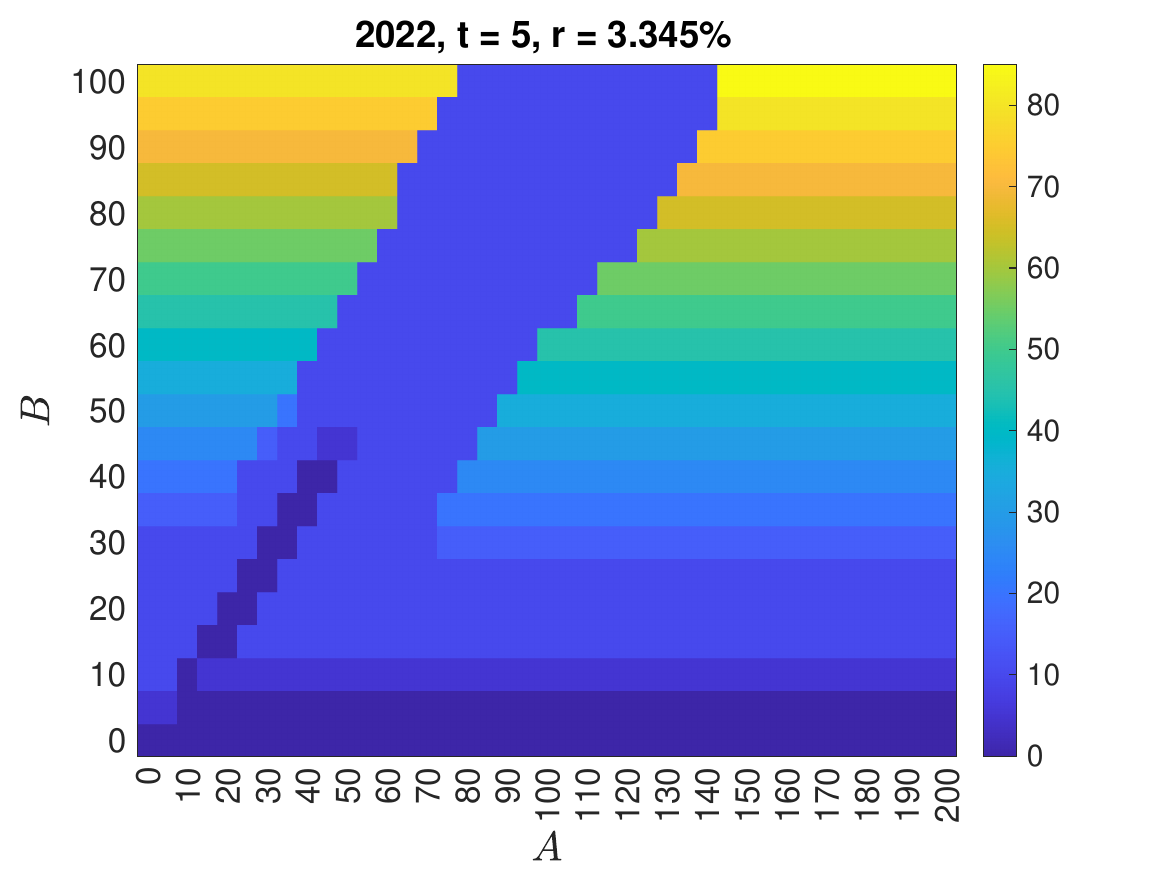} & 
    \includegraphics[width=0.48\textwidth]{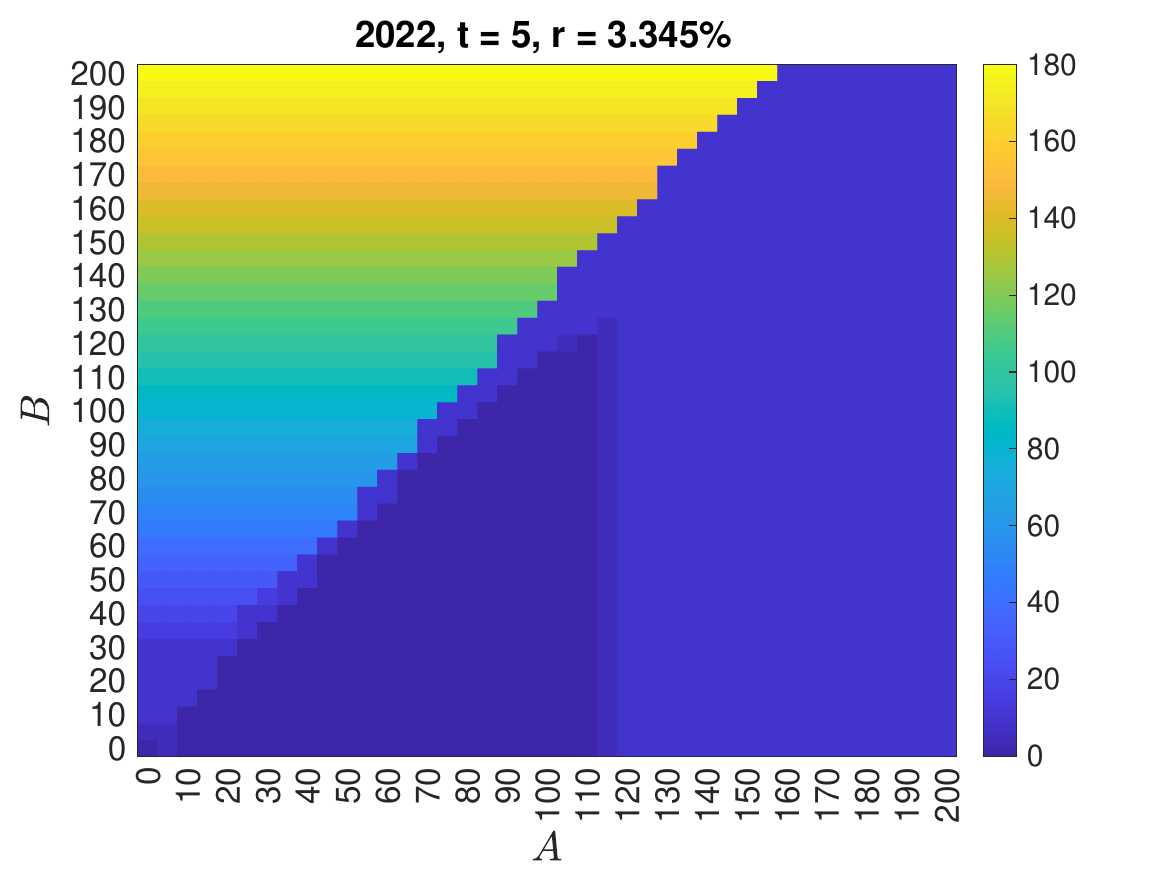} \\
    \hspace{20pt} Panel i) & \hspace{20pt} Panel l) \\
    \includegraphics[width=0.48\textwidth]{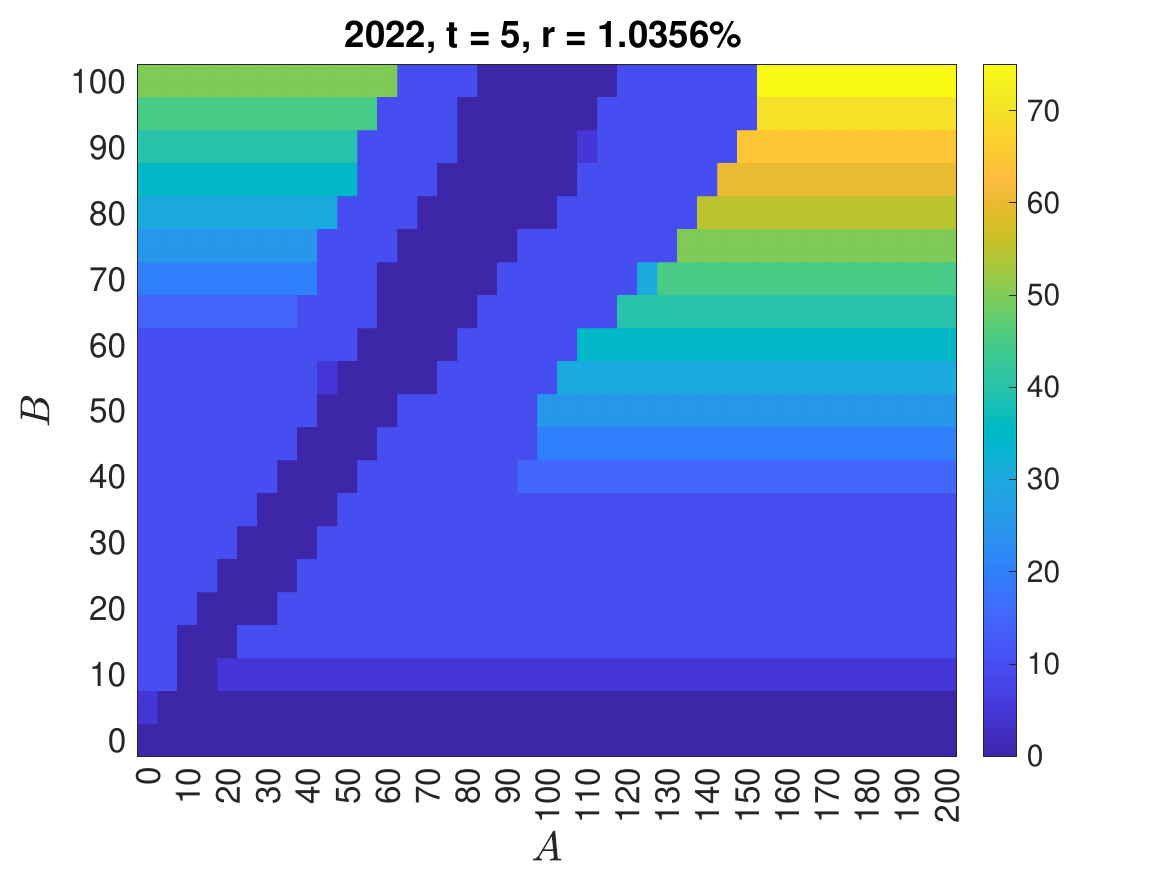} &
    \includegraphics[width=0.48\textwidth]{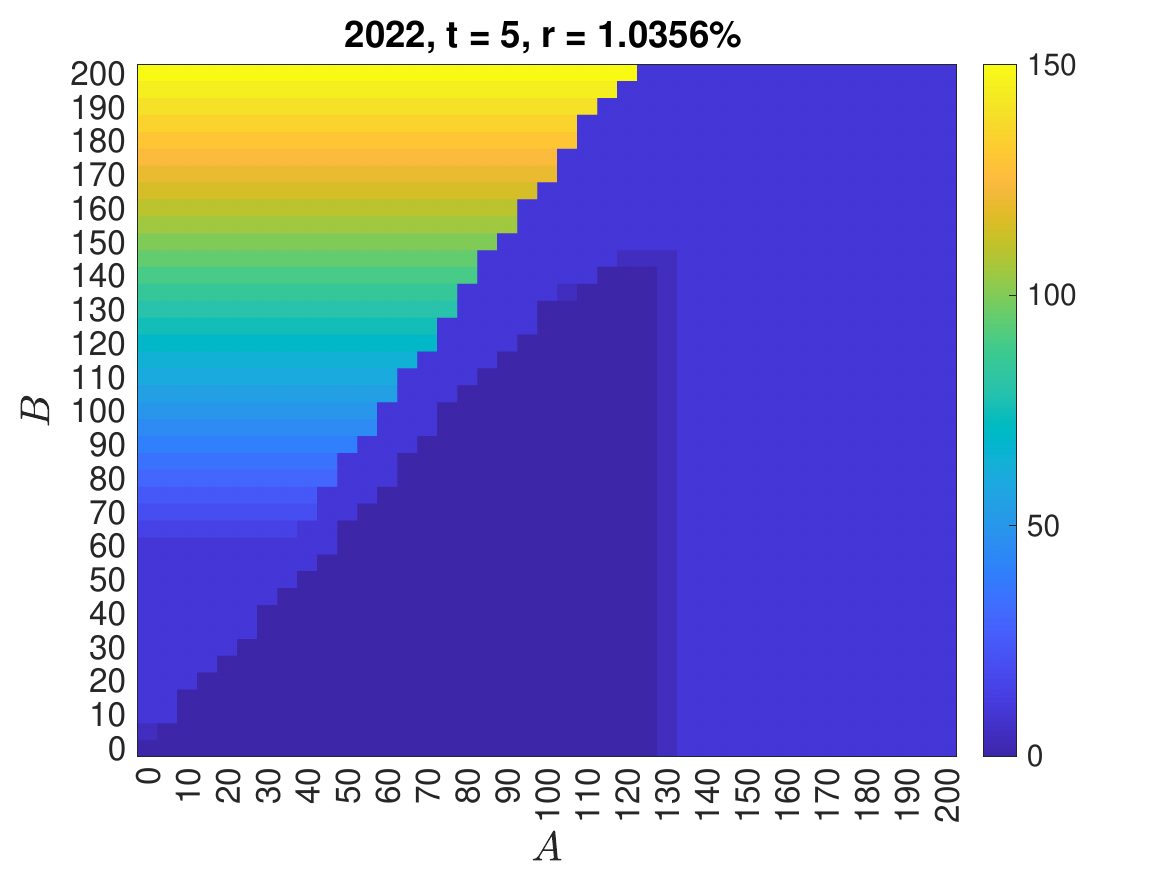}
    \end{tabular}
    \caption{Optimal withdrawal policies at $t=5$ for different level of the interest rate. $T=10$ and model parameters as of 12/30/2022. Parameters: $m=3$, $n_A = 40$, $n_B = 20$ for the panels with no step up feature, $n_B = 40$ for the panels with the step up feature.}
    \label{fig:optimalWithdrawals_2010}
\end{figure}

In this section, we investigate numerically the optimal withdrawal strategies. For each interest rate scenario, we consider the benchmark GMWB annuity, with and without the step-up feature, at five years after inception. We consider the optimal withdrawal strategies associated to the three central values among the $5(m+1)$ possible values of $r$ along its binomial tree discretization (see Section \ref{sec:numericalAlgorithm}).
In the 2021 scenario, where $r_0 = -0.67\%$, the three central nodes at $t=5$ are $r = 2.79\%$, $r=0.48\%$, $r = -1.82\%$, representing three benchmark situations: the interest rate is positive/almost zero/negative. For the 2022 scenario, where $r_0 = 2.19\%$, the three central nodes are $r = 5.65\%$, $r=3.35\%$ and $r = 1.04\%$. For each node, we consider the $n_A \times n_B$ matrix containing the optimal withdrawal decisions associated to all points in the grid $\griglia$ (see Section \ref{sec:numericalAlgorithm}).
Figures \ref{fig:optimalWithdrawals_2021} and \ref{fig:optimalWithdrawals_2010} display the optimal withdrawal strategies in the 2021 and 2022 market scenarios, respectively.

\subsubsection{No step-up feature}

We can observe that, in the 2022 scenario when the interest rate is positive (see panels g, h, i), the optimal withdrawal policies are qualitatively similar to the ones shown in \cite{ChenVetzalForsyth2008} (considering a constant and positive interest rate), in \cite{GoudenegeMolentZanette2019} (considering a Hull-White model with a negligible probability of negative rates) and in \cite{GudkovIgnatievaZiveyi2019} (considering a strictly positive CIR interest rate process).

Interestingly, even in the absence of the step-up feature, the optimal withdrawal policy in the 2021 scenario is markedly different, demonstrating that a market environment with negative rates significantly alters the optimal withdrawal decisions. To the best of our knowledge, this phenomenon has never been reported in the literature.
When the interest rate is close to zero (panel b) or even negative (panel c) and the benefit account is greater than the investment account, it is not convenient anymore to withdraw in excess of $G$, since the PH can benefit from the low interest rate environment and she is better off by withdrawing only $G$, without any penalty, and cashing in more later in time. If, on the contrary, the investment account is larger than the benefit account and sufficiently far away from zero, it is optimal to withdraw more than $G$ in order to protect the investment account from the management fee $\alpha$. When the interest rate is markedly negative (panel c), we see that there is a wider region in the bottom left corner where it is optimal to withdraw less than the guaranteed amount $G$.

\subsubsection{Step-up feature}

The inclusion of the step-up feature changes significantly the optimal withdrawal policies, showing a peculiar behavior that has never been reported in the literature so far.
In the two interest rate scenarios, we can observe that, when the interest rate is positive (panels d, j, k, l), the $A \times B$ plane splits into three regions: the one on the left of the bisector (when $B>A$), a region on the top right of the bisector (when $A>B$ and $A>A^*$, where $A^*$ is a threshold that depends on the interest rate level) and a smaller region on the bottom left of the the bisector (when $A>B$ and $A \in [0, A^*]$).
In the first region ($B>A$) it is always optimal to withdraw more than  the guaranteed amount $G$. This is because when $B>A$ it is relatively unlikely that $A$ will matter for both the step-up and the terminal payoff. Therefore, the PH is better off by withdrawing from the benefit account rather than waiting until maturity, due to the positivity of interest rates. In the second region it is optimal to withdraw only the minimum guaranteed amount $G$, because the PH wants to benefit as much as possible from the step-up feature that is likely to increase the benefit account when $A>B$. In the third region it is optimal to withdraw less than the guaranteed minimum $G$ or nothing at all, because both accounts are too close to zero. Since, anyway, the investment account is worth a bit more than the benefit account, it is convenient to wait for the next step-up, taking also into account the fact that the penalty will not be relevant for small values of the accounts.

When interest rates are negative (panels c, f) the first two regions described above are replaced by a large region where it is optimal to withdraw only the guaranteed amount $G$. Indeed, in the presence of a negative interest rate, it is generally optimal to wait and restrict withdrawals to the minimum guaranteed amount $G$, in order to avoid excessive penalties in the future withdrawals or in the payoff at maturity. 

\subsubsection{Optimal withdrawals under smaller penalties/fees}

\begin{figure}
    \centering
    \begin{tabular}{p{0.47\textwidth}|p{0.47\textwidth}}
    \begin{center} With step-up, $\alpha = \beta = 10\%$ \end{center} & \begin{center} With step-up, $\alpha = \beta = 5\%$ \end{center} \\
    \hspace{20pt} Panel m) & \hspace{20pt} Panel p) \\
    \includegraphics[width=0.48\textwidth]{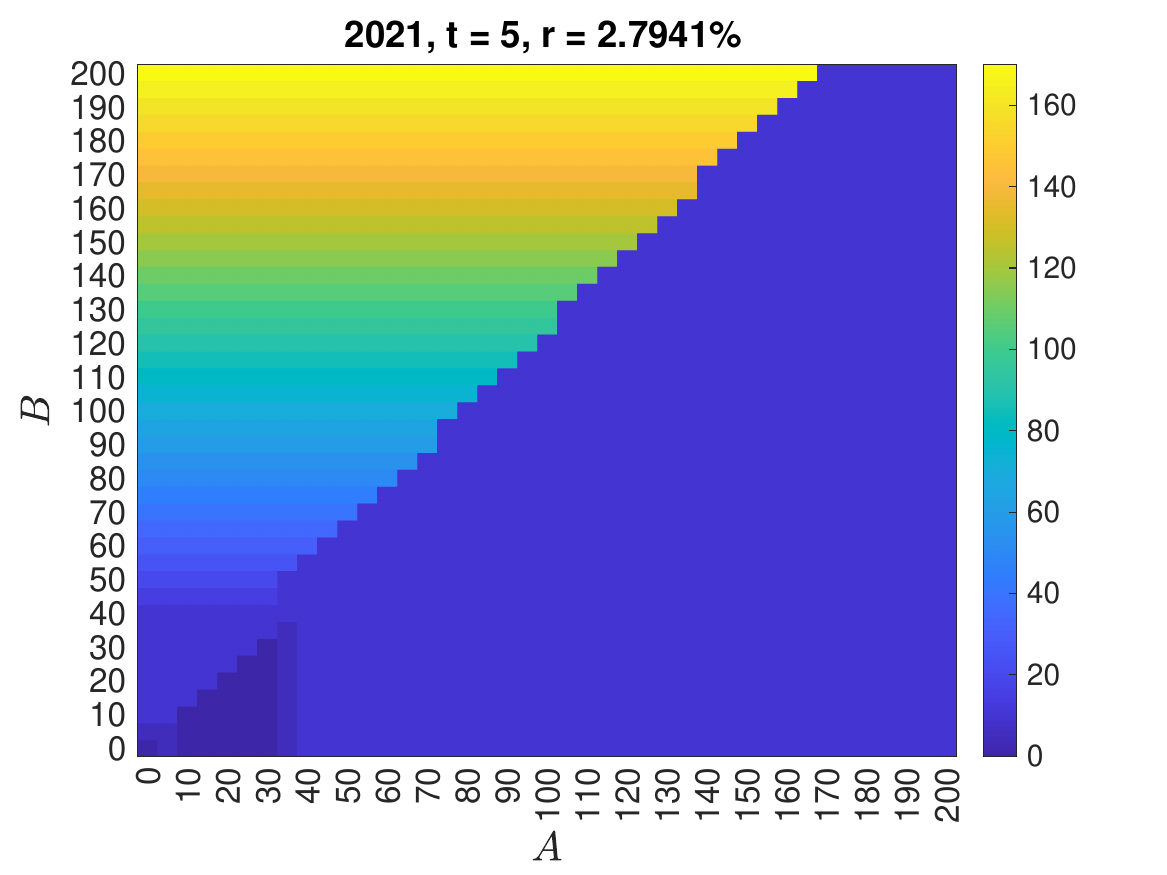} &
    \includegraphics[width=0.48\textwidth]{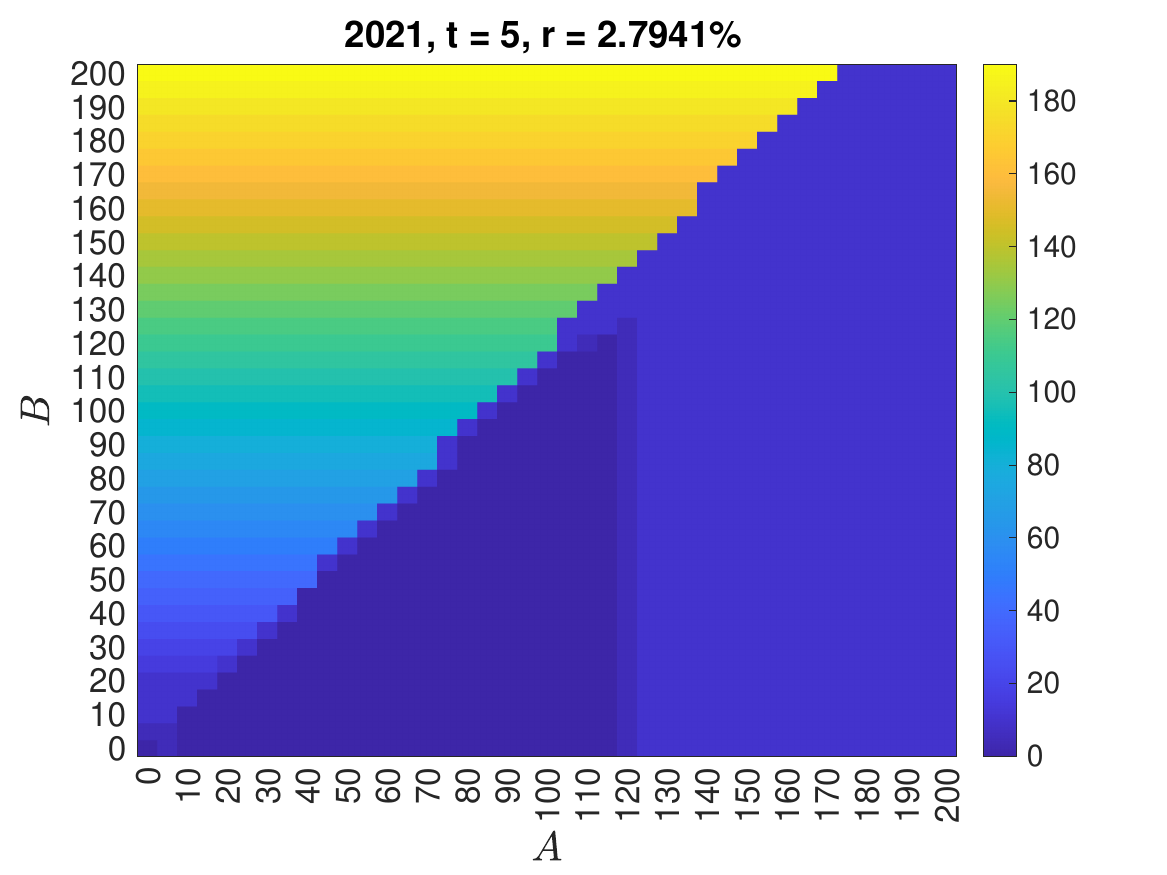} \\
    \hspace{20pt} Panel n) & \hspace{20pt} Panel q) \\
    \includegraphics[width=0.48\textwidth]{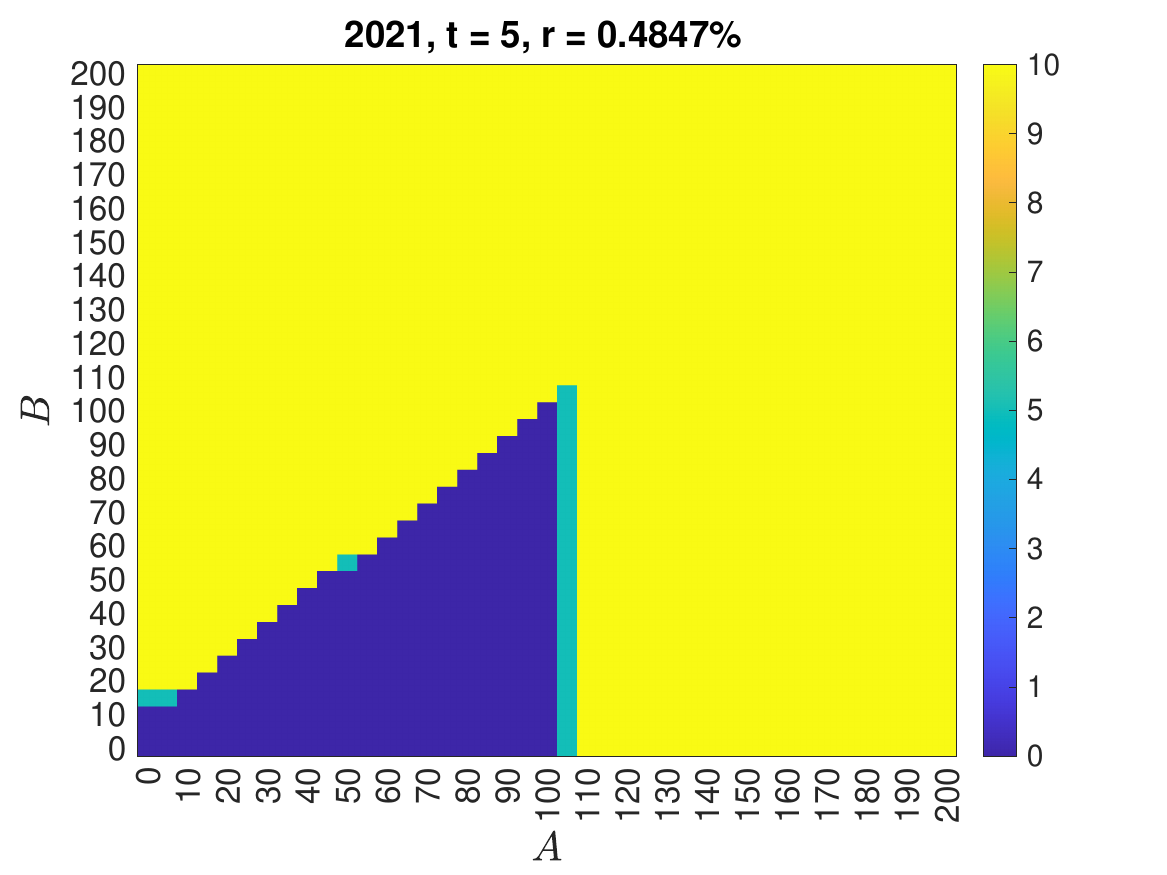} & 
    \includegraphics[width=0.48\textwidth]{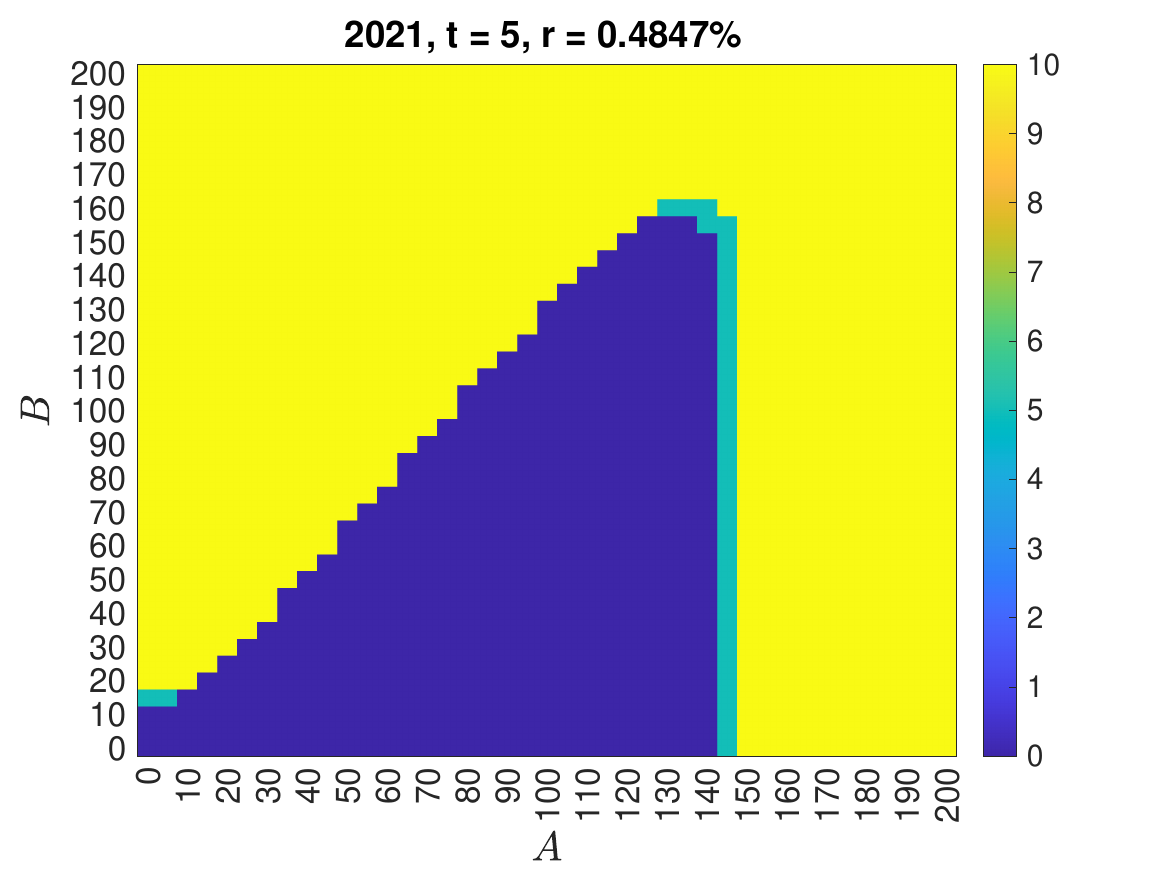} \\
    \hspace{20pt} Panel o) & \hspace{20pt} Panel r) \\
    \includegraphics[width=0.48\textwidth]{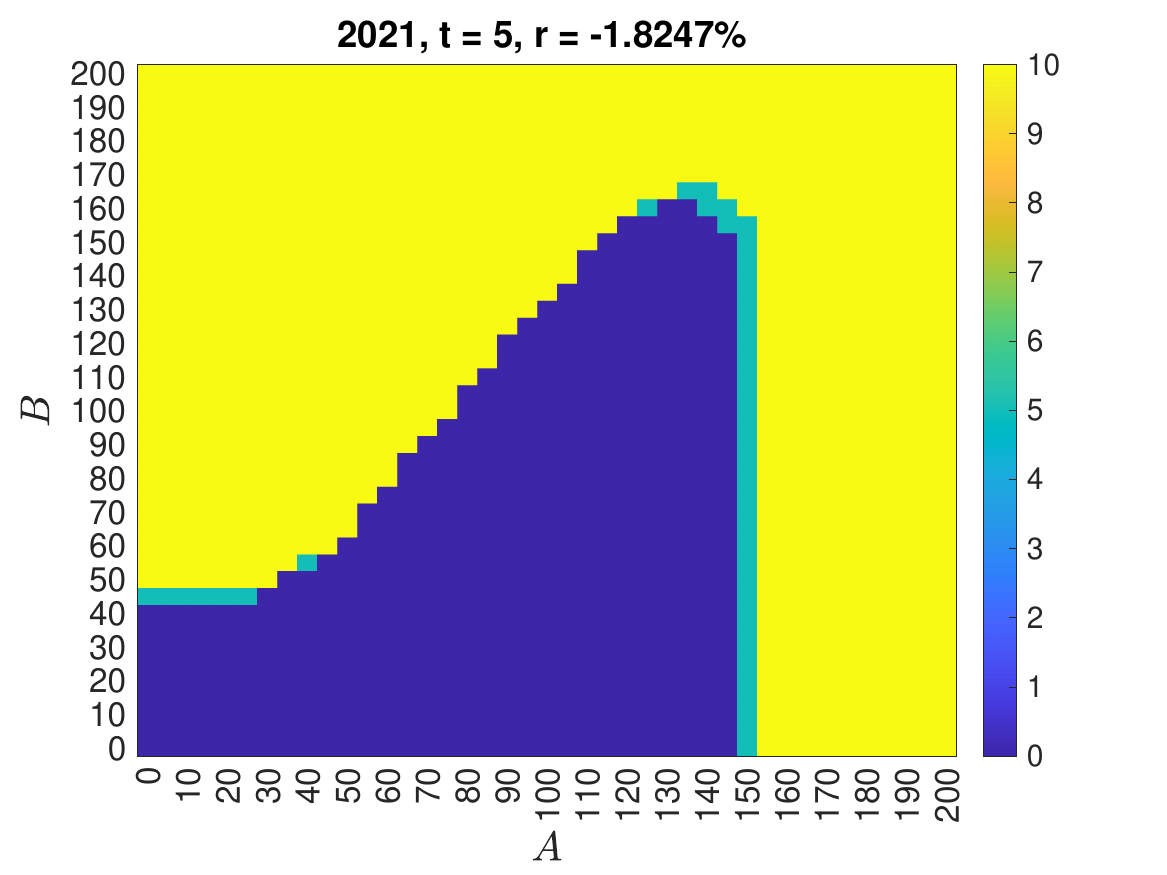} &
    \includegraphics[width=0.48\textwidth]{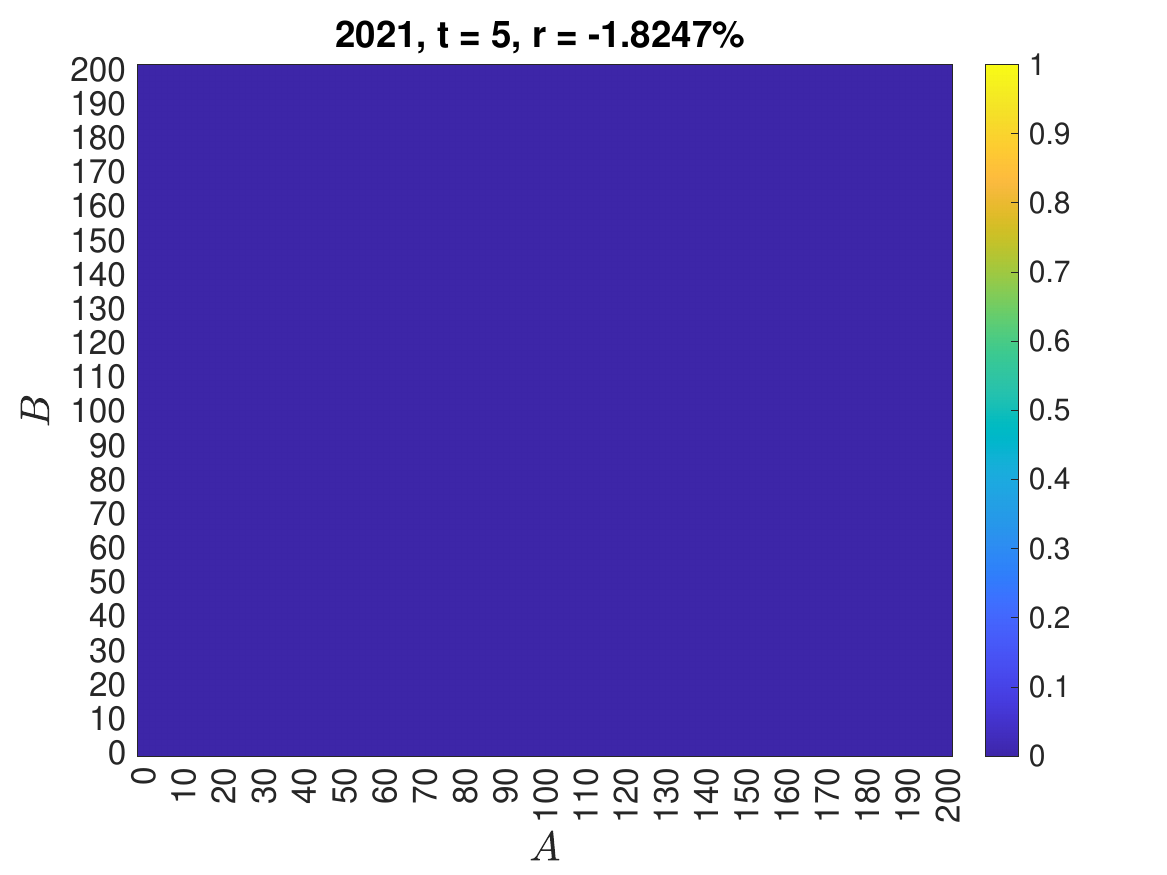}
    \end{tabular}
    \caption{Optimal withdrawal policies at $t=5$ for different level of the interest rate. $T=10$ and model parameters as as of 12/31/2021. Parameters: $m=3$, $n_A = 40$, $n_B = 40$.}
    \label{fig:optimalWithdrawals_2021_lowfees}
\end{figure}

For the sake of completeness we report in Figure \ref{fig:optimalWithdrawals_2021_lowfees} the optimal withdrawal policies in the 2021 scenario considering smaller fees and penalties: $\alpha = \beta = 10\%$ in the left panels (panels m, n, o) and $\alpha = \beta = 5\%$ in the right ones (panels p, q, r). In this case, when the interest rate is positive (panels m, p), the region associated to no withdrawal becomes significantly larger and covers parts of Figure \ref{fig:optimalWithdrawals_2021} (panel d) where it was optimal to withdraw the guaranteed amount $G$. This is explained by the lower penalties and fees, which make it attractive to wait for possible better payoffs at later dates. When the interest rate is close to zero (panels n, q) the PH either withdraws $G$ or nothing and when the interest rate is negative (panels o, r) she withdraws nothing at all.

\subsubsection{Optimal withdrawals with bonus feature}

\begin{figure}
    \centering
    \begin{tabular}{p{0.47\textwidth}|p{0.47\textwidth}}
    \begin{center} With step-up and bonus \end{center} & \begin{center} With step-up and bonus \end{center} \\
    \hspace{20pt} Panel s) & \hspace{20pt} Panel v) \\
    \includegraphics[width=0.48\textwidth]{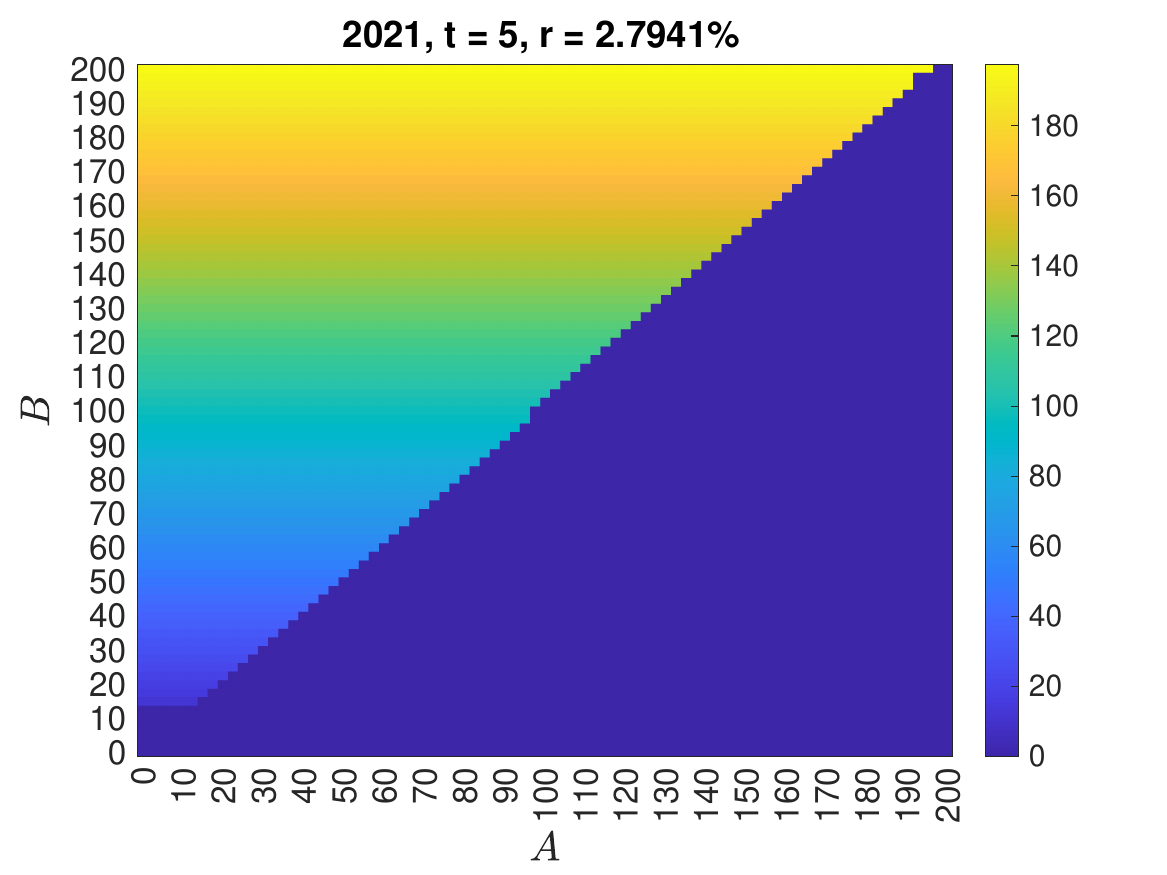} &
    \includegraphics[width=0.48\textwidth]{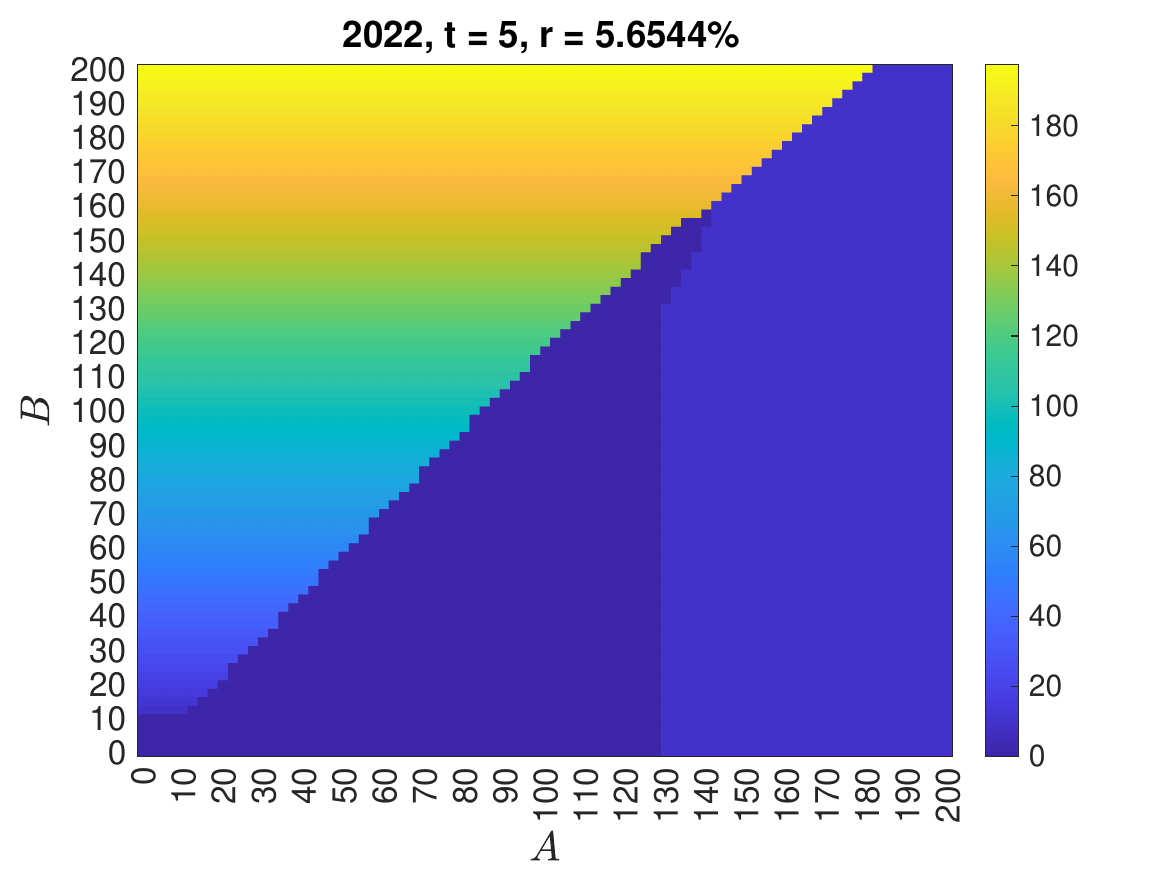} \\
    \hspace{20pt} Panel t) & \hspace{20pt} Panel w) \\
    \includegraphics[width=0.48\textwidth]{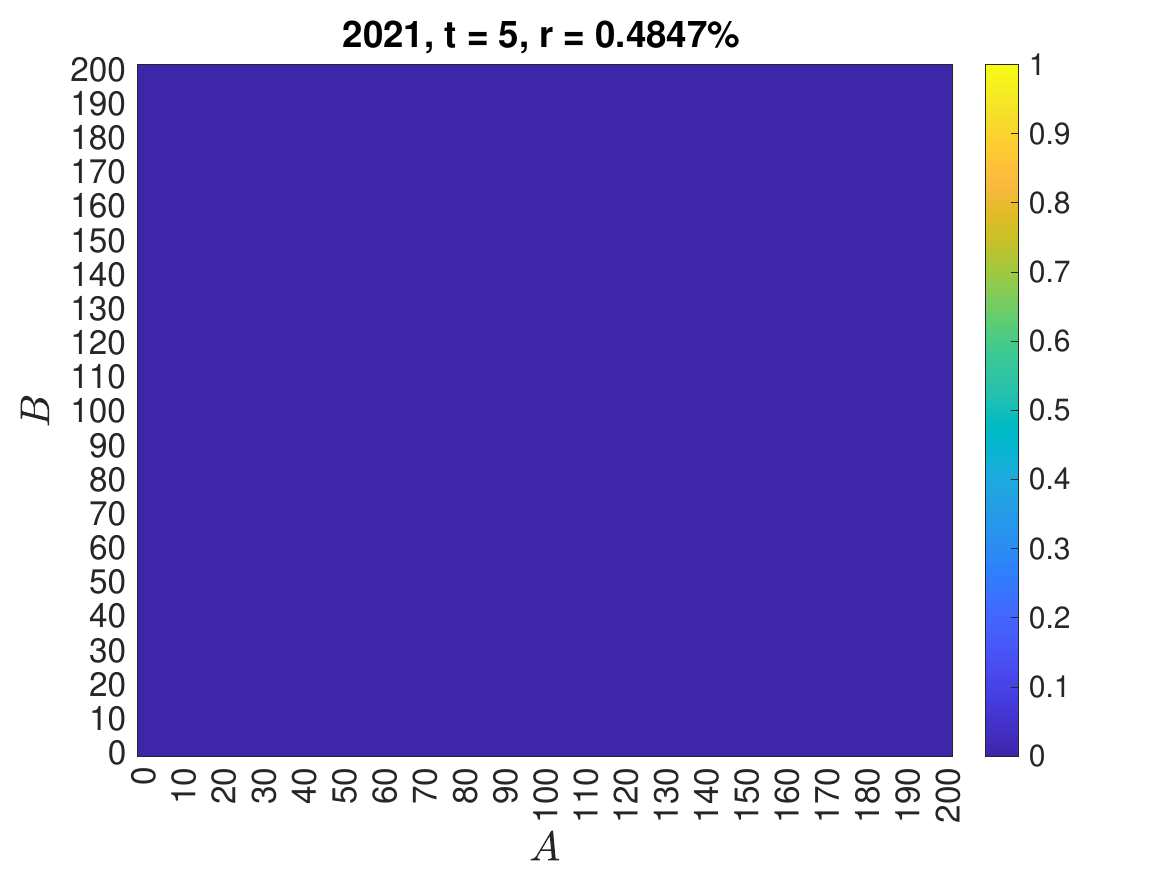} & 
    \includegraphics[width=0.48\textwidth]{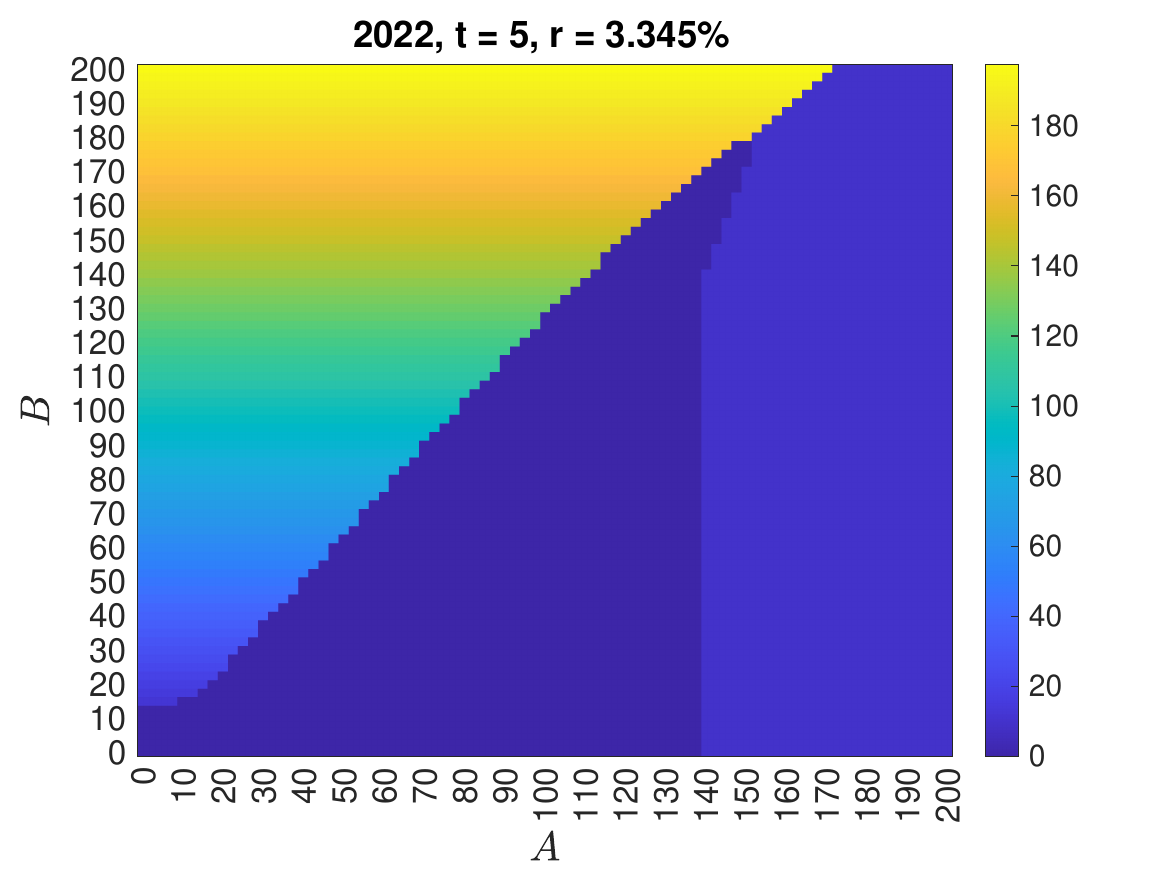} \\
    \hspace{20pt} Panel u) & \hspace{20pt} Panel x) \\
    \includegraphics[width=0.48\textwidth]{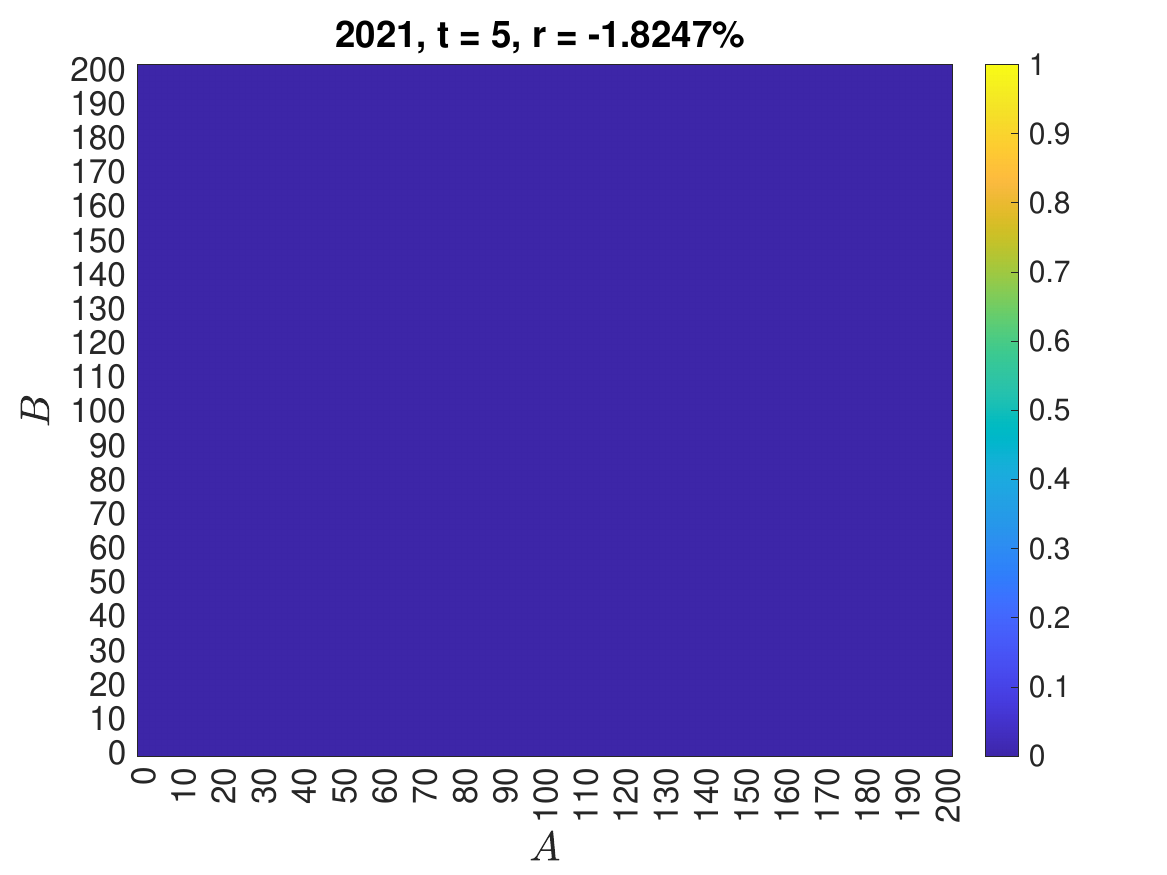} &
    \includegraphics[width=0.48\textwidth]{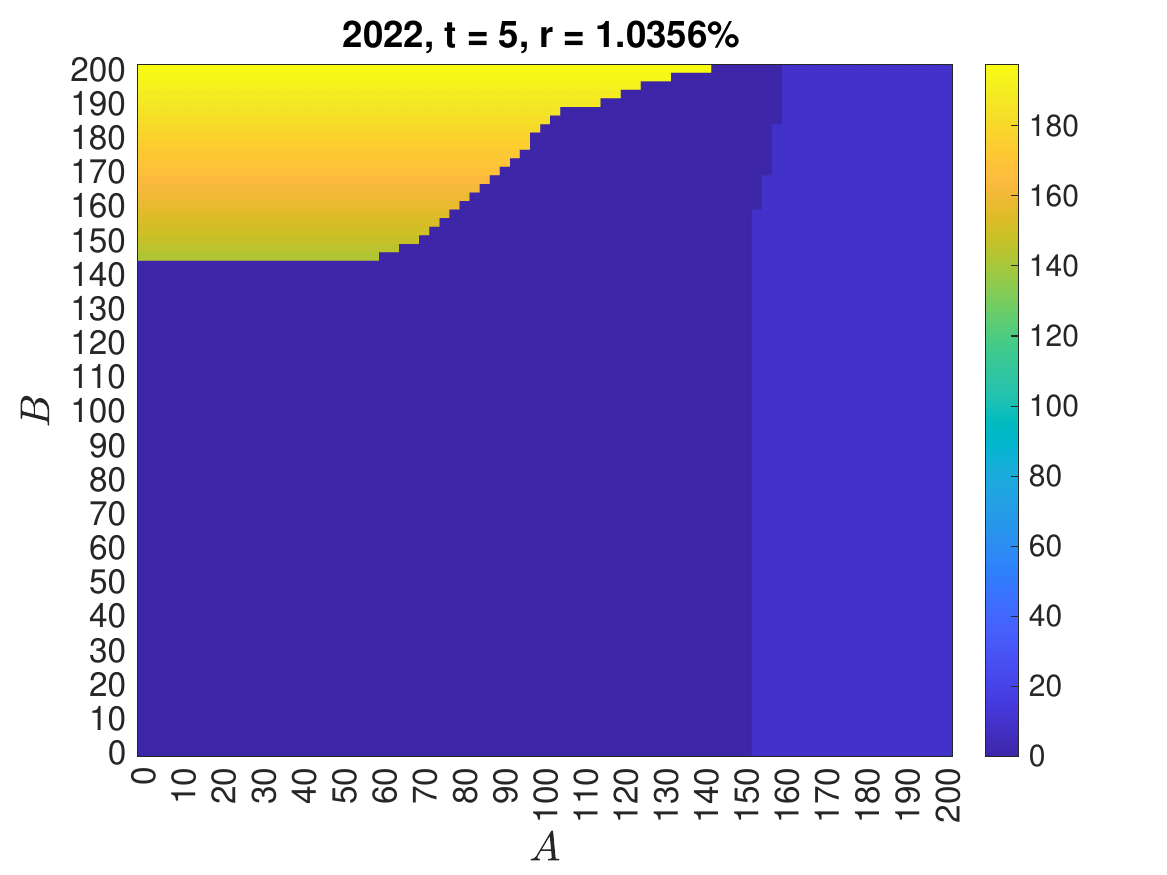}
    \end{tabular}
    \caption{Optimal withdrawal policies at $t=5$ of a GMWB with $b = 2.5$ for different level of the interest rate across the two different scenarios. Parameters: $m=2$, $n_A = n_B = 80$.}
    \label{fig:optimalWithdrawals_bonus}
\end{figure}

Finally, Figure \ref{fig:optimalWithdrawals_bonus} shows how the presence of the bonus feature affects optimal withdrawals. Comparing  Figure \ref{fig:optimalWithdrawals_bonus} with the right panels of Figures \ref{fig:optimalWithdrawals_2021} (panels d, e, f) and \ref{fig:optimalWithdrawals_2010} (panels j, k, l), we see that the inclusion of the bonus feature removes all regions where it was optimal to withdraw only the guaranteed amount $G$ and where it is now optimal to make no withdrawal at all, in order to gain the bonus. Indeed, when the interest rate is low (or even negative) (panels t, u) there is no incentive to withdraw since by waiting the PH can receive the bonus, while the discounting does not penalize future cashflows.

\section{Conclusions}\label{sec:conclusion}

In the present paper, we analyzed GMWB  annuities in the context of a stochastic model of a financial market with interest rate risk. We considered a comprehensive specification of the contract, allowing for several features including dynamic withdrawals, the possibility of surrender, step-up and bonus features, and taking also into account mortality risk. Despite the presence of correlation between market and interest rate risks, the structure of our model enables us to consider the interest rate process as the only relevant driving process, exploiting the explicit description of the joint probability distribution. We framed the valuation problem into a dynamic programming setting and we developed a reliable and flexible algorithm for its numerical solution. We have analyzed the determinants of the fair value of GMWB annuities and the corresponding optimal withdrawal strategies in relation to two interest rate scenarios, characterized respectively by low/negative rates (at 12/31/2021) and positive rates (at 12/30/2022).

Our results show that low (possibly negative) interest rates profoundly affect the valuation of GMWB annuities and their optimal withdrawal policies. In particular, we have shown that selling at par GMWB annuities in a low interest rate environment requires setting very large penalties for excess withdrawals and management fees. Low/negative interest rates pose serious challenges to insurance companies, as mid to long maturity products with protective features are either extremely costly to replicate or necessitate large penalties and fees that might make those products unattractive for retail investors.

Among the possible directions of further research, we believe that the extension of our framework to regime-switching models is particularly promising (see, e.g., \cite{Costabile2017,IgnatievaSongZiveyi2016,Kang_et_al2022}). Indeed, the different interest rate behavior from the 2021 scenario to the 2022 scenario can be interpreted as the outcome of a regime-switch, corresponding to a change in the monetary policy. The extension to regime-switching interest rate models can therefore take into account macroeconomic risks that are not reflected in simple diffusive models for the interest rate.

\bibliographystyle{alpha-abbrvsort}
\bibliography{bib.bib}

\newcommand{\etalchar}[1]{$^{#1}$}
\begin{thebibliography}{DDMR22}

\bibitem[AGWZ18]{AlonsoGarciaWoodZiveyi2018}
J.~Alonso-Garc\'ia, O.~Wood, and J.~Ziveyi.
\newblock Pricing and hedging guaranteed minimum withdrawal benefits under a
  general {L\'evy} framework using the {COS} method.
\newblock {\em Quantitative Finance}, 18(6):1049--1075, 2018.

\bibitem[AF15]{AzimzadehForsyth2015}
Y.~Azimzadeh and {P.A.} Forsyth.
\newblock The existence of optimal bang-bang controls for {GMxB} contracts.
\newblock {\em SIAM Journal on Financial Mathematics}, 6:117--139, 2015.

\bibitem[BMZ22]{BacinelloMaggistroZoccolan2022}
A.R. Bacinello, R.~Maggistro, and I.~Zoccolan.
\newblock Optimal withdrawal strategies in {GLWB} variable annuities.
\newblock Working paper (available on SSRN), 2022.

\bibitem[BMM16]{BacinelloMillossovichMontealegre2016}
A.R. Bacinello, P.~Millossovich, and A.~Montealegre.
\newblock The valuation of {GMWB} variable annuities under alternative fund
  distributions and policyholder behaviours.
\newblock {\em Scandinavian Actuarial Journal}, 5:446--465, 2016.

\bibitem[BMOP11]{BacinelloMillossovichOlivieriPitacco2011}
A.R. Bacinello, P.~Millossovich, A.~Olivieri, and E.~Pitacco.
\newblock Variable annuities: a unifying valuation approach.
\newblock {\em Insurance: Mathematics and Economics}, 49(3):285--297, 2011.

\bibitem[BZ19]{BacinelloZoccolan2019}
A.R. Bacinello and I.~Zoccolan.
\newblock Variable annuities with a threshold fee: valuation, numerical
  implementation and comparative static analysis.
\newblock {\em Decisions in Economics and Finance}, 42:21--49, 2019.

\bibitem[BR22]{BattauzRotondi2021}
A.~Battauz and F.~Rotondi.
\newblock American options and stochastic interest rates.
\newblock {\em Computational Management Science}, 19:567--604, 2022.

\bibitem[BKR08]{BauerKlingRuss2008}
D.~Bauer, A.~Kling, and J.~Russ.
\newblock A universal pricing framework for guaranteed minimum benefits in
  variable annuities.
\newblock {\em ASTIN Bulletin}, 38(2):621--651, 2008.

\bibitem[BHM14]{BernardHardyMackay2014}
C.~Bernard, M.~Hardy, and A.~Mackay.
\newblock State-dependent fees for variable annuity guarantees.
\newblock {\em ASTIN Bulletin}, 44:559--585, 2014.

\bibitem[Ber05]{Bertsekas2005}
D.~Bertsekas.
\newblock {\em Dynamic Programming and Optimal Control}.
\newblock Athena Scientific, 2005.

\bibitem[Bif05]{Biffis2005}
E.~Biffis.
\newblock Affine processes for dynamic mortality and actuarial valuations.
\newblock {\em Insurance: Mathematics and Economics}, 37:443--468, 2005.

\bibitem[BM06]{BrigoMercurio2006}
D.~Brigo and F.~Mercurio.
\newblock {\em Interest Rate Models: Theory and Practice}.
\newblock Springer, second edition, 2006.

\bibitem[Cha14]{Chang2014}
C.-K. Chang.
\newblock A dimension-reduction algorithm for the valuation of surrender
  options in {EIA} contracts with stochastic interest rates.
\newblock {\em Mathematics and Computers in Simulation}, 97(1):39--52, 2014.

\bibitem[CF08]{ChenForsyth2008}
Z.~Chen and P.A. Forsyth.
\newblock A numerical scheme for the impulse control formulation for pricing
  variable annuities with a guaranteed minimum withdrawal benefit {(GMWB)}.
\newblock {\em Numerische Mathematik}, 109:535--569, 2008.

\bibitem[CVF08]{ChenVetzalForsyth2008}
Z.~Chen, K.~Vetzal, and P.A. Forsyth.
\newblock The effect of modelling parameters on the value of {GMWB} guarantees.
\newblock {\em Insurance: Mathematics and Economics}, 47:165--173, 2008.

\bibitem[CT04]{ContTankov}
R.~Cont and P.~Tankov.
\newblock {\em Financial Modelling with Jump Processes}.
\newblock Chapman \& Hall - CRC, Boca Raton (FL), 2004.

\bibitem[Cos17]{Costabile2017}
M.~Costabile.
\newblock A lattice-based model to evaluate variable annuities with guaranteed
  minimum withdrawal benefits under a regime-switching model.
\newblock {\em Scandinavian Actuarial Journal}, 3:231--244, 2017.

\bibitem[DKZ08]{DaiKwokZong2008}
M.~Dai, Y.K. Kwok, and J.~Zong.
\newblock Guaranteed minimum withdrawal benefit in variable annuities.
\newblock {\em Mathematical Finance}, 18(4):595--611, 2008.

\bibitem[DYL15]{DaiYangLiu2015}
T.S. Dai, S.S. Yang, and L.-C. Liu.
\newblock Pricing guaranteed minimum/lifetime withdrawal benefits with various
  provisions under investment, interest rate and mortality risks.
\newblock {\em Insurance: Mathematics and Economics}, 64:364--379, 2015.

\bibitem[DDMR22]{DeAngelisDeMarchisMartireRusso2022}
P.~{De Angelis}, R.~{De Marchis}, A.~Martire, and E.~Russo.
\newblock A flexible lattice framework for valuing options on assets paying
  discrete dividends and variable annuities embedding {GMWB} riders.
\newblock {\em Decisions in Economics and Finance}, 45:415--446, 2022.

\bibitem[Del14]{Delong2014}
L.~Delong.
\newblock Pricing and hedging of variable annuities with state-dependent fees.
\newblock {\em Insurance: Mathematics and Economics}, pages 24--33, 2014.

\bibitem[DKL{\etalchar{+}}13]{Dhaene_et_al2013}
J.~Dhaene, A.~Kukush, E.~Luciano, W.~Schoutens, and B.~Stassen.
\newblock On the (in-)dependence between financial and actuarial risks.
\newblock {\em Insurance: Mathematics and Economics}, 52:522--531, 2013.

\bibitem[DXK19]{DongXuXwok2019}
B.~Dong, W.~Xu, and Y.K. Kwok.
\newblock Willow tree algorithms for pricing guaranteed minimum withdrawal
  benefits under jump-diffusion and {CEV} models.
\newblock {\em Quantitative Finance}, 19(10):1741--1761, 2019.

\bibitem[DJR14]{DonnellyJaimungalRubisov2014}
R.~Donnelly, S.~Jaimungal, and D.H. Rubisov.
\newblock Valuing guaranteed withdrawal benefits with stochastic interest rates
  and volatility.
\newblock {\em Quantitative Finance}, 14(2):369--382, 2014.

\bibitem[Eat83]{Eaton1983}
M.~Eaton.
\newblock {\em Multivariate Statistics: A Vector Space Approach}.
\newblock Wiley, 1983.

\bibitem[FS10]{FollmerSchweizer2010}
H.~F\"ollmer and M.~Schweizer.
\newblock The minimal martingale measure.
\newblock In R.~Cont, editor, {\em Encyclopedia of Quantitative Finance}, pages
  1200--1204. Wiley, 2010.

\bibitem[GMZ19]{GoudenegeMolentZanette2019}
L.~Goudenege, A.~Molent, and A.~Zanette.
\newblock Pricing and hedging {GMWB} in the {H}eston and in the
  {B}lack--{S}choles with stochastic interest rate models.
\newblock {\em Computational Management Science}, 16:217--248, 2019.

\bibitem[GMZ21]{GoudenegeMolentZanette2021}
L.~Goudenege, A.~Molent, and A.~Zanette.
\newblock Gaussian process regression for pricing variable annuities with
  stochastic volatility and interest rate.
\newblock {\em Decisions in Economics and Finance}, 44:57--72, 2021.

\bibitem[GIZ19]{GudkovIgnatievaZiveyi2019}
N.~Gudkov, K.~Ignatieva, and J.~Ziveyi.
\newblock Pricing of guaranteed minimum withdrawal benefits in variable
  annuities under stochastic volatility, stochastic interest rates and
  stochastic mortality via the componentwise splitting method.
\newblock {\em Quantitative Finance}, 19(3):501--518, 2019.

\bibitem[HK16]{HuangKwok16}
Y.T. Huang and Y.K. Kwok.
\newblock Regression-based {M}onte {C}arlo methods for stochastic control
  models: variable annuities with lifelong guarantees.
\newblock {\em Quantitative Finance}, 16(6):905--928, 2016.

\bibitem[HT16]{HuangKong2016}
{Y.T.} Huang and {Y.K.} Tong.
\newblock Regression-based {Monte} {Carlo} methods for stochastic control
  models: variable annuities with lifelong guarantees.
\newblock {\em Quantitative Finance}, 16(6):905--928, 2016.

\bibitem[ISZ16]{IgnatievaSongZiveyi2016}
K.~Ignatieva, A.~Song, and J.~Ziveyi.
\newblock Pricing and hedging of guaranteed minimum benefits under
  regime-switching and stochastic mortality.
\newblock {\em Insurance: Mathematics and Economics}, 70:286--300, 2016.

\bibitem[ISZ18]{IgnatievaSongZiveyi2018}
K.~Ignatieva, A.~Song, and J.~Ziveyi.
\newblock Pricing and hedging guaranteed minimum withdrawal benefits under a
  general {L\'evy} framework using the {COS} method.
\newblock {\em ASTIN Bulletin}, 48(1):139--169, 2018.

\bibitem[KSZZ22]{Kang_et_al2022}
B.~Kang, Y.~Shen, D.~Zhu, and J.~Ziveyi.
\newblock Valuation of guaranteed minimum maturity benefits under generalised
  regime-switching models using the {F}ourier {C}osine method.
\newblock {\em Insurance: Mathematics and Economics}, 105:96--127, 2022.

\bibitem[KZ18]{KangZiveyi2018}
B.~Kang and J.~Ziveyi.
\newblock Optimal surrender of guaranteed minimum maturity benefits under
  stochastic volatility and interest rates.
\newblock {\em Insurance: Mathematics and Economics}, 79(1):43--56, 2018.

\bibitem[Kir23]{Kirkby2023}
J.L. Kirkby.
\newblock Hybrid equity swap, cap, and floor pricing under stochastic interest
  by {Markov} chain approximation.
\newblock {\em European Journal of Operational Research}, 305:961--978, 2023.

\bibitem[KNC17]{KirkbyNguyenCui2017}
J.L Kirkby, D.~Nguyen, and Z.~Cui.
\newblock A unified approach to {Bermudan} and barrier options under stochastic
  volatility models with jumps.
\newblock {\em Journal of Economic Dynamics and Control}, 80:75--100, 2017.

\bibitem[LS01]{LongstaffSchwartz2001}
F.A. Longstaff and E.S. Schwartz.
\newblock Valuing american options by simulation: a simple least-square
  approach.
\newblock {\em The Review of Financial Studies}, 14(1):113--147, 2001.

\bibitem[LS15]{LuoShevchenko2015}
X.~Luo and P.~Shevchenko.
\newblock Valuation of variable annuities with guaranteed minimum withdrawal
  and death benefits via stochastic control optimization.
\newblock {\em Insurance: Mathematics and Economics}, 62:5--15, 2015.

\bibitem[MS06]{MilevskySalisbury2006}
M.A. Milevsky and T.S. Salisbury.
\newblock Financial valuation of guaranteed minimum withdrawal benefits.
\newblock {\em Insurance: Mathematics and Economics}, 38:21--38, 2006.

\bibitem[NR90]{nelsonRamaswamy90}
D.B. Nelson and K.~Ramaswamy.
\newblock Simple binomial processes as diffusion approximations in financial
  models.
\newblock {\em The Review of Financial Studies}, 3(3):393--430, 1990.

\bibitem[OP15]{OlivieriPitacco2015}
A.~Olivieri and E.~Pitacco.
\newblock {\em Introduction to Insurance Mathematics: Technical and Financial
  Features of Risk Transfers}.
\newblock Springer, second edition, 2015.

\bibitem[PLK12]{PengLeungKwok2012}
J.~Peng, K.S. Leung, and Y.K. Kwok.
\newblock Pricing guaranteed minimum withdrawal benefits under stochastic
  interest rates.
\newblock {\em Quantitative Finance}, 12(6):933--941, 2012.

\bibitem[Pit04]{Pitacco2004}
E.~Pitacco.
\newblock Survival models in a dynamic context: a survey.
\newblock {\em Insurance: Mathematics and Economics}, 35:279--298, 2004.

\bibitem[RT19]{RussoTorri2019}
V.~Russo and G.~Torri.
\newblock Calibration of one-factor and two-factor {Hull--White} models using
  swaptions.
\newblock {\em Computational Management Science}, 16:275--295, 2019.

\bibitem[SL16]{ShevchenkoLuo2016}
P.V. Shevchenko and X.~Luo.
\newblock A unified pricing of variable annuity guarantees under the stochastic
  control framework.
\newblock {\em Risks}, 4(22), 2016.

\bibitem[SL17]{ShevchenkoLuo2017}
P.V. Shevchenko and X.~Luo.
\newblock Valuation of variable annuities with guaranteed minimum withdrawal
  benefit under stochastic interest rate.
\newblock {\em Insurance: Mathematics and Economics}, 76:104--117, 2017.

\bibitem[WZ22]{WeiZhu22}
W.~Wei and D.~Zhu.
\newblock Generic improvements to least squares monte carlo methods with
  applications to optimal stopping problems.
\newblock {\em European Journal of Operational Research}, 298(3):1132--1144,
  2022.

\bibitem[YD13]{YangDai2013}
S.S. Yang and T.-S. Dai.
\newblock A flexible tree for evaluating guaranteed minimum withdrawal benefits
  under deferred life annuity contracts with various provisions.
\newblock {\em Insurance: Mathematics and Economics}, 52:231--242, 2013.

\end{thebibliography}

\newpage

\appendix

\section{}
\subsection{The binomial discretization for the interest rate process}\label{sec:discretization}

Following the work of \cite{nelsonRamaswamy90} on lattice discretizations of diffusions, we consider the following binomial discretization of the interest rate process $r=(r_t)_{t \geq 0}$ defined by
$$ \mathrm{d} r_t = (\theta(t) -a r_t ) \mathrm{d} t + \sigma_r \mathrm{d} W_t^r,
\qquad r_0 \in \mathbb{R}.$$
Let $\Delta t \:= n / T$ be the time step of a uniform discretization of the interval $[0,T]$. The binomial tree we consider for $r$ is a discrete-time stochastic process $\tilde{r}=(\tilde{r}_t)_{t = 0, \Delta t, \ldots, n\Delta t}$ such that
$$ \tilde{r}_{t + \Delta t} = 
\begin{cases}
\tilde{r}_t + \Delta r & \text{with probability }\pi_t, \\
    \tilde{r}_t - \Delta r & \text{with probability }1-\pi_t,
\end{cases}
$$
where $\Delta r := \sigma_r \sqrt{\Delta t}$ and
$$\pi_t = 
\begin{cases}
\frac{1}{2} + \frac{\theta(t) - a \tilde{r}_t}{2 \sigma_r} \sqrt{\Delta t}, & \text{if } 0 \leq \frac{1}{2} + \frac{\theta(t) - a \tilde{r}_t}{2 \sigma_r} \sqrt{\Delta t} \leq 1, \\
    0, & \text{if } \frac{1}{2} + \frac{\theta(t) - a \tilde{r}_t}{2 \sigma_r} \sqrt{\Delta t} < 0, \\
    1, & \text{if } \frac{1}{2} + \frac{\theta(t) - a \tilde{r}_t}{2 \sigma_r} \sqrt{\Delta t} > 1.
\end{cases}
$$
As discussed in the example following \cite[Theorem 1]{nelsonRamaswamy90}, since $r_t$ is normally distributed and the first two moments of $\tilde{r}_{t + \Delta t} - \tilde{r}_t$ match the ones of $\mathrm{d} r_t$ as the time step $\Delta t$ goes to zero, the process $\tilde{r}$ converges in distribution to its continuous-time counterpart $r$ as $n\to+\infty$.

\subsection{Proof of Proposition \ref{prop:ConditionalDistribution}}

As shown in \cite[Subsection 3.3.1]{BrigoMercurio2006}, it holds that
$$ r_s = r_{n-1}e^{-a(s-(n-1))} + \int_{n-1}^s e^{-a(s-u)} \theta(u) \mathrm{d} u + \sigma_r \int_{n-1}^s e^{-a(s-u)} \mathrm{d} W_u^r,
\qquad\text{ for all }s\geq n-1.$$
In particular, if $\Delta t$ represents the time between two consecutive anniversary dates $n$ and $n-1$,
\begin{equation}\label{eq:r_n} 
r_n = r_{n-1}e^{-a\Delta t} + \int_{n-1}^n e^{-a(n-u)} \theta(u) \mathrm{d} u + \sigma_r \int_{n-1}^n e^{-a(n-u)} \mathrm{d} W_u^r
\end{equation}
and
\begin{align}
    \int_{n-1}^n r_s \mathrm{d} s & = r_{n-1}\int_{n-1}^n e^{-a(s-(n-1))} \mathrm{d} s + \int_{n-1}^n \int_{n-1}^s e^{-a(s-u)} \theta(u) \mathrm{d} u \,\mathrm{d} s  + \sigma_r \int_{n-1}^n \int_{n-1}^s e^{-a (s-u)} \mathrm{d} W_u^r \,\mathrm{d} s \nonumber\\
    & = r_{n-1} \frac{1-e^{-a \Delta t}}{a} + \int_{n-1}^n \int_{n-1}^s e^{-a(s-u)} \theta(u) \mathrm{d} u \, \mathrm{d} s + \sigma_r \int_{n-1}^n \frac{1-e^{-a(n-u)}}{a} \mathrm{d} W_u^r,
\label{eq:r_integral}
\end{align}
where we have applied stochastic Fubini's theorem on the Wiener integral.
We now compute the joint characteristic function of $( \int_{n-1}^{n} r_s \mathrm{d} s , r_n ,W_n^S-W_{n-1}^S)$. For any $(u_1,u_2,u_3)\in\mathbb{R}^3$, using equations \eqref{eq:r_n} and \eqref{eq:r_integral}, we have that
\begin{align*}
& \mathbb{E} \left[ \exp \left( \im u_1 \int_{n-1}^n r_s \mathrm{d} s + \im u_2 r_n + \im u_3 (W_n^S -W_{n-1}^S) \right) \bigg| \mathcal{F}_{n-1} \right] \\
&\quad = \exp \left( \im u_1 \left( r_{n-1} \frac{1-e^{-a \Delta t}}{a} + \int_{n-1}^n \int_{n-1}^s e^{-a(s-u)} \theta(u) \mathrm{d} u \, \mathrm{d} s \right) \right.\\
&\qquad\qquad\left. +  \im u_2 \left( r_{n-1} e^{- a \Delta t} + \int_{n-1}^n e^{-a(n-u)} \theta(u) \mathrm{d} u \right) \right) \\
&\quad \times \mathbb{E} \left[ \exp \left( \im u_1 \sigma_r \int_{n-1}^n \frac{1-e^{-a(n-u)}}{a} \mathrm{d} W_u^r + \im u_2 \sigma_r \int_{n-1}^n e^{-a(n-u)} \mathrm{d} W_u^r \right. \right. \\
    & \qquad \qquad\quad +\left. \left. \im u_3 \int_{n-1}^n ( \rho \mathrm{d} W_u^r + \sqrt{1-\rho^2} \mathrm{d} W_u^{\perp r} ) \right) \bigg| \mathcal{F}_{n-1}  \right],
\end{align*}
where we have used a Cholesky decomposition to write $W_t^S = \rho W_t^r + \sqrt{1-\rho^2} W_t^{\perp r}$, for all $t\geq0$, where $W^{\perp r}$ is a standard Brownian motion independent of $W^r$. By independence, the last conditional expectation can be computed as follows:
\begin{align*}
 &\mathbb{E} \left[ \left. \exp \left( \im  \int_{n-1}^n \biggl( u_1 \sigma_r \frac{1-e^{-a(n-u)}}{a} + u_2 \sigma_r e^{-a(n-u)} + u_3 \rho \biggr)  \mathrm{d} W_u^r \right) \right| \mathcal{F}_{n-1} \right] \cdot \\
& \times \mathbb{E} \left[ \left. \exp \Bigl( \im u_3 \sqrt{1-\rho^2} \bigl(W^{\perp r}_n-W^{\perp r}_{n-1}\bigr) \Bigr) \right| \mathcal{F}_{n-1}  \right]	\\
&= \exp \left( - \frac{1}{2} \int_{n-1}^n \biggl( u_1 \sigma_r \frac{1-e^{-a(n-u)}}{a} + u_2 \sigma_r e^{-a(n-u)} + u_3 \rho \biggr)^2 \mathrm{d}u  \right) 
\exp \left( -\frac{u_3^2}{2} (1-\rho^2) \Delta t \right).
\end{align*}
Let us define
\begin{align*}
    \mu_{1,n-1} &:= r_{n-1} \frac{1-e^{-a \Delta t}}{a} + \int_{n-1}^n \int_{n-1}^s e^{-a(s-u)} \theta(u) \mathrm{d} u \, \mathrm{d} s, \\
    \mu_{2,n-1} &:= r_{n-1} e^{- a \Delta t} + \int_{n-1}^n e^{-a(n-u)} \theta(u) \mathrm{d} u, \\
    \mu_{3,n-1} &:= 0
\end{align*}
and
\begin{align*}
    \sigma_{11} &:= \frac{\sigma_r^2}{a^2} \int_{n-1}^n (1 - e^{-a (n-u)} )^2 \mathrm{d} u,\\
    \sigma_{22} &:= \sigma_r^2 \int_{n-1}^n e^{-2a(n-u)} \mathrm{d} u, \\
    \sigma_{33} &:= \Delta t, \\
    \sigma_{12} &:= \frac{\sigma_r^2}{a} \int_{n-1}^n e^{-a(n-u)}(1-e^{-a(n-u)}) \mathrm{d} u, \\
    \sigma_{13} &:= \rho \frac{\sigma_r}{a} \int_{n-1}^n ( 1 - e^{-a(n-u)} ) \mathrm{d} u, \\
    \sigma_{23} &:= \rho \sigma_r \int_{n-1}^n e^{-a(n-u)} \mathrm{d} u.
\end{align*}
The explicit expressions of these parameters appearing in the statement of the proposition can be obtained by explicitly computing the integrals, with the function $\theta:\mathbb{R}_+\to\mathbb{R}$ given as in \eqref{eqn:thetat}.
Using this notation, the joint characteristic function of $( \int_{n-1}^{n} r_s \mathrm{d} s , r_n ,W_n^S - W_{n-1}^S )$ has the form
$$ \exp \Bigl( \im \, \mu_{n-1}^{\top} \mathbf{u} - \frac{1}{2} \mathbf{u}^{\top} \Sigma \mathbf{u} \Bigr), $$
where
$$ \mu_{n-1} := \left[ \begin{array}{c}
    \mu_{1,n-1} \\ \mu_{2,n-1} \\ 0
    \end{array} \right], \quad \Sigma := \left[ \begin{array}{ccc}
    \sigma_{11} & \sigma_{12} & \sigma_{13} \\
    \sigma_{12} & \sigma_{22} & \sigma_{23} \\
    \sigma_{13} & \sigma_{23} & \sigma_{33}
    \end{array} \right] \text{ and } \mathbf{u} := \left[ \begin{array}{c}
    u_1 \\ u_2 \\ u_3
    \end{array} \right], $$
thus proving that \eqref{eq:joint_distribution} holds. Given this result, the fact that the distribution of $W^S_n-W^S_{n-1}$ conditionally on $\mathcal{F}_{n-1}\vee\sigma(\int_{n-1}^{n} r_s \mathrm{d} s , r_n )$ is given by \eqref{eq:cond_dsitribution} is standard (see for instance  \cite{Eaton1983}). 
The only thing that needs to be checked is the existence of $\Sigma_{22}^{-1}$, where
$$ \Sigma_{22} := \left[ \begin{array}{cc}
    \sigma_{11} & \sigma_{21} \\
    \sigma_{21} & \sigma_{22}
    \end{array} \right]. $$
To this effect, we compute 
\begin{align*}
\det \Sigma_{22} 
& =\sigma_{11} \sigma_{22} - \sigma_{12}^2
= \frac{\sigma_r^4}{2a^4}e^{-2a \Delta t} (e^{a \Delta t}-1 ) \bigl( 2 - 2e^{a \Delta t} + a \Delta t ( 1 + e^{a \Delta t} ) \bigr).
\end{align*}
Recalling that $a>0$, the fact that $\det\Sigma_{22}>0$ follows by noting that, for any $x>0$, it always holds that $e^x-1>0$ and $2-2e^x + x(1 + e^x) >0$.
\qed

\section{Supplementary Material}

\begin{figure}
    \centering
    \includegraphics[width=0.48\textwidth]{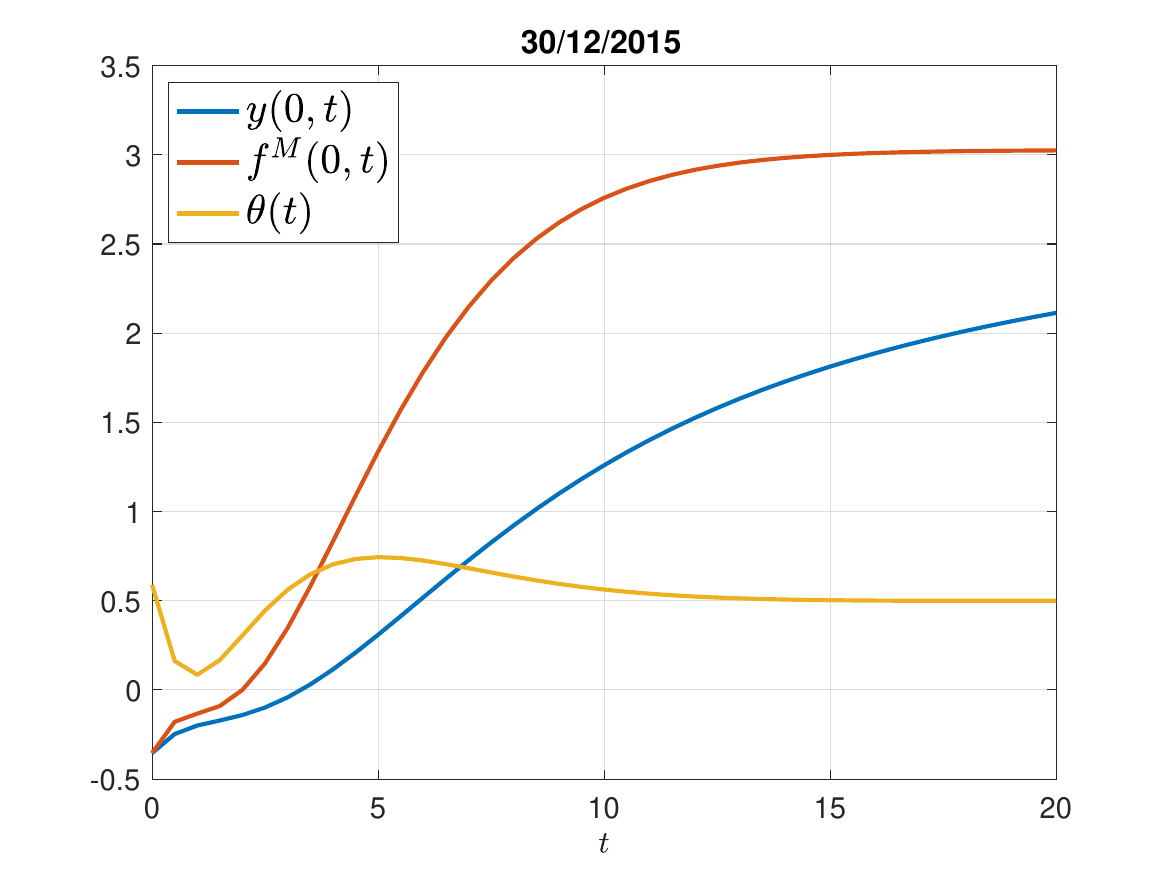}
    \caption{Yield curve, forwaed rate curve and long-run mean in percentage points as of 12/30/\textbf{2015}.}\label{fig:thetaAndFwd_2015}
\end{figure}

We provide here the details and the results of the valuation of general GMWB annuities within a third interest rate scenario. More precisely, we consider here an intermediate scenario between the ones analyzed in the paper considering the market as of the end of 2015\footnotemark\footnotetext{The parameters of the instantaneous forward curve as of 12/30/2015 are $\beta_0 = 3.0270$, $\beta_1 = -3.3810$, $\beta_2 = 37.9604$, $\beta_3 = -43.1112$, $\tau_1 = 1.5904$, $\tau_2 = 1.6993$.}, when the current interest rate level was mildly negative ($r(0) = -0.29\%$) as in 2021, but the outlook was much more positive as in 2022. Indeed, looking at Figure \ref{fig:thetaAndFwd_2015}, we can see the interest rate was expected to become positive after a couple of years and increase even further afterwards.

Table \ref{tab:optimalFees2015} parallels Tables 6 and 7 of the paper. As we can see, in this new scenario it is possible to let the withdrawal penalty take the standard value of $\beta = 10\%$ while getting a reasonable fair management fee of $\alpha = 6.60\%$. As for the decomposition of the fair value of the GMWB annuity described in Table 8 of the paper for the other two scenarios, in the 2015 one we get again an intermediate situation in which CB (the present value of ten annual constant cashflows equal to $G$) accounts for the 91.09\%, GMWB (the fair value of the annuity without the step-up feature) for the 3.59\% and the step-up feature for the 5.31\%.

Table \ref{tab:optimalFeesNOMortalityRisk_2015} parallels Table 9 of the paper, showing that in this intermediate scenario, like in the 2021 one, accounting for mortality risk delivers a small increase of the fair values.

Tables \ref{tab:sensitivity2015} and \ref{tab:sensitivity_a_2015} parallel Tables 10, 11 and 12 of the paper displaying the sensitivities of the fair value of the GMWB annuity with respect to market model parameters. As expected, the impact of $\sigma_S$ and $\sigma_r$ is again of first order whereas the one of $\rho$ and $a$ is of second order.

\begin{figure}
    \centering
    \begin{tabular}{p{0.47\textwidth}|p{0.47\textwidth}}
    \begin{center} Without step-up \end{center} & \begin{center} With step-up \end{center} \\
    \includegraphics[width=0.48\textwidth]{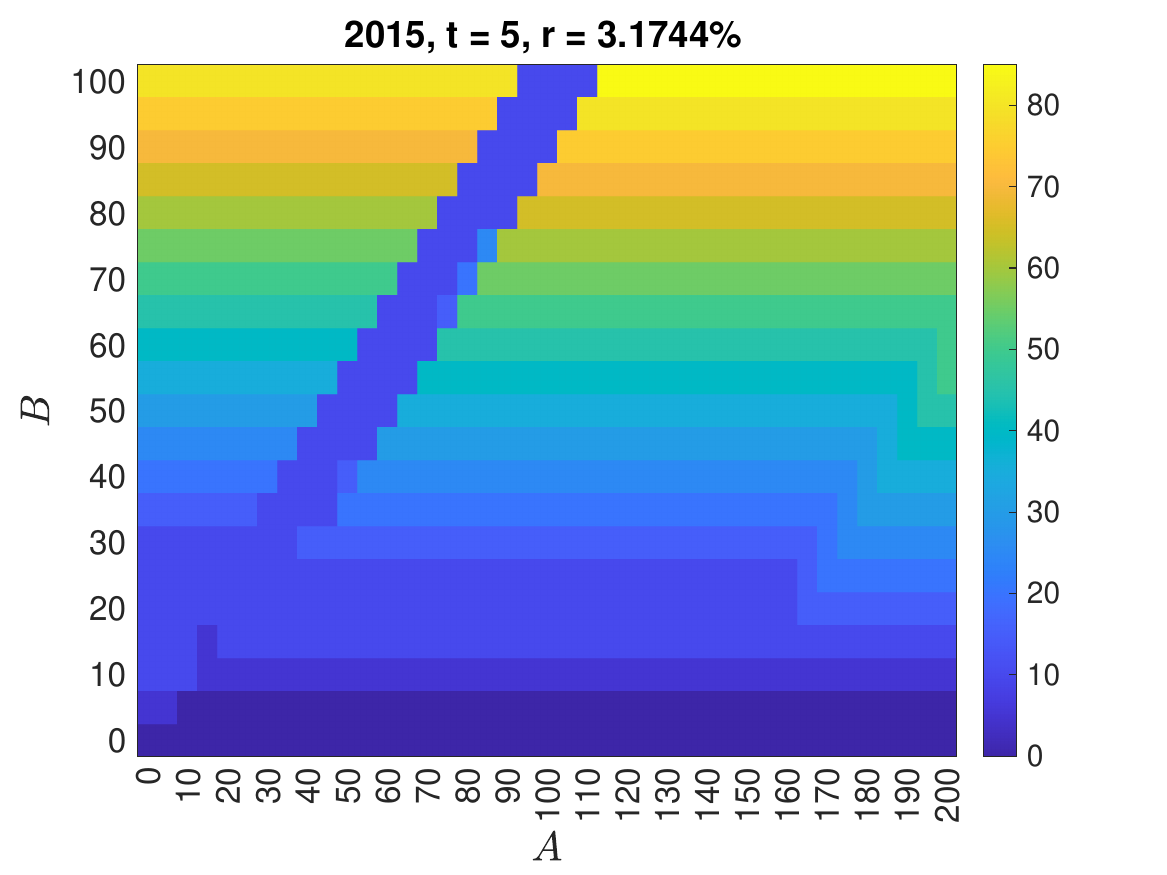} &
    \includegraphics[width=0.48\textwidth]{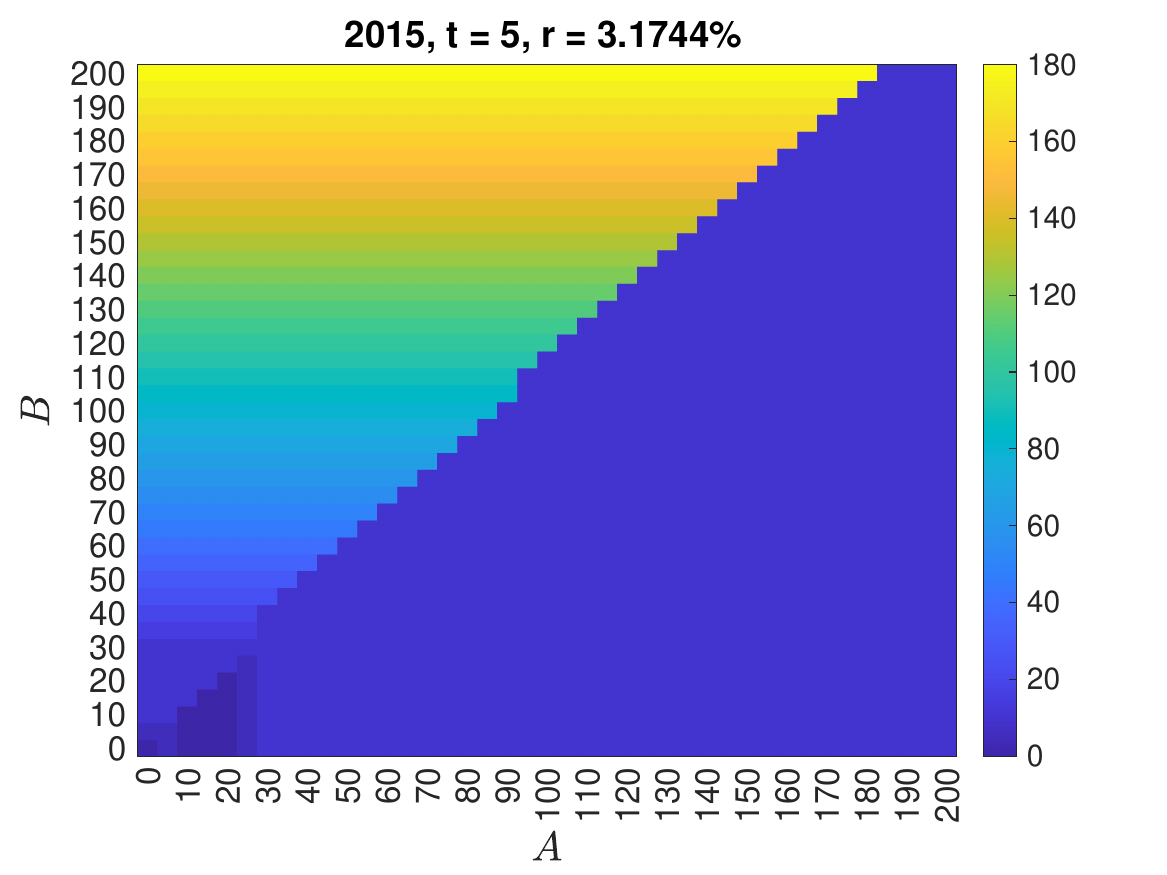} \\
    \includegraphics[width=0.48\textwidth]{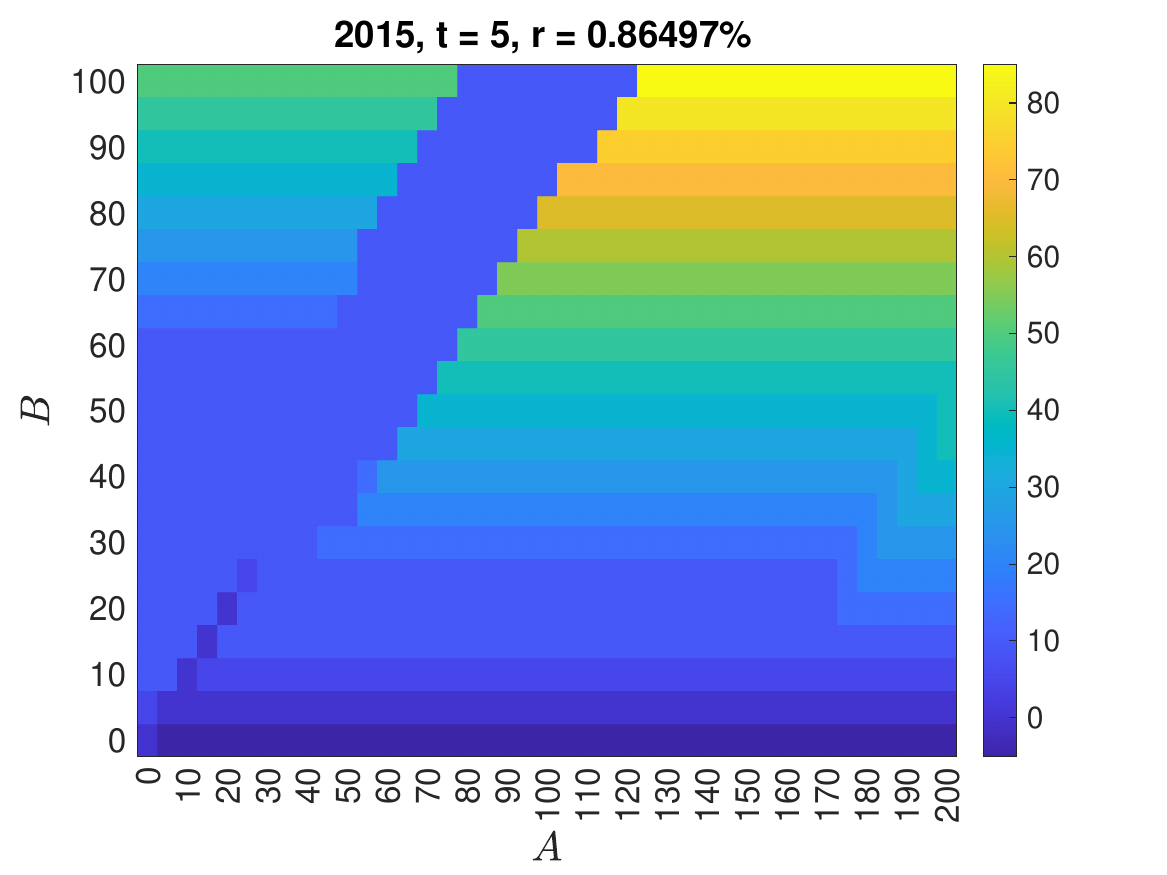} & 
    \includegraphics[width=0.48\textwidth]{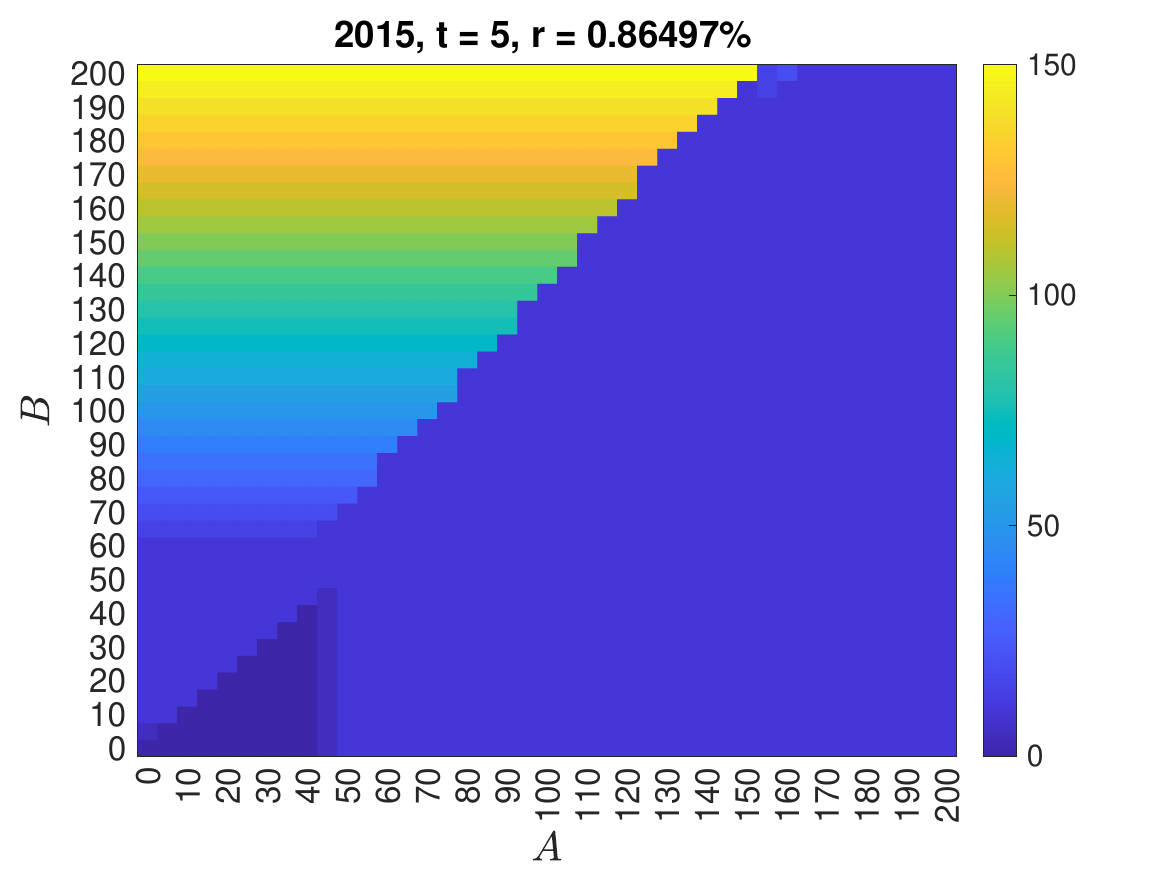} \\
    \includegraphics[width=0.48\textwidth]{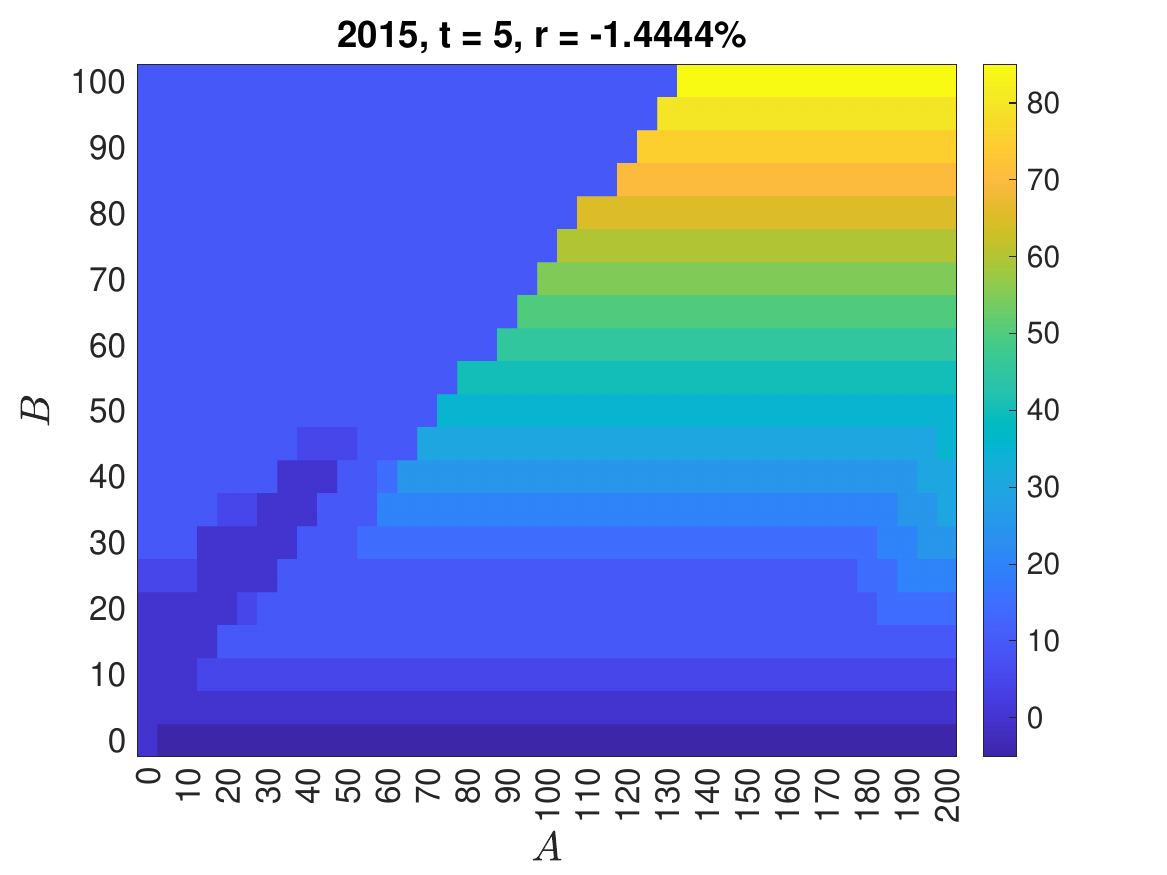} &
    \includegraphics[width=0.48\textwidth]{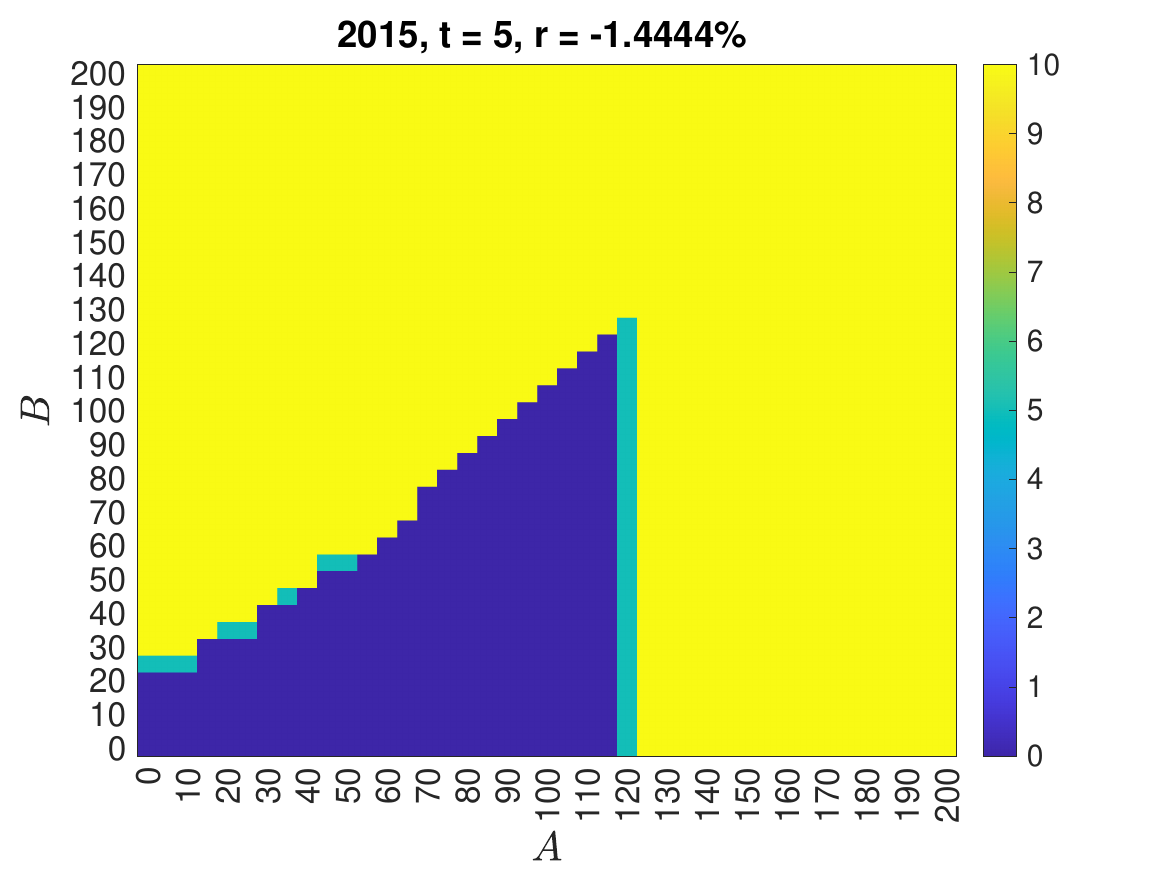}
    \end{tabular}
    \caption{Optimal withdrawal policies at $t=5$ for different level of the interest rate. $T=10$ and model parameters as of 12/30/\textbf{2015}. Parameters: $m=3$, $n_A = 40$, $n_B = 20$ for the panels with no step-up feature, $n_B = 40$ for the panels with the step-up feature.}
    \label{fig:optimalWithdrawals_2015}
\end{figure}

\begin{figure}
    \centering
    \begin{tabular}{p{0.47\textwidth}}
    \begin{center} 2015 \end{center} \\
    \includegraphics[width=0.47\textwidth]{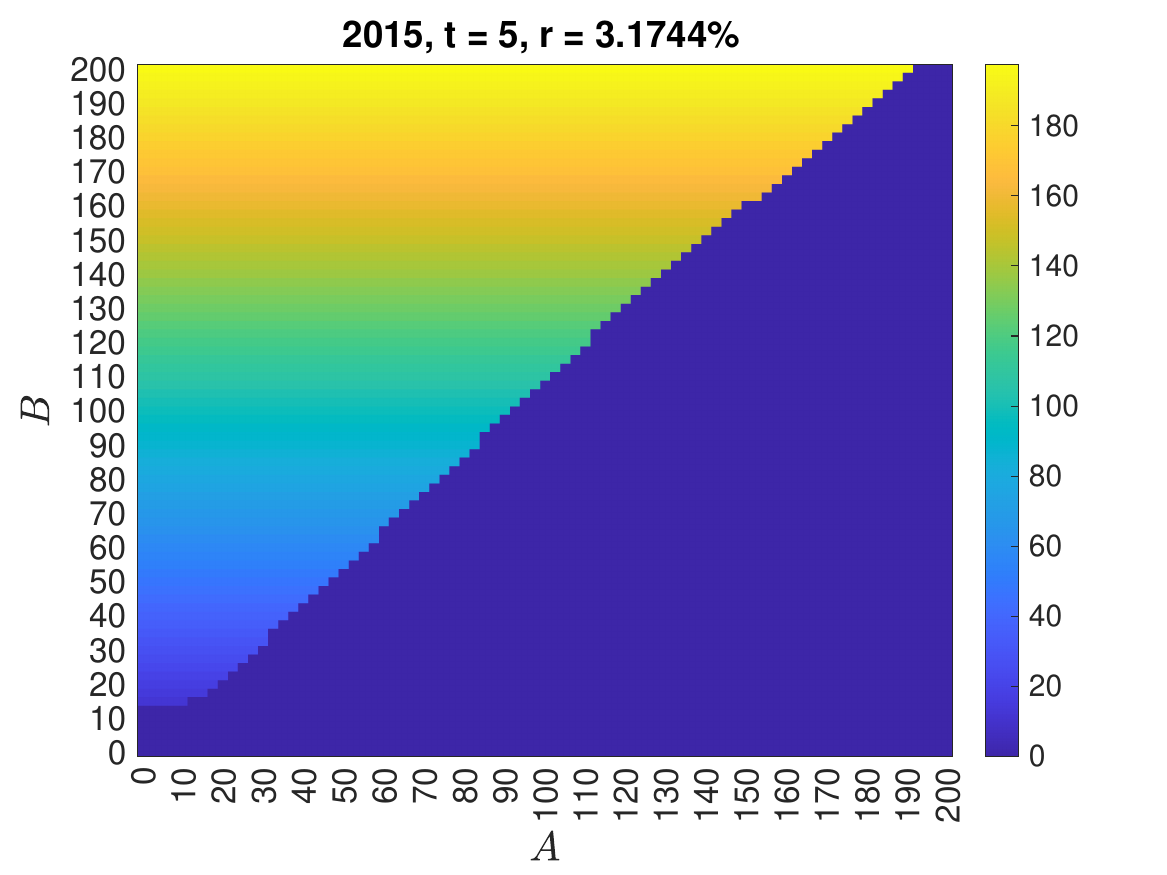}\\
    \includegraphics[width=0.47\textwidth]{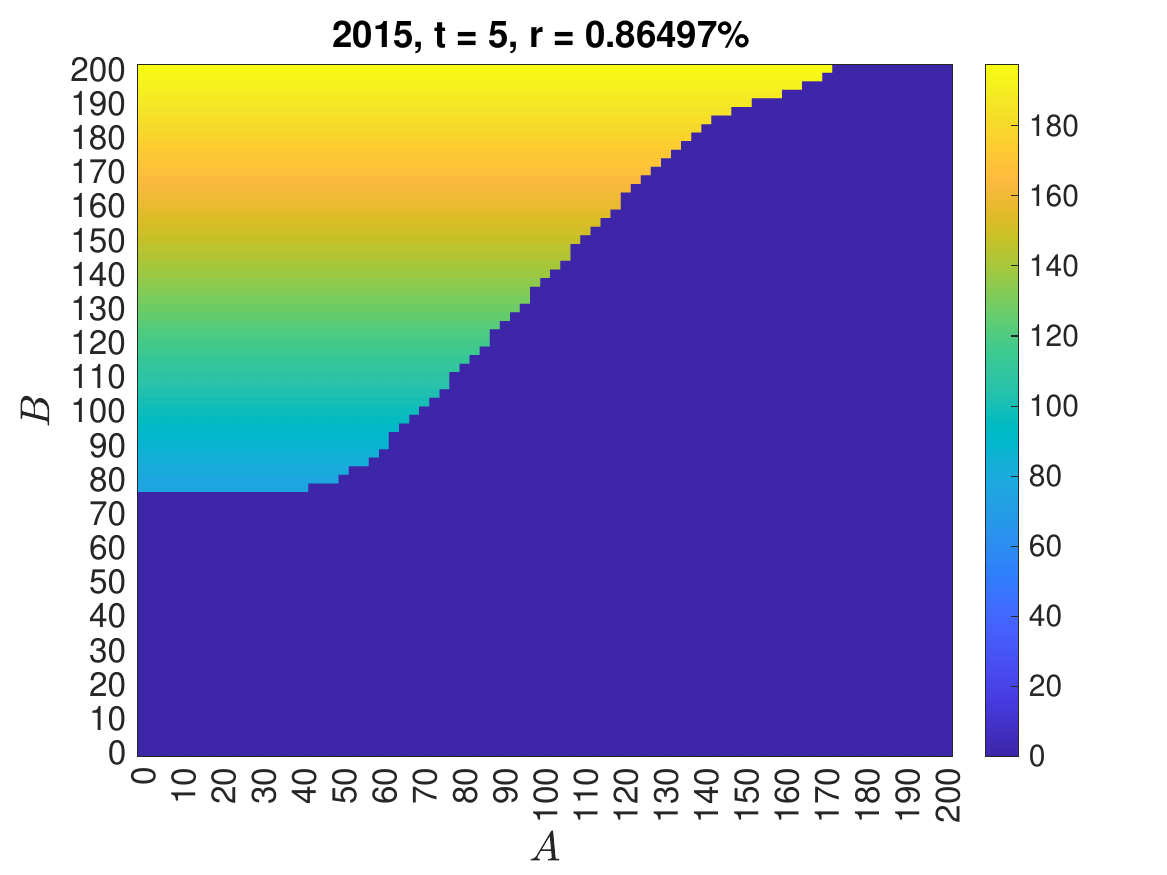}\\
    \includegraphics[width=0.47\textwidth]{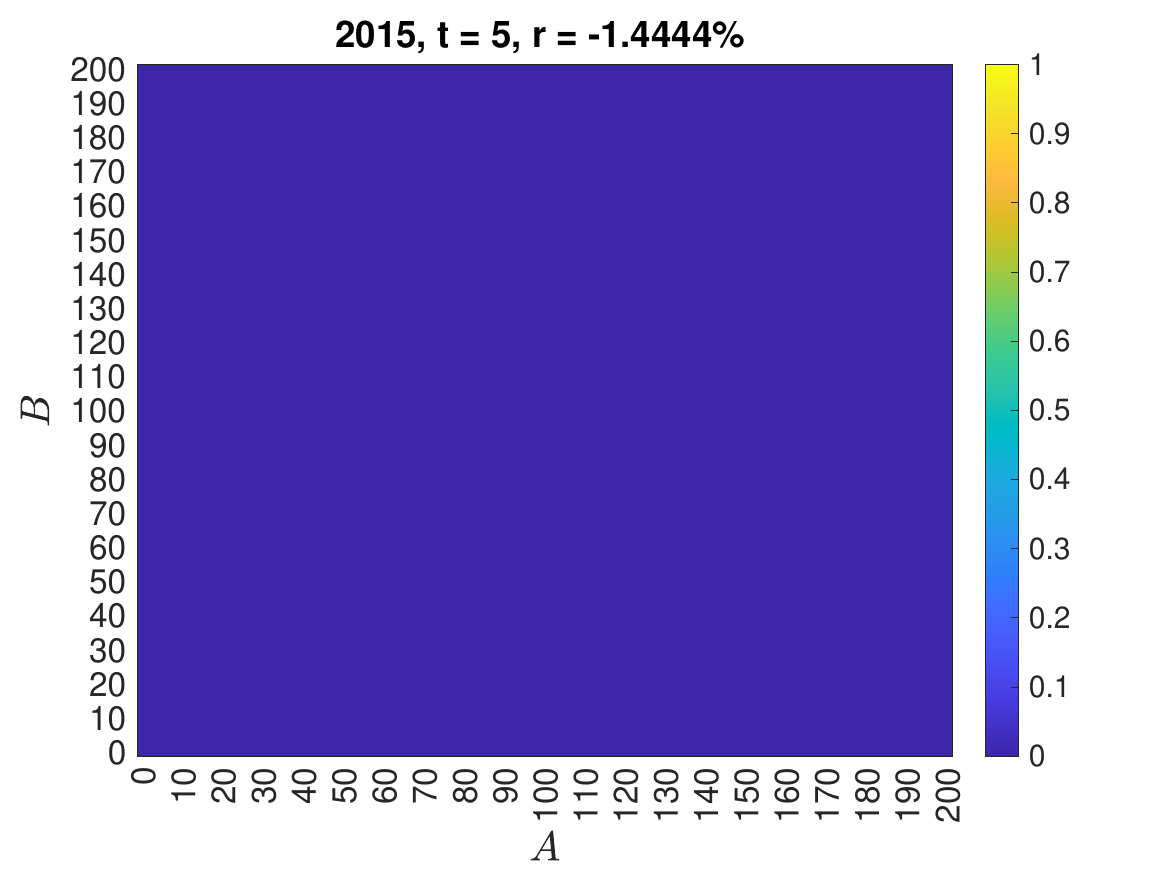}\\
    \end{tabular}
    \caption{Optimal withdrawal policies at $t=5$ of a GMWB with $b = 2.5$ for different level of the interest rate across the two different scenarios. Parameters: $m=2$, $n_A = n_B = 80$.}
    \label{fig:optimalWithdrawals_bonus_2015}
\end{figure}

Finally, Figures \ref{fig:optimalWithdrawals_2015} and \ref{fig:optimalWithdrawals_bonus_2015} parallels Figures 4, 5 and 7 of the paper displaying optimal withdrawal strategies for the 2015 scenario. Interestingly, we notice that optimal strategies are qualitatively similar to the ones of the 2021 scenarios for low levels of the interest rates and to the ones of the 2022 scenario for larger values. Therefore, this 2015 scenario looks indeed like the midpoint between the other two.

\begin{table}[]
    \centering
        \resizebox{\textwidth}{!}{\begin{tabular}{c|c|c|c|c|c|c|c|}
        \multicolumn{3}{c|}{2015: Without} & \multicolumn{5}{c|}{$\alpha$} \\ \cline{4-8}
        \multicolumn{3}{c|}{step-up} & 0\% & 2.5\% & 5\% & 7.5\% & 10\% \\ \hline
        \multirow{15}{*}{$\beta$} & \multirow{2}{*}{0\%} & S & 107.74 & 102.94 & 99.86 & 98.01 & 96.96 \\ 
        & & S+S & 111.18 & 106.59 & 104.06 & 102.94 & 102.57 \\ 
        & & D & 115.31 & 110.27 & 107.45 & 105.90 & 105.07 \\ \cline{2-8}
        & \multirow{3}{*}{5\%} & S & 107.45 & 102.61 & 99.49 & 97.61 & 96.54 \\ 
        & & S+S & 109.97 & 104.92 & 101.87 & 100.29 & 99.62 \\ 
        & & D & 111.71 & 106.47 & 103.60 & 102.02 & 101.16 \\ \cline{2-8}
        & \multirow{3}{*}{10\%} & S & 107.16 & 102.28 & 99.13 & 97.21 & 96.11 \\ 
        & & S+S & 109.21 & 103.83 & 100.41 & 98.41 & 97.36 \\ 
        & & D & 109.31 & 104.03 & 101.16 & 99.50 & 98.58 \\ \cline{2-8}
        & \multirow{3}{*}{15\%} & S & 106.90 & 101.98 & 98.79 & 96.84 & 95.72 \\ 
        & & S+S & 107.71 & 102.31 & 99.51 & 97.51 & 96.28 \\ 
        & & D & 107.91 & 102.77 & 99.71 & 97.95 & 96.94 \\ \cline{2-8}
        & \multirow{3}{*}{20\%} & S & 106.66 & 101.7 & 98.47 & 96.49 & 95.34 \\ 
        & & S+S & 107.05 & 101.98 & 98.60 & 96.92 & 95.65 \\ 
        & & D & 107.20 & 102.09 & 98.89 & 97.00 & 95.92 \\ \hline
        \end{tabular}

        \begin{tabular}{c|c|c|c|c|c|c|c|}
        \multicolumn{3}{c|}{2015: With} & \multicolumn{5}{c|}{$\alpha$} \\ \cline{4-8}
        \multicolumn{3}{c|}{step-up} & 0\% & 2.5\% & 5\% & 7.5\% & 10\% \\ \hline
        \multirow{15}{*}{$\beta$} & \multirow{2}{*}{0\%} & S & 117.28 & 110.93 & 106.65 & 103.62 & 101.51 \\ 
        & & S+S & 120.51 & 115.44 & 111.85 & 109.31 & 107.53 \\ 
        & & D & 126.98 & 119.73 & 114.97 & 111.80 & 109.66 \\ \cline{2-8}
        & \multirow{3}{*}{5\%} & S & 116.37 & 110.05 & 105.68 & 102.87 & 100.82 \\ 
        & & S+S & 117.87 & 112.68 & 109.01 & 106.40 & 104.51 \\ 
        & & D & 122.34 & 115.17 & 110.53 & 107.56 & 105.53 \\ \cline{2-8}
        & \multirow{3}{*}{10\%} & S & 115.33 & 109.20 & 104.92 & 102.14 & 100.14 \\ 
        & & S+S & 115.67 & 110.37 & 106.57 & 103.90 & 101.96 \\ 
        & & D & 118.64 & 111.81 & 107.52 & 104.73 & 102.81 \\ \cline{2-8}
        & \multirow{3}{*}{15\%} & S & 114.34 & 108.31 & 104.20 & 101.36 & 99.55 \\ 
        & & S+S & 114.85 & 108.74 & 104.95 & 102.24 & 100.32 \\ 
        & & D & 115.94 & 109.65 & 105.66 & 102.92 & 101.04 \\ \cline{2-8}
        & \multirow{3}{*}{20\%} & S & 113.56 & 107.53 & 103.43 & 100.74 & 98.98 \\ 
        & & S+S & 113.78 & 107.60 & 103.72 & 101.04 & 99.13 \\ 
        & & D & 114.07 & 108.29 & 104.38 & 101.70 & 99.81 \\ \hline
        \end{tabular}}
        \newline
        \vspace*{0.5 cm}
        \newline
        \begin{tabular}{c|c|c|}
            \multicolumn{2}{c|}{2015: Without step-up} & $\alpha = 6.60\%$ \\ \hline
             \multirow{3}{*}{$\beta = 10\%$} & S & 97.79  \\
             & S+S & 98.51 \\
             & D & 100.00 \\ \hline
        \end{tabular}
        \quad
        \begin{tabular}{c|c|c|}
            \multicolumn{2}{c|}{2015: With step-up} & $\alpha = 6.60\%$ \\ \hline
             \multirow{3}{*}{$\beta = 10\%$} & S & 103.03  \\
             & S+S & 104.78 \\
             & D & 105.61 \\ \hline
        \end{tabular}
        
    \caption{Fair value of a GMWB annuity at $t=0$ for different values of  $\alpha$ and $\beta$. Rows labelled by S (resp. S+S / D) refer to the case of static withdrawals (resp. static withdrawals and surrender / dynamic withdrawals). $T=10$ and market model parameters as of 12/30/\textbf{2015}. Parameters: $m=2$, $n_A = 30$, $n_B = 10$ for the panels with no step up feature, $n_B = 30$ for the panels with the step up feature.}
    \label{tab:optimalFees2015}
\end{table}

\begin{table}[]
\small
\centering
    \begin{tabular}{c|c|c|}
            \multicolumn{2}{c|}{2015: Without step-up,} & \multirow{2}{*}{$\alpha = 6.60\%$} \\
            \multicolumn{2}{c|}{no mortality risk} & \\ \hline
             \multirow{2}{*}{$\beta = 10\%$} & S & 98.21  \\
             & D & 100.42 \\ \hline
        \end{tabular}
        \quad
        \begin{tabular}{c|c|c|}
            \multicolumn{2}{c|}{2015: With step-up,} & \multirow{2}{*}{$\alpha = 6.60\%$} \\
            \multicolumn{2}{c|}{no mortality risk} & \\ \hline
             \multirow{2}{*}{$\beta = 10\%$} & S & 103.77  \\
             & D & 106.02 \\ \hline
        \end{tabular}
    \caption{Fair values of a GMWB annuity at $t=0$, with no mortality risk. Parameters as in Table \ref{tab:optimalFees2015}.}
    \label{tab:optimalFeesNOMortalityRisk_2015}
\end{table}

\begin{table}[]
    \centering
        \resizebox{\textwidth}{!}{\begin{tabular}{c|c|c|c|c|c|c|c|}
        \multicolumn{3}{c|}{2015: Without} & \multicolumn{5}{c|}{$\sigma_S$} \\ \cline{4-8}
        \multicolumn{3}{c|}{step-up} & 10\% & 15\% & 20\% & 25\% & 30\% \\ \hline
        \multirow{10}{*}{$\sigma_r$} & \multirow{2}{*}{1\%} & S & 95.15 & 96.04 & 97.32 & 98.63 & 99.63 \\ 
        & & D & 96.55 & 97.26 & 98.45 & 99.92 & 101.31 \\ \cline{2-8}
        & \multirow{2}{*}{1.5\%} & S & 95.30 & 96.26 & 97.53 & 98.76 & 99.67 \\ 
        & & D & 97.23 & 97.93 & 99.09 & 100.47 & 101.78 \\ \cline{2-8}
        & \multirow{2}{*}{2\%} & S & 95.52 & 96.53 & 97.79 & 98.95 & 99.78 \\ 
        & & D & 98.16 & 98.86 & 100.00 & 101.35 & 102.58 \\ \cline{2-8}
        & \multirow{2}{*}{2.5\%} & S & 95.79 & 96.85 & 98.08 & 99.18 & 99.93 \\ 
        & & D & 99.37 & 100.09 & 101.18 & 102.49 & 103.65 \\ \cline{2-8}
        & \multirow{2}{*}{3\%} & S & 96.11 & 97.20 & 98.4 & 99.43 & 100.11 \\ 
        & & D & 100.79 & 101.54 & 102.58 & 103.83 & 104.96 \\ \hline
        \end{tabular}
        \begin{tabular}{c|c|c|c|c|c|c|c|}
        \multicolumn{3}{c|}{2015: With} & \multicolumn{5}{c|}{$\sigma_S$} \\ \cline{4-8}
        \multicolumn{3}{c|}{step-up} & 10\% & 15\% & 20\% & 25\% & 30\% \\ \hline
        \multirow{10}{*}{$\sigma_r$} & \multirow{2}{*}{1\%} & S & 96.89 & 99.37 & 102.73 & 106.71 & 111.12 \\ 
        & & D & 97.53 & 100.64 & 104.60 & 109.04 & 113.64 \\ \cline{2-8}
        & \multirow{2}{*}{1.5\%} & S & 96.98 & 99.56 & 102.79 & 106.80 & 111.01 \\ 
        & & D & 98.17 & 101.10 & 104.86 & 109.19 & 113.62 \\ \cline{2-8}
        & \multirow{2}{*}{2\%} & S & 97.09 & 99.70 & 103.03 & 106.78 & 111.11 \\ 
        & & D & 99.10 & 101.92 & 105.61 & 109.87 & 114.21 \\ \cline{2-8}
        & \multirow{2}{*}{2.5\%} & S & 97.36 & 99.90 & 103.13 & 106.70 & 111.21 \\ 
        & & D & 100.33 & 103.09 & 106.72 & 110.91 & 115.18 \\ \cline{2-8}
        & \multirow{2}{*}{3\%} & S & 97.45 & 100.07 & 103.38 & 107.11 & 111.15 \\ 
        & & D & 101.78 & 104.50 & 108.10 & 112.23 & 116.43 \\ \hline
        \end{tabular}}
        \newline
        \vspace*{1 cm}
        \newline
        \resizebox{\textwidth}{!}{\begin{tabular}{c|c|c|c|c|c|c|c|}
        \multicolumn{3}{c|}{2015: Without} & \multicolumn{5}{c|}{$\sigma_r$} \\ \cline{4-8}
        \multicolumn{3}{c|}{step-up} & 1\% & 1.5\% & 2\% & 2.5\% & 3\% \\ \hline
        \multirow{10}{*}{$\rho$} & \multirow{2}{*}{-90\%} & S & 95.66 & 95.71 & 95.69 & 95.66 & 95.67 \\ 
        & & D & 96.81 & 97.48 & 98.27 & 99.36 & 100.69 \\ \cline{2-8}
        & \multirow{2}{*}{-50\%} & S & 96.14 & 96.21 & 96.25 & 96.32 & 96.42 \\ 
        & & D & 97.30 & 97.95 & 98.79 & 99.92 & 101.27 \\ \cline{2-8}
        & \multirow{2}{*}{0} & S & 96.75 & 96.87 & 97.02 & 97.22 & 97.45 \\ 
        & & D & 97.91 & 98.55 & 99.44 & 100.62 & 102.00 \\ \cline{2-8}
        & \multirow{2}{*}{50\%} & S & 97.32 & 97.53 & 97.79 & 98.08 & 98.40 \\ 
        & & D & 98.45 & 99.09 & 100.00 & 101.18 & 102.58 \\ \cline{2-8}
        & \multirow{2}{*}{90\%} & S & 97.76 & 98.08 & 98.41 & 98.75 & 99.10 \\ 
        & & D & 98.80 & 99.52 & 100.42 & 101.59 & 103.00 \\ \hline
        \end{tabular}
        \begin{tabular}{c|c|c|c|c|c|c|c|}
        \multicolumn{3}{c|}{2015: With} & \multicolumn{5}{c|}{$\sigma_r$} \\ \cline{4-8}
        \multicolumn{3}{c|}{step-up} & 1\% & 1.5\% & 2\% & 2.5\% & 3\% \\ \hline
        \multirow{10}{*}{$\rho$} & \multirow{2}{*}{-90\%} & S & 102.40 & 102.35 & 102.40 & 102.15 & 102.24 \\ 
        & & D & 100.79 & 102.56 & 104.15 & 105.77 & 107.47 \\ \cline{2-8}
        & \multirow{2}{*}{-50\%} & S & 102.49 & 102.46 & 102.51 & 102.37 & 102.41 \\ 
        & & D & 101.79 & 103.07 & 104.37 & 105.81 & 107.40 \\ \cline{2-8}
        & \multirow{2}{*}{0} & S & 102.53 & 102.69 & 102.76 & 102.77 & 102.80 \\ 
        & & D & 103.22 & 103.94 & 104.95 & 106.22 & 107.70 \\ \cline{2-8}
        & \multirow{2}{*}{50\%} & S & 102.68 & 102.86 & 103.03 & 103.13 & 103.36 \\ 
        & & D & 104.60 & 104.86 & 105.61 & 106.72 & 108.10 \\ \cline{2-8}
        & \multirow{2}{*}{90\%} & S & 102.83 & 103.01 & 103.25 & 103.39 & 103.71 \\ 
        & & D & 105.96 & 105.91 & 106.43 & 107.39 & 108.68 \\ \hline
        \end{tabular}}
    \caption{Fair value of a GMWB annuity at $t=0$ for different values of $\sigma_S$ and $\sigma_r$ (resp. $\sigma_r$ and $\rho$) in the top (resp. bottom) panel. Rows labelled by S (resp. D) refer to the case of static (resp. dynamic) withdrawals. When static withdrawals are considered, rows labelled by (MC) display the Monte Carlo estimate with 100000 paths and the 95\% confidence interval. $T=10$, $\alpha = 6.60\%$, $\beta = 10\%$ and remaining parameters as in Table \ref{tab:optimalFeesNOMortalityRisk_2015}, as of 12/30/\textbf{2015}. Parameters: $m=2$, $n_A = 30$, $n_B = 10$ for the left panel with no step-up feature, $n_B = 30$ for the right panel  with the step-up feature.}\label{tab:sensitivity2015}
\end{table}

\begin{table}[]
    \centering
    \resizebox{\textwidth}{!}{\begin{tabular}{c|c|c|c|c|c|}
         2015: Without & \multicolumn{5}{c|}{$a$} \\ \cline{2-6}
         step-up & 5\% & 10\% & 15\% & 20\% & 25\% \\ \hline
         S & 97.73 & 97.79 & 97.84 & 97.90 & 97.98 \\
         D & 100.60 & 100.00 & 99.61 & 99.38 & 99.25 \\ \hline
    \end{tabular}
    \begin{tabular}{c|c|c|c|c|c|}
         2015: With & \multicolumn{5}{c|}{$a$} \\ \cline{2-6}
         step-up & 5\% & 10\% & 15\% & 20\% & 25\% \\ \hline
         S & 102.79 & 103.03 & 103.12 & 103.13 & 103.13 \\
         D & 106.17 & 105.61 & 105.26 & 105.06 & 104.94 \\ \hline
    \end{tabular}}   
    \caption{Fair value of a GMWB at $t=0$ for different values of $a$. Parameters as in Table \ref{tab:optimalFeesNOMortalityRisk_2015}.}
    \label{tab:sensitivity_a_2015}
\end{table}

\begin{table}
    \centering
    \begin{tabular}{c|c|c|c|c|c|}
         scenario & $\alpha^*$ & $\beta^*$ & no step-up & step-up only & step-up + bonus  \\ \hline
         2015 & 6.60\% & 10\% & 100.00 & 105.61 & 117.10 \\ \hline
    \end{tabular}
    \caption{Fair value of a GMWB annuity with $b=2.5$, dynamic withdrawals. Parameters: $m=2$, $n_A = n_B = 80$.}
    \label{tab:pricesWithBonus_2015}
\end{table}

\end{document}